\documentclass[useAMS,usenatbib]{mn2e}
\usepackage{graphicx}
\usepackage[hang]{subfigure}
\usepackage{ amssymb }
\usepackage{fmtcount}

\def\Msun{\hbox{$\rm\thinspace M_{\odot}$}}
\def\asec{$^{\prime\prime}$}

\title[$z\sim6$ LBGs: ages, masses and SFRs]{The ages, masses and star-formation rates of spectroscopically confirmed $\bmath{z\sim6}$ galaxies in CANDELS}
\author[E. Curtis-Lake et al.]{E. Curtis-Lake$^{1}$\thanks{Email: efcl@roe.ac.uk},
R.J. McLure$^{1}$, J.S. Dunlop$^{1}$, M. Schenker$^{2}$, A.B. Rogers$^{1}$,
\newauthor T. Targett$^{1}$, M. Cirasuolo$^{1}$, O. Almaini$^{3}$, M.L.N. Ashby$^{4}$, E.J. Bradshaw$^{3}$, 
\newauthor S.L. Finkelstein$^{5}$, M. Dickinson$^{6}$, R.S. Ellis$^{2}$, S.M. Faber$^{7}$, G.G. Fazio$^{4}$, 
\newauthor H.C. Ferguson$^{8}$, A. Fontana$^{9}$, N.A. Grogin$^{8}$, W.G. Hartley$^{3}$, D.D. Kocevski$^{7}$, 
\newauthor A.M. Koekemoer$^{8}$, K. Lai$^{7}$, B.E. Robertson$^{10}$, E. Vanzella$^{11}$, S.P. Willner$^{4}$ \\
$^{1}$ SUPA\thanks{Scottish Universities Physics Alliance},
Institute for Astronomy, University of Edinburgh, 
Royal Observatory, Edinburgh EH9 3HJ\\
$^{2}$ Department of Astronomy, California Institute of Technology, Pasadena, CA 91125, USA\\
$^{3}$ School of Physics \& Astronomy, University of Nottingham, University Park, Nottingham NG7 2RD\\
$^{4}$ Harvard-Smithsonian Center for Astrophysics, 60 Garden Street, Cambridge, MA 02138, USA\\
$^{5}$ Hubble Fellow, The University of Texas at Austin, Austin, TX 78712\\
$^{6}$ National Optical Astronomy Observatory, PO Box 26732, Tucson, AZ 85726, USA\\
$^{7}$ University of California Observatories/Lick Observatory, University of California, Santa Cruz, CA 95064, USA\\
$^{8}$ Space Telescope Science Institute, 3700 San Martin Drive, Baltimore, MD 21218, USA\\
$^{9}$ INAF - Osservatorio Astronomico di Roma, via di Frascati 33, 00040 Monteporzio Catone, Italy\\
$^{10}$ Department of Astronomy, University of Arizona, 933 North Cerry Avenue, Tucson, AZ 85721, USA\\
$^{11}$ INAF Osservatorio Astronomico di Trieste, Via G. B. Tiepolo 11, 34131 Trieste, Italy}
\begin{document}

\date{}

\pagerange{\pageref{firstpage}--\pageref{lastpage}} \pubyear{2002}

\maketitle

\label{firstpage}
\def\Msun{\hbox{$\rm\thinspace M_{\odot}$}}

\vspace*{-2cm}

\begin{abstract}
We report the results of a study exploring the stellar populations of 13
luminous (L $>$ 1.2L$^*$), spectroscopically confirmed, galaxies in the
redshift interval $5.5<z<6.5$, all with {\it Hubble Space Telescope} (HST) WFC3/IR and
{\it Spitzer} IRAC imaging from the HST/CANDELS and {\it Spitzer}/SEDS surveys. 
Based on fitting the observed photometry with galaxy spectral energy distribution (SED) templates covering a
wide range of different star-formation histories, including exponentially
increasing star-formation rates and a self consistent treatment of Lyman-alpha emission,
we find that the derived stellar masses
lie within the range $10^9$ M$_{\odot} < M_* < 10^{10}$ M$_{\odot}$ and
are robust to within a factor of two. In contrast, we confirm previous
reports that the ages of the stellar populations are poorly constrained.
Although the best-fitting models for 3/13 of the sample have ages of
$\gtrsim$ 300 Myr, the degeneracies introduced by dust extinction mean
that only two of these objects actually {\it require} a $\gtrsim$ 300 Myr
old stellar population to reproduce the observed photometry. Moreover,
when considering only smoothly-varying star-formation histories, we
observe a clear tension between the data and models such that a galaxy SED
template with an old age is often chosen in order to try and fit objects
with blue UV-slopes but red UV-to-optical colours. To break this tension
we explore SED fitting with two-component models (burst plus on-going
star-formation), thereby relaxing the requirement that the current star
formation rate and assembled stellar mass must be coupled, and allow for
nebular line+continuum emission. On average, the inclusion of nebular
emission leads to lower stellar-mass estimates (median offset
0.18 dex), moderately higher specific star-formation rates, and allows for
a wider range of plausible stellar ages. However, based on our SED
modelling, we find no strong evidence for extremely young ages in our
sample (i.e. $<50$ Myr). Finally, considering all of the different
star-formation histories explored, we find that the median best-fitting
ages are of the order $\simeq 200-300$ Myr and that the objects with the tightest constraints indicate ages in the range 50-200 Myr.
\end{abstract}


\newpage

\section{Introduction}

A combination of near-Infrared and \textit{Spitzer} \citep{Werner2004} Infrared Array Camera (IRAC; \citealt{Fazio2004}) observations have shown that many high-redshift ($z\geq5$) Lyman-break galaxies (LBGs) have red rest-frame ultraviolet-optical colours (e.g. \citealt{Yan2005,Eyles2005,Ono2010a,Labbe2010}).  If this red colour is taken to indicate the strength of the Balmer break, then old stellar ages ($>$ 300 Myr) are inferred for these objects, indicating a formation redshift at $z>8$.
This property was first reported for two $i-$drop galaxies in \cite{Eyles2005}, with their spectral energy distribution (SED) fitting indicating ages in the range 250-650 Myr, and stellar masses of order 20\% of current-day $L^*$ galaxies.  A later analysis of a larger sample showed that 40\% of their sample displayed evidence for substantial 4000\AA/Balmer spectral breaks \citep{Eyles2007}. More recently, \cite{RichardJohan2011} reported the discovery of a lensed galaxy at $z=6.027$ which also appears to have a strong Balmer-break, consistent with a mature stellar population. If confirmed, a substantial population of galaxies at $z>6$ with mature stellar populations has important consequences for star formation at $z>8$ and its contribution to reionization.

SED-fitting techniques, however, often employ templates that only account for stellar emission (eg. \citealt{Bruzual2003,Maraston2005}) and recently several groups have investigated how the inclusion of nebular emission in the fitting can affect the derived parameters (e.g. \citealt{Robertson2010a,Schaerer2010a,Labbe2010,Ono2010a}). For example, \cite{Labbe2010} reported difficulties in reconciling the observed Balmer break with a very blue ultraviolet (UV) continuum slope ($\beta\sim-3$; $f_{\lambda}\propto \lambda^{\beta}$) observed in a stack of the photometry for the faintest galaxies in a sample of $z\sim7$ LBGs ($H_{160, AB}>27.5$) taken from the Hubble Ultra-deep Field (HUDF) and Early Release Science (ERS) field.  They suggest that episodic star formation may be required to match the colours of these objects (being simultaneously blue in the rest-frame UV and red in colours spanning the 4000\AA/Balmer break), as well as the contribution of nebular emission lines that are not included in the standard population synthesis codes.  An illustration of the potentially dramatic effect that nebular emission can have on derived stellar masses and ages is provided by \cite{Ono2010a}, in their fitting of stacked multi-wavelength observations of Lyman alpha emitters (LAEs) at $z\sim5.7$ and $z\sim6.6$ within the Subaru/ XMM-Newton Deep Survey (SXDS) field.  The derived masses from models with maximum nebular emission (escape fraction of ionising photons, $f_{esc}=0$), are more than an order-of-magnitude lower than those based on only stellar emission (e.g. $M_*\sim3\times10^7\,\mathrm{M}_{\odot}$ compared to $M_*\sim5\times10^8\,\mathrm{M}_{\odot}$ for the $z\sim5.7$ stack) and the derived ages are also much younger ($\sim3$ Myr compared to $\sim300$ Myr). 

Recent results from cosmological hydrodynamic simulations of reionisation predict that Lyman-break selected galaxies have young ages, low intrinsic reddening and sub-solar metallicities at high redshifts (e.g. $\sim50-150$ Myr at $6<z<8$; \citealt{Finlator2010}) with the distribution broadening to older ages at later times (e.g. 200-600 Myr at $z\sim4$; \citealt{Finlator2006}).  The results from \cite{Dayal2009}, in which Ly$\alpha$ emission is incorporated into a cosmological smoothed-particle hydrodynamic (SPH) simulation, also support these predictions.  By matching models to observed Ly$\alpha$ and 
UV-luminosity functions at $5.7<z<7.6$, the brightest LAEs, which are shown in their simulation to be a subset of the LBG 
population \citep{DayalPratika2011}, are expected to have intermediate ages ($\sim200$ Myr), with the spread in age increasing to lower luminosities (ages from a few Myr to 300 Myr).

Given that it is difficult to constrain the star-formation history (SFH) of high-redshift galaxies from the available photometric data, many studies investigating the physical properties of high-redshift galaxies use small template sets (including restricted ranges in reddening and metallicity) when SED fitting to avoid introducing too many degeneracies. In this way, if the template sets are consistent, the results for samples at different redshifts can be compared to one another to search for any evolution in the typical galaxy properties.  The study by \cite{Yabe2009} illustrates the importance of matching the template sets when comparing the physical properties of their $z\sim5$ LBGs to similar samples at different redshifts, where differences between the template sets affect the distributions of derived star-formation rates (SFRs) and ages by up to a factor of 10.  They do show, however, that derived masses are fairly insensitive to changes in template SFHs.  

This approach is limited in its usefulness, however, if the galaxies evolve in a parameter space that is not allowed to vary in the template set, such as in dust content, metallicity or even in the SFHs.  Another limitation is in the search for trends in physical properties within the individual samples (eg. SFR vs. stellar mass), as any apparent trends may be imposed by the template set itself, as discussed by \cite{McLure2011}.  When comparing to other galaxy populations, for example to determine how high-redshift LBGs are linked to lower redshift populations, it is clearly desirable to have robust parameter determination that is independent of the adopted fitting method.

Tracking the evolution of these trends gives us the best indication of the typical mass-assembly route for these galaxies.  In particular, tracking the SFR (or UV luminosity) of galaxies of a given mass over time can rule out certain SFHs.  \cite{Stark2009} tracked the evolution of the physical properties of LBGs from $z\sim6$ to $z\sim4$, finding little evolution in the stellar mass-UV luminosity relation  ($M_*-L_{\mathrm{UV}}$) over this redshift range.  They suggested that this is best explained by the different redshift bins being populated by distinct populations with short periods of star formation, as opposed to the population remaining in the LBG phase, constantly forming stars and building up in stellar mass over the entire redshift range.  This view is also supported by the clustering analysis of star-forming galaxies performed by \cite{Lee2009a},  which suggested a typical duration of star formation $<400$ Myr ($1\sigma$) at $z\sim4$ and $z\sim5$.  

The alternative explanation, however, is that these galaxies exhibit rising SFHs, a feature predicted by the models of \cite{Finlator2010} and supported by the observation that high-redshift galaxies selected to have constant number density display UV luminosities which increase with time \citep{Papovich2010a}.  Moreover, an exponentially increasing SFR would naturally produce little evolution in the observed specific SFR (sSFR), which current data suggests remains fairly constant at $\sim$2.5 Gyr$^{-1}$ from $z\sim2.5-7$ \citep{Gonzalez2011,McLure2011}.  However, these observations provide a tension with current models of galaxy formation and evolution which predict sSFRs that continue to rise with increasing redshift proportional to $(1+z)^{2.25}$ \citep{Neistein2008,Dave2011}.

A different approach to determining the physical properties by SED fitting is used by \cite{Schaerer2010a}, who allow for a large range of SFHs, dust extinction, metallicity and nebular emission.  They show that a wide range of ages and SFRs are acceptable within the confidence contours of their fits, although they also have the added complications of uncertainty in photometric redshift and Ly$\alpha$ flux  inherent to photometrically-selected dropout samples compiled from the literature.  

The new \textit{Hubble Space Telescope} (HST) WFC3 data taken as part of the Cosmic Assembly Near-infrared (IR) Deep Survey (CANDELS, \citealt{GroginNormanA.2011,KoekemoerAntonM.2011}), combined with mid-IR data from the \textit{Spitzer} Extended Deep Survey (SEDS) allows us to place tighter constraints on the rest-frame UV and optical fluxes of high-redshift galaxies.  Motivated by the newly-available data, in this paper we explore the potential to accurately constrain the physical properties of high-redshift galaxies based on a sample of spectroscopically confirmed LBGs in the redshift range $5.5<z<6.5$.

The structure of the paper is as follows.  In section 2 we describe the sample as well as the determination of the Ly$\alpha$ equivalent widths (EWs) from the available spectroscopy. In Section 3 we outline the photometric data used to construct the sample SEDs, as well as the deconfusion of the IRAC images.  In Section 4 the observed characteristics of the objects are examined and compared to a more complete photometrically selected sample.  The SED fitting method is outlined in Section 5 along with a summary of the different template sets used.  Section 6 gives the results from fitting with the smoothly varying SFHs starting from more restricted template sets to compare results directly to the literature, then with a wider range of models and parameters.  Section 7 details the results of fitting with simple two-component templates both with and without nebular emission.  Section 8 provides a discussion of the derived physical parameters and our conclusions are presented in Section 9.  Throughout this paper we assume a cosmology with $H_0=70$ km s$^{-1}$Mpc$^{-1}$, $\Omega_m=0.3$, $\Omega_{\Lambda}=0.7$.  All magnitudes are quoted in the AB system \citep{Oke1983}.

\section{Spectroscopic data}

\subsection{Spectroscopic samples in the UDS and GOODS-S fields}

The properties of the 13 LBGs used in this paper are summarised in Table~\ref{GOODSobjects}.  The fundamental criteria for inclusion in the sample were that LBG-selected objects had $z_{spec}>5.5$ and m$_z<26.5$ and were covered by both WFC3 near-IR and deep {\it Spitzer} 3.6$\mu$m and 4.5$\mu$m imaging.  Objects were drawn from separate two sources.

Eleven of the objects are taken from the spectroscopic sample of \cite{Vanzella2009} within the GOODS-S field, one of the five fields covered by CANDELS.  This particular study is chosen because it currently provides the most complete sample of spectroscopically targeted LBGs at $z\sim6$ with spectra of high enough quality to determine Ly$\alpha$ properties.  

The objects were selected by requiring $z_{spec}>5.5$, m$_{z}<26.5$ as well as requiring $>5\sigma$ detections in the WFC3 $H-$band and a detection in both the IRAC 3.6$\mu$m and 4.5$\mu$m bands.  The first two criteria allowed us to select UV-luminous LBGs in a narrow redshift range while the IRAC detections ensured that the SED could be sampled red-wards of the 4000\AA/Balmer-break, essential for constraining masses.  This selection left us with 4 objects selected from the $V_{606}$-band dropout sample, and 7 from the $i_{775}$-band dropouts.  One object meeting the above requirements was rejected due to it's proximity to a nearby star that would make deconfusion of the IRAC bands incredibly uncertain.  Furthermore, three objects that were bright enough in $z_{850}$ were not bright enough in the $H-$band for use as reliable priors in the IRAC deconfusion, and so were also left out of the sample along with one final object that lacked a detection at 4.5$\mu$m.

We also required the spectroscopic redshift quality flag to be A or B, although after visual inspection, two absorption systems were included which had a quality flag of C in \cite{Vanzella2009}.  All but one of the absorption systems with $z_{spec}>5.5$ were assigned a quality flag of C by \cite{Vanzella2009}, given the difficulty of unambiguously assigning the observed continuum break to Ly$\alpha$ absorption.  However, in the interest of examining the properties of a sample of $z\sim6$ objects covering a full span in intrinsic properties, we have kept them in our sample.  We note that the high quality WFC3 near-infrared plus {\it Spitzer} IRAC data provide photometric redshifts for these objects which are in excellent agreement with the spectroscopic redshifts, indicating that they are robust.

This sample is supplemented by two objects, taken from a sample of spectroscopically confirmed $z>6$ LBGs in the Ultra Deep Survey (UDS) as presented in \cite{Curtis-Lake2011}, which also fall within the area targeted by CANDELS.  The objects' parent sample of LBGs was selected using a photometric-redshift analysis (see \citealt{McLure2009} for details) and the spectroscopic sample is described in \cite{Curtis-Lake2011}.

\begin{table*}
 \centering
  \caption{Full sample of objects for SED fitting in the GOODS-S and UDS fields taken from Vanzella et al. (2009) and Curtis-Lake et al. (2012) respectively. The first column provides the ID used to identify the objects throughout this work.  The second column gives the GOODS/UDS ID incorporating the J2000 coordinates.  Columns 3-5 give the spectroscopic redshift, object class (emission line (em) or absorption (abs) system) and redshift quality flag (A-C, see text for details).  Columns 6 and 7 give the total $z-$band apparent magnitude along with the luminosity in units of L$^*$ (taken to be $(6.3\pm0.7)\times10^{40}$ erg s$^{-1}$ \AA$^{-1}$ from the UV luminosity function at $z\sim6$, McLure et al. 2009).  Column 8 gives the paper reference; Vanzella et al. 2009 (V09) or Curtis-Lake et al. 2012 (CL12).  The final column gives the broad-band calibrated EW (see Section 2.3 for details) for each of the emission line objects where there was visible continuum red-wards of Ly$\alpha$ in the spectrum.}
  \label{GOODSobjects}
  \begin{tabular}{@{}llccccccc@{}}
  \hline
  \hline
  Our ID & GOODS/UDS ID  & $z_{spec}$ & Class & Quality & m$_z$    & L (/ L*) & Reference & EW$_{rest}$ (/\AA)\\
 \hline
  2   & J033215.90-274123.9        & 5.57  & em  & B & 25.84 & 1.9 & V09  & \phantom{0}6.9 $\pm$ 1.1\phantom{0}\\
  5   & J033225.61-275548.7        & 5.79  & em  & A & 24.58 & 6.4 & V09  & \phantom{0}6.5 $\pm$ 1.1\phantom{0}\\
  6   & J033246.04-274929.7        & 5.79  & em  & A & 25.98 & 1.8 & V09  &           31.0 $\pm$ 4.0\phantom{0}\\
  8   & J033240:01-274815.0        & 5.83  & em  & A & 25.23 & 3.6 & V09  &           14.8 $\pm$ 2.0\phantom{0}\\
  13  & J033228.19-274818.7        & 5.94  & em  & B & 26.43 & 1.2 & V09  &           24.8 $\pm$ 4.9\phantom{0}\\
  18  & J033224.80-274758.8        & 5.99  & em  & B & 26.10 & 1.7 & V09  &                                \--- \\
  23  & J033222.28-275257.2        & 6.20  & abs & C & 26.17 & 1.6 & V09  &                                n/a\\
  24  & J033224.40-275009.9        & 5.50  & abs & C & 25.67 & 2.2 & V09  &                                n/a\\
  25  & J033237.63-275022.4        & 5.52  & em  & A & 26.03 & 1.6 & V09  & \phantom{0}8.2 $\pm$ 1.5\phantom{0}\\
  26  & J033211.93-274157.1        & 5.58  & em  & B & 26.35 & 1.2 & V09  &                                \---\\
  27  & J033245.23-274909.9        & 5.58  & em  & B & 25.94 & 1.7 & V09  &           31.6 $\pm$ 7.0\phantom{0}\\
  157 & UUDS\_J021800.90-051137.8  & 6.03  & em  & A & 24.85 & 5.3 & CL12 &           54.7 $\pm$ 11.3\\
  248 & UUDS\_J021735.34-051032.6  & 6.12  & em  & A & 25.13 & 4.2 & CL12 &           46.8 $\pm$ 8.4\phantom{0}\\           
\hline
\hline
\end{tabular}
\end{table*}

\subsection{Ly$\bmath{\alpha}$ properties}

We measure the Ly$\alpha$ EW for each object from the spectra.  The main aim is to determine a robust estimate that can be used to add the Ly$\alpha$ flux to the models before SED fitting.  As such we chose a method that links the EW to the multi-wavelength photometry, using an approach similar to that described in \cite{Curtis-Lake2011}, to which the reader is referred for a fuller explanation.  Essentially, the spectrum is first flux-calibrated by integrating over the filter profile that encompasses Ly$\alpha$ and scaling it to match the photometry in that band.  To prevent sky-line residuals dominating the integrated filter flux, the continuum is modelled by a simple step function with the continuum estimated from clean regions of the spectrum red-wards of Ly$\alpha$.  The error in the continuum estimate is folded into the final error on the derived EW.  

The continuum at Ly$\alpha$ is then estimated using either Subaru NB921 narrow-band imaging \citep{SobralDavid2011}, providing photometry just red-wards of Ly$\alpha$ for the two objects in the UDS, or HST WFC3 F125W imaging for the objects in the GOODS-S, assuming a flat slope in $F_{\nu}$ ($F_{\lambda}\propto\lambda^{-2}$) in each case.  Although, in principle, Ly$\alpha$ EWs can be derived directly from spectra without the need to flux calibrate, the S/N in the measured continuum is low for these faint objects, and linking the measured EW with the photometry allows us to reliably add Ly$\alpha$ to the models in a self-consistent fashion.

Two of the objects from the \cite{Vanzella2009} sample (IDs 18 and 26) show no visible continuum red-ward of Ly$\alpha$ and as a consequence any attempt at flux calibration is very uncertain.  For these objects the filters that have Ly$\alpha$ contributing to the filter flux are not used in the SED fitting.

\section{Photometric data}
In this section we provide a brief description of the various optical, near-IR and mid-IR imaging datasets which were used to construct the SEDs of our sample of $z~\simeq~6$ LBGs.  In all cases the optical and near-IR HST photometry was measured using 0.6\asec$-$diameter circular apertures, with the WFC3/IR data in the $F125W (J_{125})$ and $F160W (H_{160})$ filters corrected for missing flux relative to the optical ACS bands. The ground-based photometry was measured in circular apertures with diameters of either 2.0\asec\,(Subaru and UKIDSS) or 1.2\asec\,(HAWK-I) depending on seeing (FWHM $\simeq 0.8$\asec and $\simeq 0.5$\asec respectively), in order to match the enclosed flux of the $H_{160}$ photometry. Obtaining accurate photometry from the confused Spitzer IRAC imaging was a 
more complicated process and is discussed further below. A full list of the photometry adopted during the SED fitting analysis of each individual object can be found in Tables 1 and 2 of the Appendix.

\subsection{Optical imaging}
For the eleven objects which originate from the Vanzella et al. (2009) sample, the optical photometry was measured from the publicly available HST ACS imaging of the GOODS-S field in the $F435W, F606W, F775W$ and $F850LP$ filters (GOODSv2.0; Giavalisco et al. 2004). For the two objects in the UKIDSS UDS field the optical photometry was measured from the deep Subaru imaging of the UDS in the $BVRi^{\prime}z^{\prime}$ filters (Furusawa et al. 2008).

\subsection{Near-IR imaging}
The primary near-IR photometry was measured from the publicly-available CANDELS imaging of the GOODS-S and UDS fields with WFC3/IR in the $J_{125}$ and $H_{160}$ filters (Grogin et al. 2011; Koekemoer et al. 2011). In addition, ground-based $K-$band photometry was measured for the two UDS objects using data release 8 (DR8) of the UKIDSS UDS and $K-$band photometry for 7/11 of the GOODS-S objects was measured from the first epoch of the HAWK-I imaging programme of the CANDELS GOODS-S and UDS fields (HANDELS; P.I. A. Fontana). 

\subsection{Spitzer IRAC imaging}
A key element of this study is the crucial information on the LBG stellar masses supplied by the available Spitzer IRAC data. For each of the LBGs the photometry at $3.6\mu$m and $4.5\mu$m was measured from either the publicly available reductions (v3.0) of the IRAC imaging obtained as part of the GOODS survey (proposal ID 194, Dickinson et al., in preparation), or the Spitzer Extended Deep Survey (SEDS; Ashby et al. 2012, in prep.) imaging of the UDS.

\subsubsection{Deconfusion analysis}
Due to the depth of the available IRAC imaging, and the comparatively broad IRAC PSF (FWHM $\simeq 1.7$\asec), the $3.6\mu$m and $4.5\mu$m data in the GOODS-S and UDS CANDELS fields is heavily confused. As a result, it is necessary to employ some form of deconfusion analysis to allow accurate photometry to be extracted beyond the natural confusion limit. 

In this study we employ the deconfusion software developed by McLure et al. (2011), which exploits the high spatial resolution WFC3/IR $H_{160}$ imaging to provide normalized templates of each object in the field and then, via a transfer function, produces a synthetic IRAC image. Through a matrix inversion procedure, the code then simultaneously fits the amplitude (or total flux) of each template in order to produce the optimal reproduction of the observed IRAC image. Based on the best-fitting model it is then possible to extract photometry at $3.6\mu$m and $4.5\mu$m in circular 0.6\asec$-$diameter apertures, matched to the WFC3/IR photometry in the $H_{160}$ filter. Full details of the deconfusion algorithm are provided in Appendix A of McLure et al. (2011).

\section{Observed colours and sample bias}

\begin{figure*}
  \centering
  \subfigure{\includegraphics[width=2.1in,trim=2cm 13cm 2cm 5cm,clip]{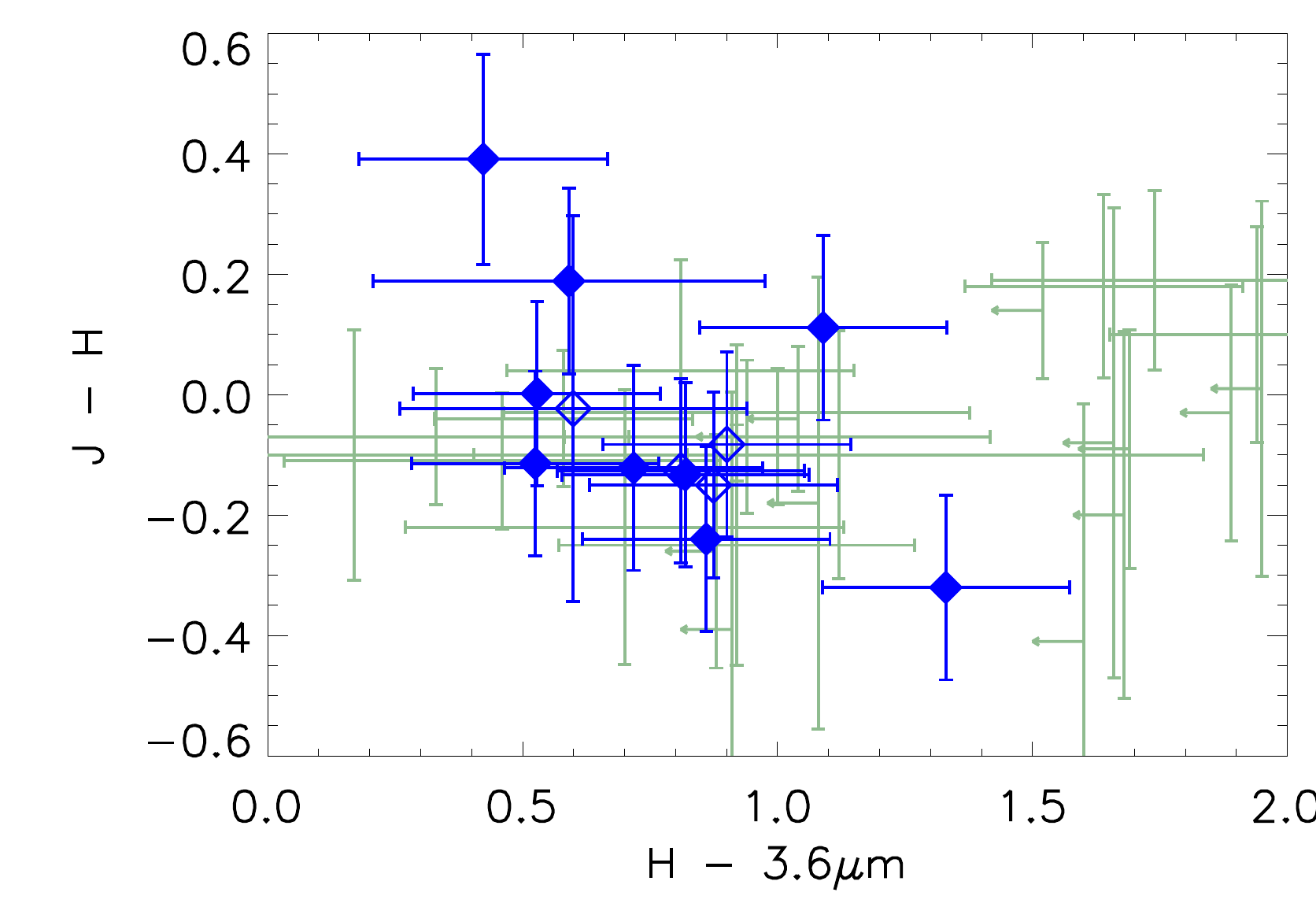}}
  \subfigure{\includegraphics[width=2.1in,trim=2cm 13cm 2cm 5cm,clip]{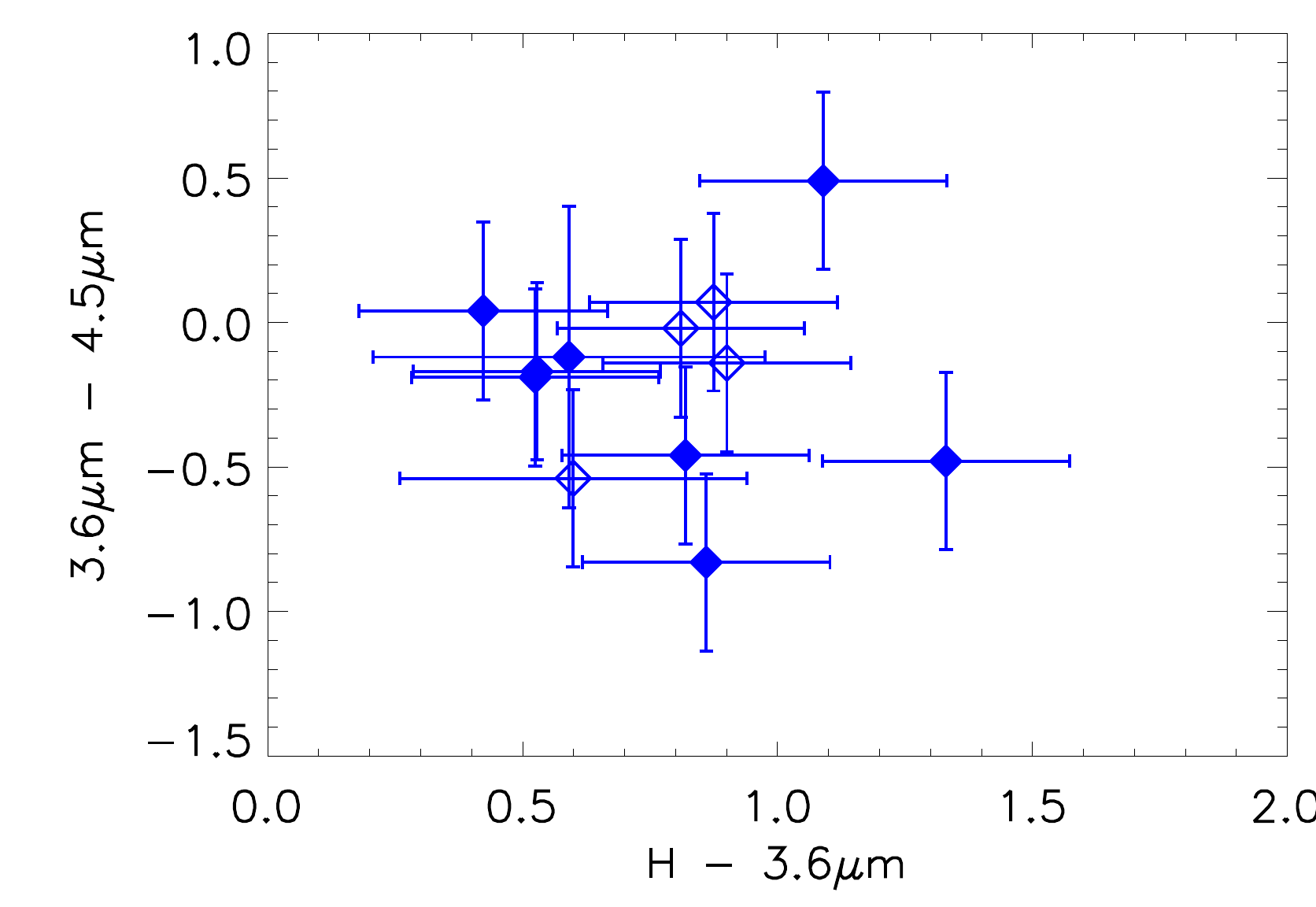}}
  \subfigure{\includegraphics[width=2.1in,trim=2cm 13cm 2cm 5cm,clip]{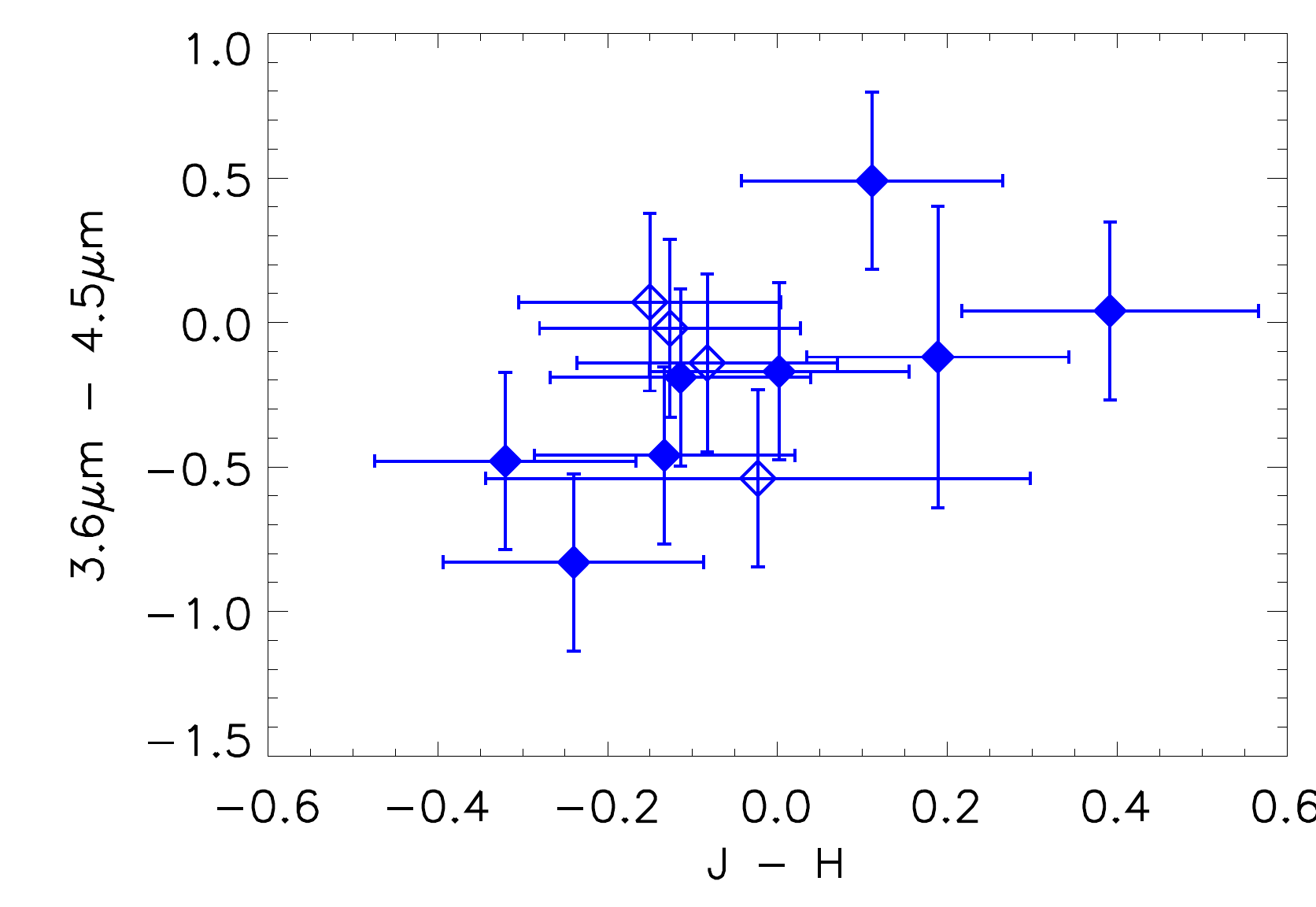}}
  \caption{The colour-colour space spanned by the sample, in which the rest-frame near-UV slope is described by the $J-H$ colour, $H-3.6\mu$m colours span the Balmer-break region and 3.6 $-$ 4.5$\mu$m colours describe the slope in the rest-frame optical.  The filled diamonds are the objects that were originally confused in the IRAC images and the open symbols show the objects that are relatively isolated.  A photometrically-selected $6.0 < z_{phot} < 6.5$ sample of LBGs taken from McLure et al. (2011) is plotted in green in the left-hand panel.  Objects from this sample without detections at 3.6$\mu$m have their $H-3.6\mu$m colours plotted as upper limits.}
  \label{fig:obsColours}
\end{figure*}

The available photometry samples three key regions of the object SEDs; the rest-frame UV, the rest-frame optical and the 4000\AA/Balmer-break region (which is bracketed by the near-IR and {\it Spitzer} observations).  We plot the information we have on these regions in Fig.~\ref{fig:obsColours} using observed colours, with the observed rest-frame near-UV slope being described by $J-H$, the Balmer-break region (the rest-frame UV-to-optical colours) bracketed by $H-3.6\mu$m, and the slope in the rest-frame optical by 3.6$\mu$m $-$ 4.5$\mu$m.  The $J-H$ colour does not give an accurate measurement of $\beta$ in this context, as the filters sample the near-UV red-wards of the region from which $\beta$ 
is estimated (i.e. $\lambda_{\rm rest}\simeq 1500$\AA).

Fig.~\ref{fig:obsColours} (b) shows that the IRAC 3.6$\mu$m $-$ 4.5$\mu$m colours extend to quite blue colours and that this does not seem to be dependent on the colours spanning the Balmer-break.  In fig.~\ref{fig:obsColours} (c), however, we see a correlation between $J-H$ and 3.6$\mu$m $-$ 4.5$\mu$m.  Discussion of the physical interpretation of these trends is deferred to Section 8.1.

Fig.~\ref{fig:obsColours} (a) suggests a weak anti-correlation between $J-H$ and $H-3.6\mu$m.  Also plotted is a more complete $z\sim6$ sample (plotted in green, taken from \citealt{McLure2011}).  This LBG sample was selected by a photometric redshift technique within the HUDF and ERS fields and the plotted objects are those with $6.0<z_{phot}<6.5$.  We can see that the colours of the underlying sample occupy a similar colour-colour space to the majority of our sample, although the reddest $J-H$ colours are mostly found in objects which also have very red $H-3.6\mu$m colours.  A positive correlation between these colours would be expected if dust were reddening the SEDs, and the lack of red $H-3.6\mu$m colours in our sample would suggest that the spectroscopically confirmed objects, that in the most part show Ly$\alpha$ in emission, are relatively dust free.

\section{SED fitting - method}

\begin{table*}
 \centering
  \caption{Summary of the different template sets used in the SED fitting.  The first column gives the label used in the text to refer to the given template set, the second column gives a short description of the template set. Column 3 details the models included and columns 4-5 give the range of metallicities and extinction explored in each set.  The final column displays whether or not the template set includes nebular emission.}
  \scalebox{0.85}{
  \begin{tabular}{@{}lllccc@{}}
  \hline
  \hline
  Label & Description & Templates used & Metallicities & E(B-V) & Nebular\\
        &             &                & (/Z$_{\odot}$) &        & emission (Y/N)\\
  \hline
  A & Restricted template set & Constant SFR           & 0.2 & 0 & N\\
    &                         & Burst\\
    &                         & $\tau =$ 100, 500 Myr\\  
  \hline
  B & For comparison to       & Constant SFR           & 1   & 0 & N\\ 
    & Eyles et al. (2007)     & Burst\\
    &                         & $\tau =$ 50, 100, 200, 500 Myr, 1 Gyr\\
  \hline
  C & All smoothly varying SFHs & Constant SFR & 0.02, 0.2, 0.4, 1                 & 150 values from& N\\
    & plus a burst               & Burst        &                                   & 0 - 0.5\\
    &                            & $\tau$ = 50, 100, 200, 500 Myr, 1, 2, 5, 10 Gyr& & in steps of 0.003\\
    &                            & Exponentially increasing SFR, \\
    &                            & SSFR = 2, 4, 6, 8, 10, 12, 14, 16, 18, 20 Gyr$^{-1}$\\
  \hline
  D & Burst + recent constant    & Underlying burst, mass fraction & 0.02, 0.2, 0.4, 1       & 13 values from & N \\
    & star formation             & 0 - 15\% (0-2\% in steps of 0.1\%, 3-15\% in steps of 1\%) &  & 0 - 0.495 \\
    &                            & PLUS recent constant star formation                          & & in steps of 0.038\\
    &                            & with duration = 10, 30, 100 Myr\\
  \hline
  E & Burst + recent constant    &  Underlying burst, mass fraction& 0.02, 0.2, 0.4, 1       & 13 values from & Y \\
    & star formation with        &  0 - 15\% (0-2\% in steps of 0.1, 3-15\% in steps of 1)  && 0 - 0.495      & \\
    & nebular emission added     & PLUS recent constant star formation                          & & in steps of 0.038\\
    &                            & with duration = 10, 30, 100 Myr\\
  \hline
  \hline
\end{tabular}}
\label{tab:templates}
\end{table*}

We fit a range of star formation histories (SFHs) to the individual object SEDs using the LePhare photometric redshift code \citep{Ilbert2006}.  These SFHs include exponentially-decreasing star formation rates ($\tau$ models), constant star formation, instantaneous burst, exponentially-increasing (EI) star formation as well as two component models with an old burst plus ongoing constant star formation with and without full nebular emission.  The different template sets employed throughout this paper are summarised in Table~\ref{tab:templates} and all templates are built using the \cite{Bruzual2003} stellar evolutionary models.

The full set of smoothly varying SFHs (template set C) consists of $\tau$ models ($\tau$ = 50, 100, 200, 500 Myr, 1, 2, 5, 10 Gyr), instantaneous burst, constant SFR and exponentially increasing (EI) SFR models (with sSFR = 2, 4, 6, 8, 10, 12, 14, 16, 18, 20 Gyr$^{-1}$ at all ages, assuming that the stellar mass at any time is the integral under the past SFH).  Ages were allowed to vary between 10 Myr to the age of the Universe at the redshift of each object, and metallicities ranging from solar (Z$_{\odot}$) to 1/50th solar (0.02 Z$_{\odot}$) were included.  Extinction by dust was applied using the \cite{Calzetti2000} dust law for starburst galaxies in 150 steps of $E(B-V)$ varying from 0 to 0.5.

The EI models are made employing an initial seed mass with the specific star-formation rate then being constant with time.  The parameterisation, however, assumes the stellar mass at any time is the integral under the past star-formation history. As the stellar mass, in fact, decreases fairly rapidly after a burst of star formation, this assumption leads to instantaneous sSFRs somewhat higher than the parameterised value. Moreover, for low sSFR models with young ages, the seed mass can account for a significant fraction of the mass in the model and the resultant SED is indistinguishable from a large burst followed by constant star formation.  

The two restricted template sets (A and B) use subsets of the full range of smoothly-varying SFHs to test the effect of various assumptions on the derived physical parameters and to check consistency with previous studies.  In particular template set B includes models consistent with those adopted by \cite{Eyles2005,Eyles2007}, who provided some of the strongest evidence to date for old stellar populations at high redshift.

The burst+constant star formation rate models (set D) were produced in a similar manner to the method applied by \cite{Eyles2005,Eyles2007}, with the same range of metallicities as the smoothly varying SFHs, and star-forming fractions varying from 0-15\% by mass in stars.  This upper limit in the mass of the star-forming component was chosen to provide a minimum flux contribution to the 3.6$\mu$m band of 10\%.  The star-forming fractions were characterised by constant star formation of durations 10, 30 and 100 Myr.  The age of the burst varied between the age of the duration of star-formation for that model, and the age of the Universe at the redshift of each object. Extinction was applied using a coarser grid, with 13 steps in $E(B-V)$ varying from 0 to 0.495.

The final template set (E) adds nebular emission, based on the \cite{Robertson2010a} model, to the burst+constant SFR models (D) according to the prescription outlined in \cite{McLure2011}. Intergalactic medium (IGM) absorption short-ward of Ly$\alpha$ is applied according to the \cite{Madau1995} prescription, and a \cite{Chabrier2003} IMF is assumed in all cases.

Lyman-$\alpha$ emission is added to each of the models using the measured EW values, using the same filter to provide the UV continuum estimate from the models as used for the original EW measurements.  The Ly$\alpha$ emission is added to the models after extinction is applied and after IGM absorption, to ensure that it is comparable to our measured value.  It is imperative to add the Ly$\alpha$ flux if fitting to the filter containing Ly$\alpha$ emission because, with the typical EW for this sample ($\sim25$\AA), it can alter the photometry by 0.2-0.5 magnitudes depending on the exact redshift.  For the two objects without reliable Ly$\alpha$ EW measurements (18 and 26), any filters with a Ly$\alpha$ flux contribution are discounted from the SED fitting.

\section{SED fitting - smoothly varying star formation histories (template sets A-C)}

\subsection{Restricted model set (template set A)}

We begin the process of investigating our ability to constrain the physical properties of the high-redshift sample by starting with the most restricted set of models, as is commonly done in the literature (eg. \citealt{Stark2009,Gonzalez2010}).  This model set is chosen taking the view that there is little evidence for much intrinsic reddening in LBGs, as the average UV continuum slopes that are observed are quite blue \citep{Bouwens2009,McLure2011,Dunlop2011}. Consequently, we begin our model fitting by restricting 
our model set to a single sub-solar metallicity (1/5th solar) with zero reddening, and consider only constant SFR, burst and $\tau$ models with $\tau=100$ and 500 Myr.  

\begin{table*}
 \centering
  \caption{Results from SED-fitting using a restricted template set (template set A) which features a single metallicity, no dust reddening and either constant, burst or $\tau=100$, 500 Myr models.  The first column gives the object ID, columns 2-4 give the best-fitting stellar mass, SFR and age respectively.  The mass and SFRs have been scaled to total using the appropriate H-band aperture corrections. The errors quoted are the 68\% confidence limits marginalised over all other parameters. The final three columns give the best-fitting model, $\chi^2$ and the number of filters used in the fit.  The typical number of degrees of freedom in the fitting for 9 photometric points is 6.}
  \scalebox{0.9}{
  \begin{tabular}{@{}lcccccc@{}}
  \hline
  \hline
  ID & log$_{10}$(M$_*$ /M$_{\odot}$) & SFR & log$_{10}$(Age /Gyr) & Model & $\chi^2$ & filters\\
   &                               & (/M$_{\odot}$ yr$^{-1}$) &  \\
\hline
       2 & \phantom{0}9.55$^{+0.05}_{-0.10}$ & \phantom{0}0.00$^{+0.00}_{-0.00}$ & 7.86$^{+0.05}_{-0.10}$ & burst & \phantom{0}3.60 & 8\\
       5 & \phantom{0}9.74$^{+0.22}_{-0.08}$ & 14.93$^{+4.85}_{-0.91}$ & 8.26$^{+0.60}_{-0.05}$ & $\tau=100$ Myr & \phantom{0}2.28 & 8\\
       6 & \phantom{0}9.45$^{+0.03}_{-0.31}$ & \phantom{0}4.70$^{+0.19}_{-1.49}$ & 8.96$^{+0.00}_{-0.70}$ & const & \phantom{0}3.54 & 7\\
       8 & \phantom{0}9.87$^{+0.10}_{-0.10}$ &           12.59$^{+0.08}_{-1.61}$ & 8.96$^{+0.00}_{-0.30}$ & const & \phantom{0}8.83 & 9\\
      13 & \phantom{0}9.35$^{+0.05}_{-0.10}$ & \phantom{0}0.00$^{+0.00}_{-0.00}$ & 7.91$^{+0.05}_{-0.10}$ & burst & \phantom{0}6.01 & 9\\
      18 & \phantom{0}8.99$^{+0.22}_{-0.24}$ & \phantom{0}2.08$^{+1.19}_{-2.08}$ & 8.31$^{+0.65}_{-0.75}$ & $\tau=100$ Myr & \phantom{0}1.43 & 8\\
      23 & \phantom{0}9.82$^{+0.10}_{-0.06}$ & \phantom{0}0.00$^{+7.33}_{-0.00}$ & 7.96$^{+0.50}_{-0.05}$ & burst & \phantom{0}1.89 & 9\\
      24 & 10.25$^{+0.03}_{-0.03}$ & \phantom{0}0.00$^{+0.00}_{-0.00}$ & 8.31$^{+0.00}_{-0.00}$ & burst & \phantom{0}9.27 & 9\\
      25 & \phantom{0}9.62$^{+0.29}_{-0.05}$ & \phantom{0}0.00$^{+7.77}_{-0.00}$ & 7.91$^{+0.95}_{-0.05}$ & burst & \phantom{0}3.84 & 9\\
      26 & \phantom{0}9.34$^{+0.11}_{-0.11}$ & \phantom{0}0.00$^{+0.00}_{-0.00}$ & 7.96$^{+0.10}_{-0.10}$ & burst & \phantom{0}4.06 & 7\\
      27 & \phantom{0}9.20$^{+0.20}_{-0.09}$ & \phantom{0}0.00$^{+5.88}_{-0.00}$ & 7.68$^{+0.93}_{-0.09}$ & burst & \phantom{0}1.25 & 9\\
     157 & \phantom{0}9.37$^{+0.16}_{-0.20}$ & 14.06$^{+1.15}_{-2.53}$ & 8.36$^{+0.20}_{-0.40}$ & const & 10.65 & 10\\
     248 & \phantom{0}9.89$^{+0.08}_{-0.14}$ & \phantom{0}9.07$^{+0.37}_{-0.22}$ & 8.81$^{+0.05}_{-0.10}$ & $\tau=500$ Myr & \phantom{0}6.27 & 10\\

  \hline
  \hline
\end{tabular}}
\label{tab:restricted}
\end{table*}

Table~\ref{tab:restricted} shows the results from fitting with this model set, with the errors determined from $\Delta\chi^2=$1, giving 68\% confidence limits marginalising over all other parameters if the degeneracies in the model set are dominating these uncertainties.  For the errors in mass and SFR, a minimum error is determined from the range of normalisations of the best-fitting models that give $\Delta\chi^2<$1.

First, we see that with this restricted template set all of the objects are assigned acceptable fits at the $2\sigma$ level ($\chi^2\lesssim14$ (12.5, 11 or 9.5) for 10 (9, 8, or 7) photometric points allowed a free choice of template, normalisation and age).  

Many of the objects are well fit by a single burst, in which case, if they have blue UV-slopes, the best-fit ages are quite low.  This is because when objects with current star-formation are fit by a burst model, the age is determined by the rest-frame UV and, in the absence of star-formation in the templates, the UV must be fit by the young stars still present in a very young burst.  The youngest stars die off over the first tens of Myr, building up a deep Balmer-break over a short amount of time.  This allows the observed red $H-3.6\mu$m colours to also be fit with young ages. Those objects that are best-fit by a burst but seem to show zero error in the SFR have no fits by constant or $\tau$ models within $\Delta\chi^2=1$.  

We see that the masses and SFRs are reasonably well defined, although some show improbably small errors, due to sparse sampling of the physical properties by this restricted model set.  Masses are constrained to within a factor of $\sim2$ and SFRs seemingly better than this, although many of the best-fitting models are burst templates which, by definition, have no on-going star formation. The ages vary between $\sim100$ Myr to $\sim1$ Gyr but the confidence intervals indicate that the ages are typically constrained to within a few hundred Myr.    

The results from this restricted model set therefore indicates that roughly a third of our sample harbour old stellar populations ($>500$ Myr), and that the SFRs and masses appear fairly well constrained and are of the order 0-20 M$_{\odot}$ yr$^{-1}$ and $M_{\star}=10^9 - 10^{10}$ M$_{\odot}$, respectively. 

\subsection{Previous analysis (template set B)}

The ages derived in the previous section agree well with those reported by \cite{Eyles2007} and four galaxies from of our sample were analysed in that work.  Here we compare our results directly to their work, using our improved photometry but a similar template set.  Table~\ref{tab:EylesComparison} shows the comparison between the results of our fitting analysis to theirs, restricting our template set to be representative of that used to derive the parameters published in Table 3 of Eyles et al. (2007); solar metallicity, no dust reddening, tau models with $\tau<1$ Gyr, an instantaneous burst and a constant SFR model.  \cite{Eyles2007} use a Salpeter IMF, so here we report their masses and SFRs re-scaled to the Chabrier IMF.

\begin{table*}
  \centering
  \caption{A comparison of the results of Eyles et al. (2007) and best-fit parameters recovered here using a restricted template set (B), which was chosen to be consistent with the templates used in that work.  Columns 1-5 list the ID, best-fitting SFH, stellar mass, age and SFR based on SED fitting with template set B.  Columns 6-10 list the values derived by Eyles et al. (2007) for the same objects.  The best-fitting masses and SFRs derived in this work are corrected to total using the appropriate aperture correction.  The reported values from Eyles et al. have been re-scaled to be consistent with a Chabrier IMF.}
  \begin{tabular}{@{}cccccccccc}
  \hline
  \hline
  \multicolumn{5}{c}{This work} &  \multicolumn{5}{c}{\cite{Eyles2007}}\\ 
  ID & model & stellar mass & age & SFR & ID & model & stellar mass & age & SFR\\
     &       & (/$10^{10} $M$_{\odot}$) & (/Myr) & (/M$_{\odot}$ yr$^{-1}$) & &  & (/$10^{10}$ M$_{\odot}$) & (/Myr) & (/M$_{\odot}$ yr$^{-1}$)\\
  \hline

  5      & const &  0.86 &571 & 21.8 & 31\_2185 & const & 1.3\phantom{0} & 640 & 23.8 \\
  6      & const & 0.31 & 905 & \phantom{0}5.2 & 23\_2897 & $\tau=$ 1 Gyr & 0.4\phantom{0} & 640 & \phantom{0}5.5  \\
  8      & $\tau=$ 1 Gyr & 1.09 & 806 & 13.3 & 23\_6714 & $\tau=$ 500 Myr & 1.7\phantom{0} & 720 & 13.3 \\
  23     & $\tau=$ 100 Myr & 1.03 & 286 & \phantom{0}9.0& 32\_4331 & $\tau=$ 70 Myr & 0.9\phantom{0} & 260 &\phantom{0}4.4 \\

  \hline
  \hline
  \end{tabular}
  \label{tab:EylesComparison}
\end{table*}

The mass constraints for objects 6 and 23 agree very well.  The masses for objects 5 and 8, which are the two objects analysed in \cite{Eyles2005}, are both lower based on our fits, although still consistent within a factor of two.  

The differences between derived masses for any of the objects, but for objects 5 and 8 in particular, can be attributed to the differences in photometry.  The CANDELS WFC3 data and deep $K-$band data constrain the rest-frame UV to much higher accuracy.  The deconfusion of the existing IRAC data is also better constrained by using the high-resolution WFC3 image priors.  

\begin{figure*}
  \centering
   \subfigure{\includegraphics[width=3in,trim=2cm 13cm 1.9cm 5cm,clip]{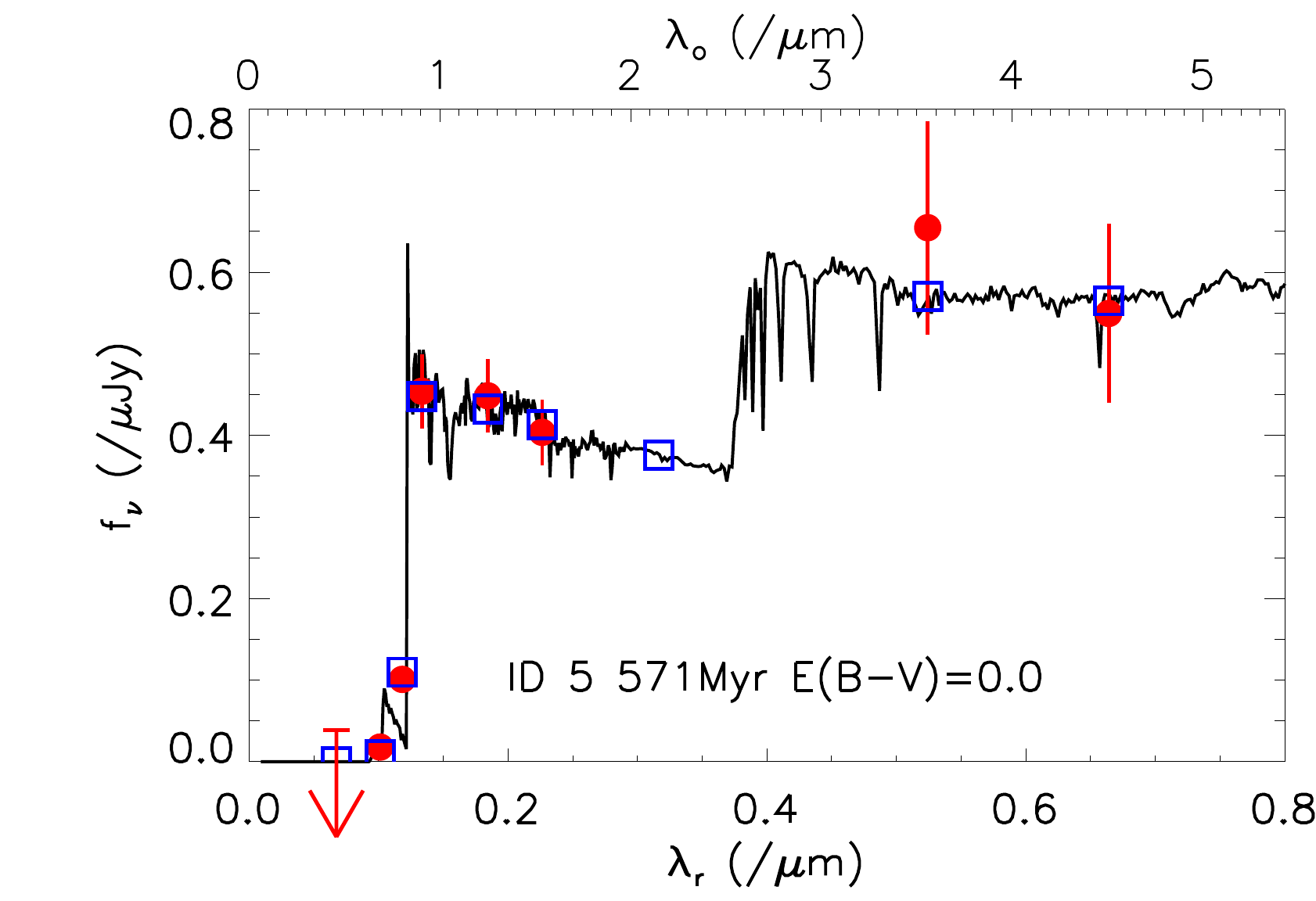}} 
   \subfigure{\includegraphics[width=3in,trim=2cm 13cm 1.9cm 5cm,clip]{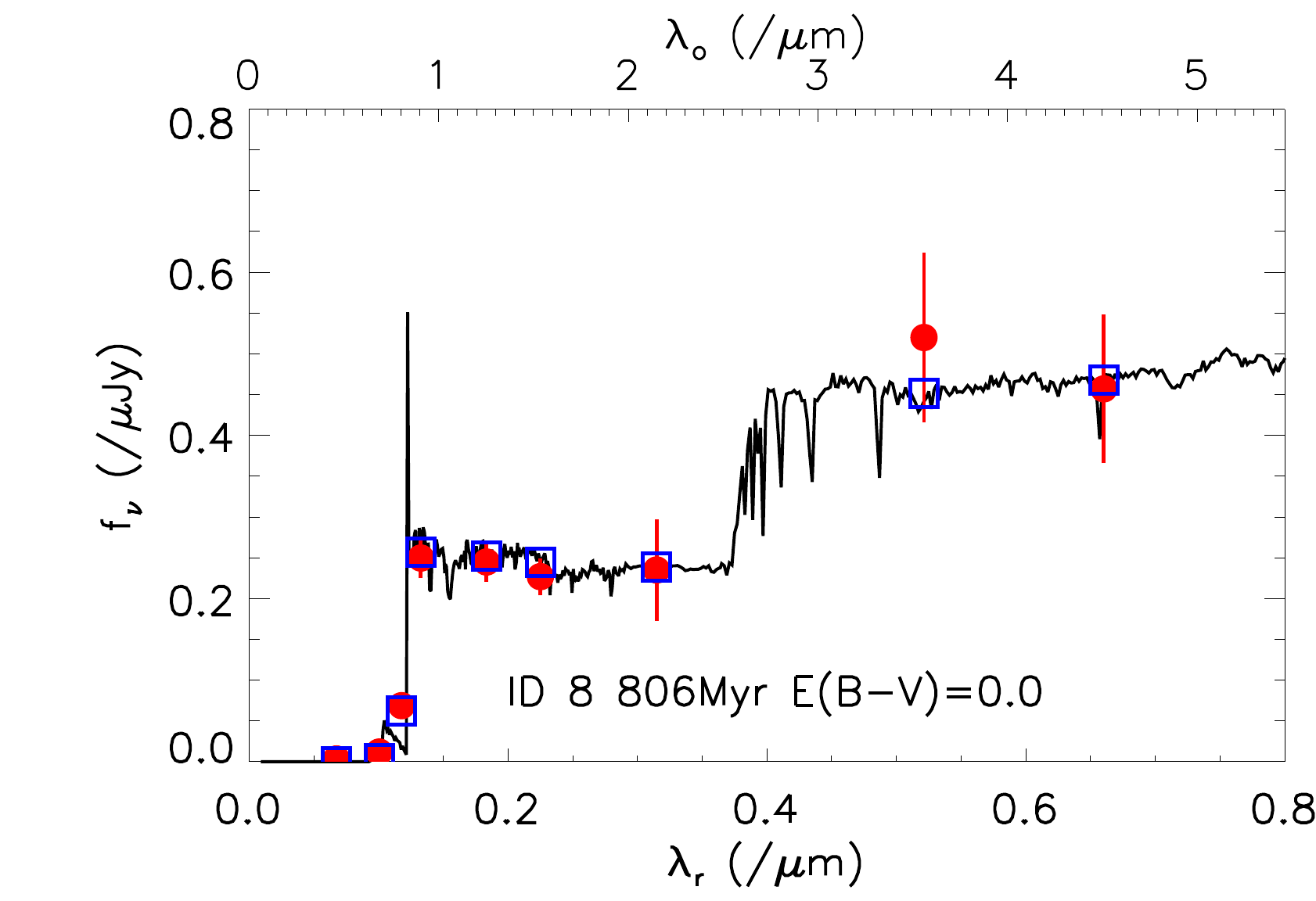}}
  \caption{The best-fitting SEDs to objects 5 and 8, with a template set restricted to solar metallicity, zero reddening models with $\tau \leq 1$ Gyr, constant SFR or an instantaneous burst.  These two objects were first analysed in Eyles et al. (2005) who suggested that they harbour old stellar populations at high redshift, with their analysis suggesting ages $>500$ Myr.  Using improved photometry, but a similarly-restricted set of templates, our results agree.}
  \label{fig:EylesComparison}
\end{figure*}

The ages are mostly consistent, although object 6 is now fit with a considerably older template.  The best-fitting ages of objects 6 and 8 are actually approaching the age of the Universe at $z\sim6$ and would have been ruled out in the \cite{Eyles2007} analysis. 

The SED fits for objects 5 and 8 are plotted in Fig.~\ref{fig:EylesComparison}.  These objects were first analysed in \cite{Eyles2005} who suggested that they harbour old stellar populations at high redshift ($>500$ Myr).  Using improved photometry but a similar template set we get consistent results, with the red rest-frame UV-to-optical colours being reproduced by a strong Balmer-break.  The SFRs derived from both studies also show good agreement.  


\subsection{All smoothly varying SFHs (template set C)}

\begin{table*}
 \centering
  \caption{Results of SED fitting with the full set of smoothly-varying SFHs, extinction and metallicities (template set C).  The first column gives the IDs of the objects.  Columns 2-7 give the best-fitting stellar mass, SFR, sSFR, extinction, metallicity and age respectively.  The sSFR for burst models ($SFR = 0\ \mathrm{M}_{\odot}\mathrm{yr}^{-1}$ by definition) are estimated using the SFR of the next best-fitting model that is not a burst.   All quoted errors are the 68\% confidence limits, marginalising over all other free parameters.  Column 8 gives the lower limit to the Age from the $1\sigma$ uncertainties from all the models except the burst.  The final two columns give the best-fitting SFH and $\chi^2$.  The typical degrees of freedom for 9 photometric points is 4.}
  \scalebox{0.9}{
  \begin{tabular}{@{}lccccccccc@{}}
  \hline
  Object & log$_{10}$(Mass) & SFR & sSFR & E(B-V) & metallicity & log$_{10}$(Age) & Age lower & Model & $\chi^2$\\
ID          & (/M$_{\odot}$)  &(/M$_{\odot}$yr$^{-1}$)& (/Gyr$^{-1}$) & & (/Z$_{\odot}$) & (/Gyr) & limit (1$\sigma$ /Myr)\\
\hline
2   & \phantom{0}9.50$^{+0.22}_{-0.11}$ & \phantom{0}0.0$^{\phantom{0}+27.6}_{\phantom{00}-0.0}$ & 1.2 & 0.005$^{+0.150}_{-0.005}$ & 0.4\phantom{0} & 7.76$^{+0.70}_{-0.25}$ & 32  & burst               & 2.79\\
5   & \phantom{0}9.84$_{-0.25}^{+0.26}$ &           19.6$_{\phantom{0}-19.6}^{\phantom{0}+40.2}$ & 2.8 & 0.000$_{-0.000}^{+0.099}$ & 0.4\phantom{0} & 8.36$_{-0.73}^{+0.60}$ & 43  & EI sSFR=2 Gyr$^{-1}$ & 2.11\\
6   & \phantom{0}9.45$_{-0.31}^{+0.16}$ & \phantom{0}4.7$_{\phantom{00}-1.5}^{\phantom{00}+2.9}$ & 1.7 & 0.000$_{-0.000}^{+0.044}$ & 0.2\phantom{0} & 8.96$_{-0.85}^{+0.00}$ & 128 & const & 3.54\\
8   & \phantom{0}9.95$_{-0.14}^{+0.12}$ &           12.2$_{\phantom{00}-2.6}^{\phantom{00}+3.9}$ & 1.3 & 0.002$_{-0.000}^{+0.032}$ & 0.2\phantom{0} & 8.96$_{-0.25}^{+0.00}$ & 509 & $\tau$=2 Gyr         & 7.06\\
13  & \phantom{0}9.14$_{-0.16}^{+0.45}$ & \phantom{0}0.0$_{\phantom{00}-0.0}^{          +132.5}$ & 2.3 & 0.163$_{-0.091}^{+0.136}$ & 1.0\phantom{0} & 7.12$_{-0.10}^{+1.84}$ & 13  & burst               & 2.82\\
18  & \phantom{0}9.02$^{+0.33}_{-0.54}$ & \phantom{0}0.0$^{\phantom{0}+45.7}_{\phantom{00}-0.0}$ & 2.7 & 0.005$^{+0.279}_{-0.005}$ & 0.02           & 7.86$^{+1.10}_{-0.84}$ & 10  & burst               & 1.23\\
23  & \phantom{0}9.90$_{-0.15}^{+0.41}$ & \phantom{0}0.0$_{\phantom{00}-0.0}^{\phantom{0}+96.5}$ & 0.3 & 0.010$_{-0.010}^{+0.210}$ & 0.02           & 8.06$_{-0.55}^{+0.85}$ & 32  & burst               & 1.75\\
24  &           10.06$_{-0.16}^{+0.64}$ & \phantom{0}0.0$_{\phantom{00}-0.0}^{          +506.5}$ & 2.3 & 0.351$_{-0.175}^{+0.020}$ & 0.2\phantom{0} & 7.02$_{-0.00}^{+1.99}$ & 10  & burst               & 1.23\\
25  & \phantom{0}9.91$_{-0.42}^{+0.18}$ &           13.6$_{\phantom{0}-13.6}^{\phantom{0}+58.0}$ & 1.7 & 0.042$_{-0.042}^{+0.153}$ & 1.0\phantom{0} & 8.71$_{-1.24}^{+0.30}$ & 29  & $\tau$=0.5 Gyr       & 2.28\\
26  & \phantom{0}9.18$^{+0.52}_{-0.37}$ & \phantom{0}0.0$^{          +137.4}_{\phantom{00}-0.0}$ & 2.4 & 0.180$^{+0.160}_{-0.160}$ & 0.4\phantom{0} & 7.22$^{+1.74}_{-0.20}$ & 10  & burst               & 1.93\\
27  & \phantom{0}9.34$^{+0.35}_{-0.28}$ & \phantom{0}3.5$^{\phantom{0}+33.6}_{\phantom{00}-3.5}$ & 1.6 & 0.000$^{+0.151}_{-0.000}$ & 0.4\phantom{0} & 8.16$^{+0.80}_{-0.82}$ & 10  & $\tau$=0.05 Gyr       & 1.02\\
157 & \phantom{0}9.43$_{-0.25}^{+0.32}$ &           15.0$_{\phantom{00}-7.8}^{\phantom{00}+6.4}$ & 5.6 & 0.002$_{-0.002}^{+0.064}$ & 0.2\phantom{0} & 8.46$_{-0.95}^{+0.50}$ & 32  & EI sSFR=4 Gyr$^{-1}$ & 8.03\\
248 & \phantom{0}9.89$^{+0.16}_{-0.14}$ & \phantom{0}9.6$^{\phantom{00}+3.6}_{\phantom{00}-3.0}$ & 1.2 & 0.000$^{+0.030}_{-0.000}$ & 0.2\phantom{0} & 8.91$^{+0.00}_{-0.35}$ & 363 & $\tau=$1 Gyr         & 6.23\\

  \hline
\end{tabular}}
\label{tab:allModel}
\end{table*}

Although all of the objects are assigned acceptable fits from the restricted template set, we now look at how the derived properties and their constraints change when we allow extinction by dust, multiple metallicities and an expanded range of SFHs (including exponentially increasing SFR models).

The best-fitting parameters and 68\% confidence limits are reported in Table ~\ref{tab:allModel}.  In this case, we require $\chi^2\lesssim11$ (9.5, 7.8 or 6) for the fit to be considered acceptable with 10 (9, 8, or 7) photometric points.  Again, all objects have been assigned an acceptable fit and the $\chi^2$ values are considerably improved for most objects.  In allowing for dust in the fitting to these objects, the uncertainties in SFR and age increase, as expected due to the known degeneracies in these parameters.  The masses derived from the restricted model set are, however, proven to be quite robust.  

\begin{figure*} 
   \centering
   \subfigure[$\tau$ models]{\includegraphics[width=2.3in,trim=2cm 13cm 1.5cm 5cm,clip]{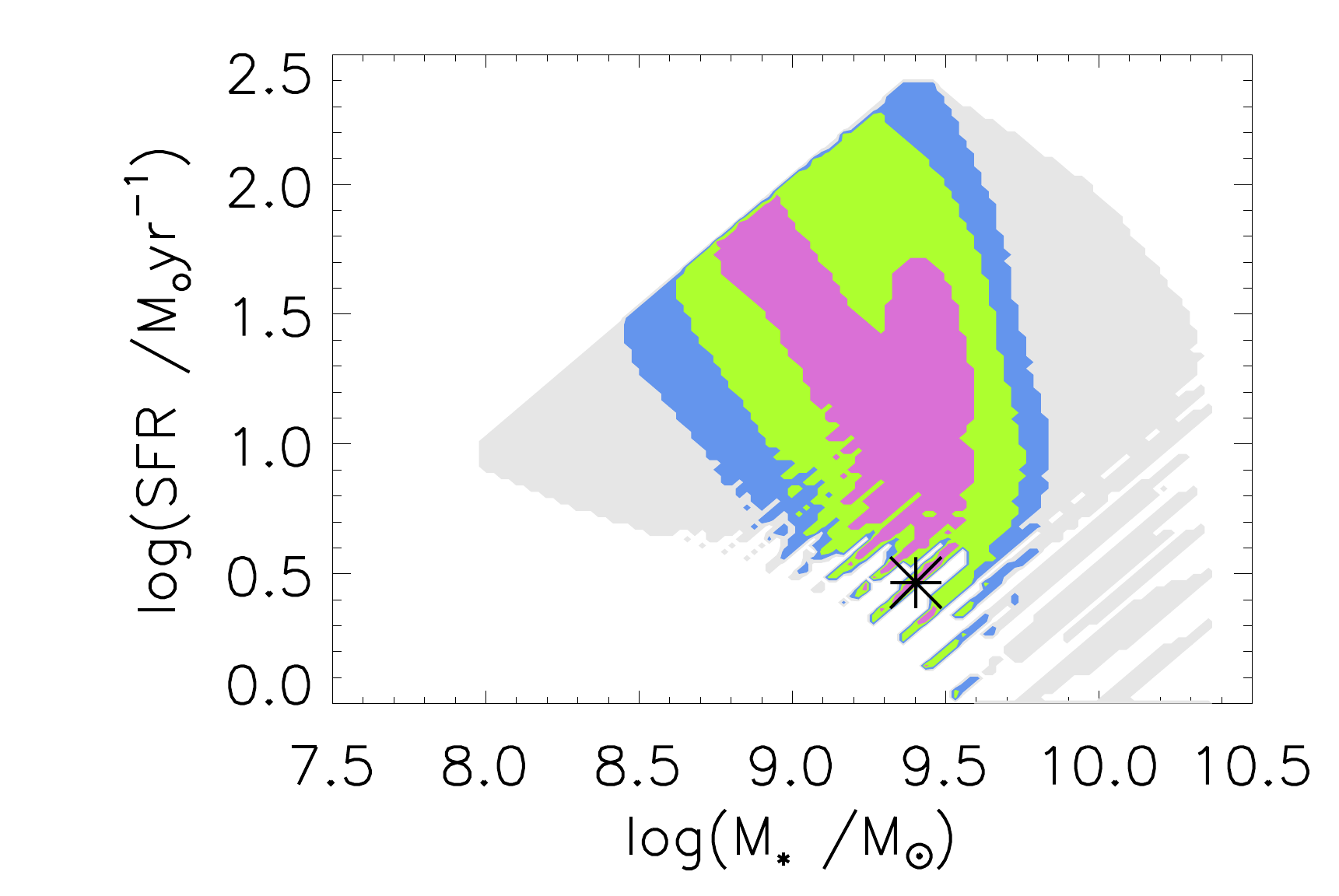}}
   \subfigure[Const SFR models]{\includegraphics[width=2.3in,trim=2cm 13cm 1.5cm 5cm,clip]{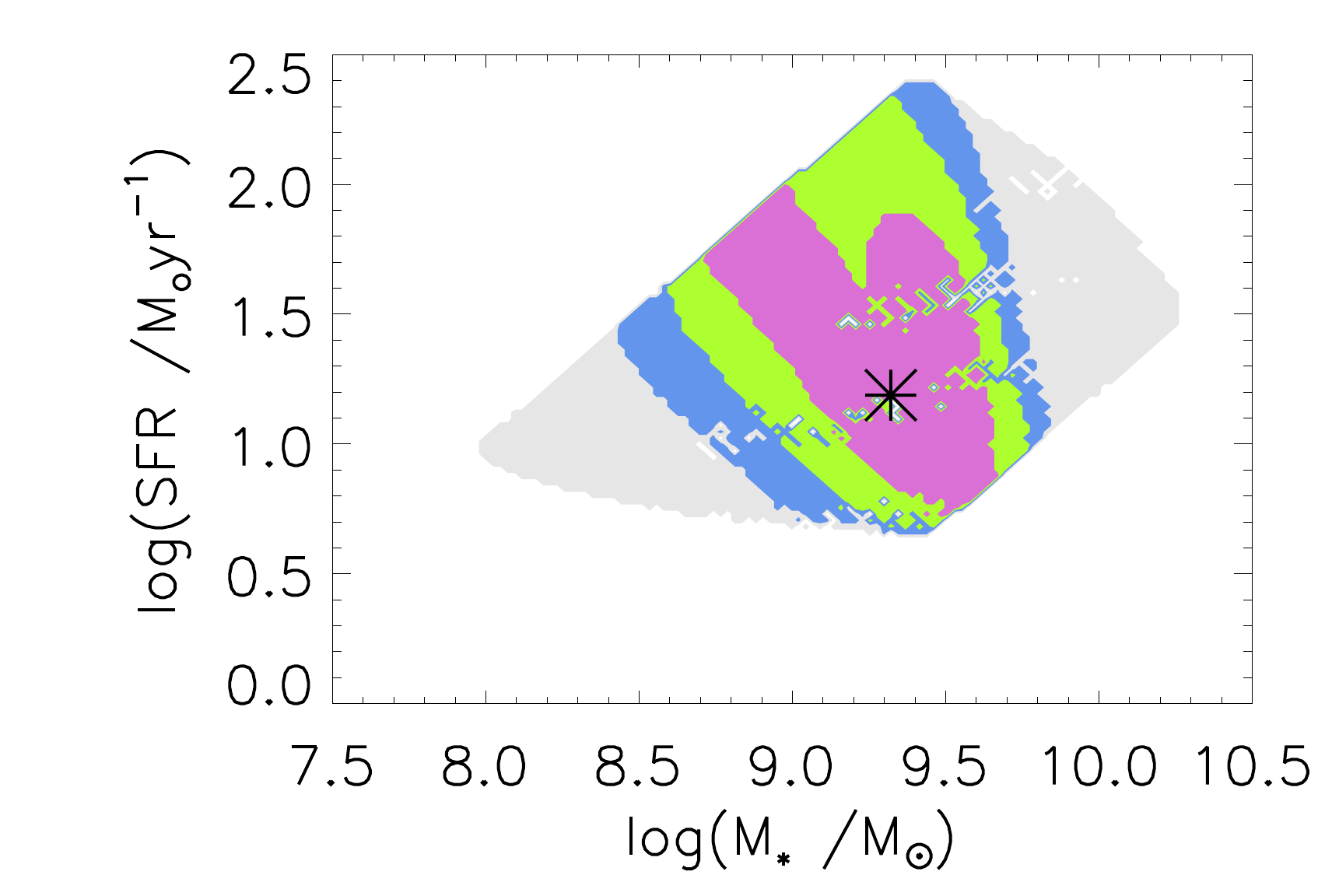}} 
   \subfigure[Exponentially increasing SFR models]{\includegraphics[width=2.3in,trim=2cm 13cm 1.5cm 5cm,clip]{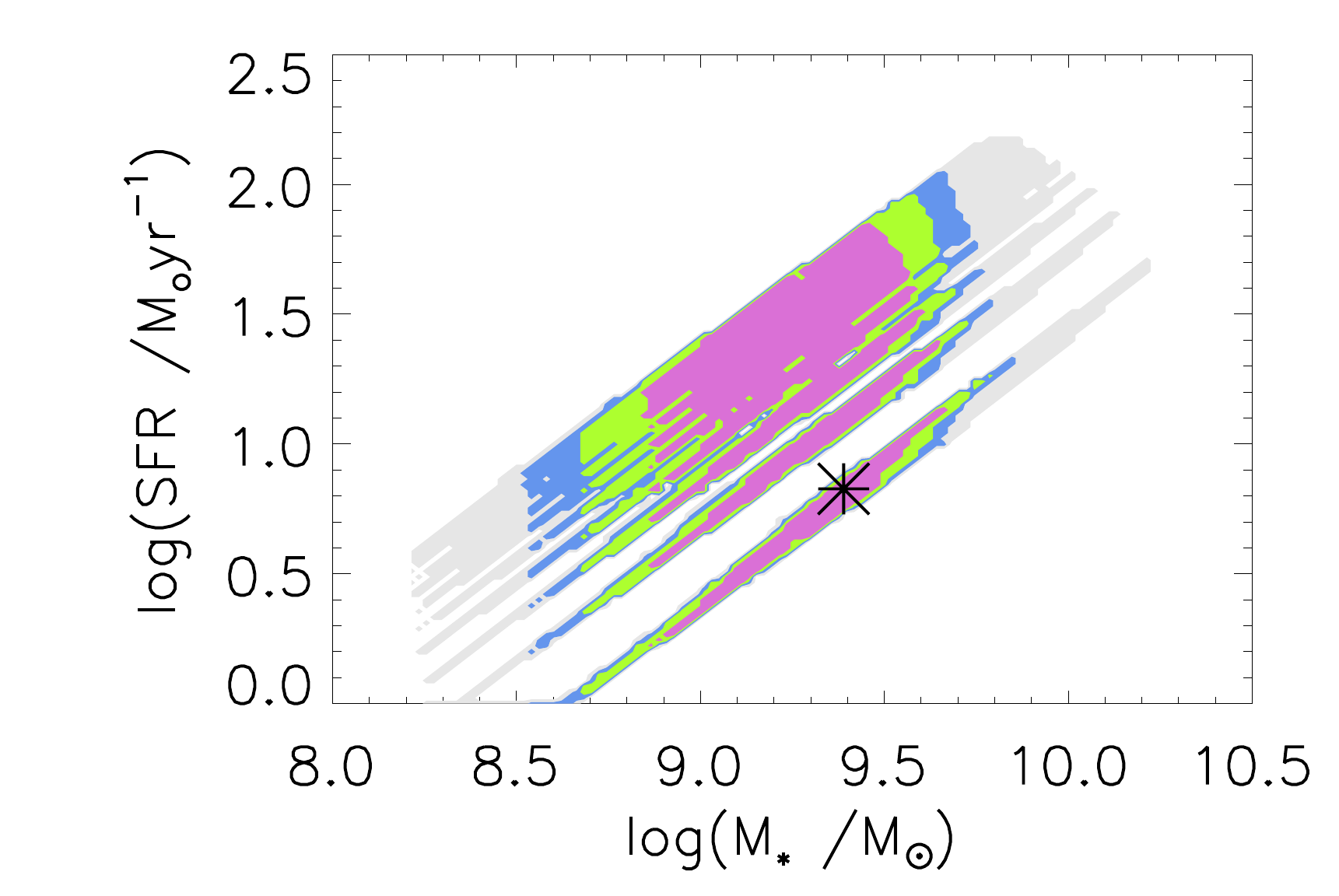}}
   \subfigure[$\tau$ models]{\includegraphics[width=2.3in,trim=2cm 13cm 1.5cm 5cm,clip]{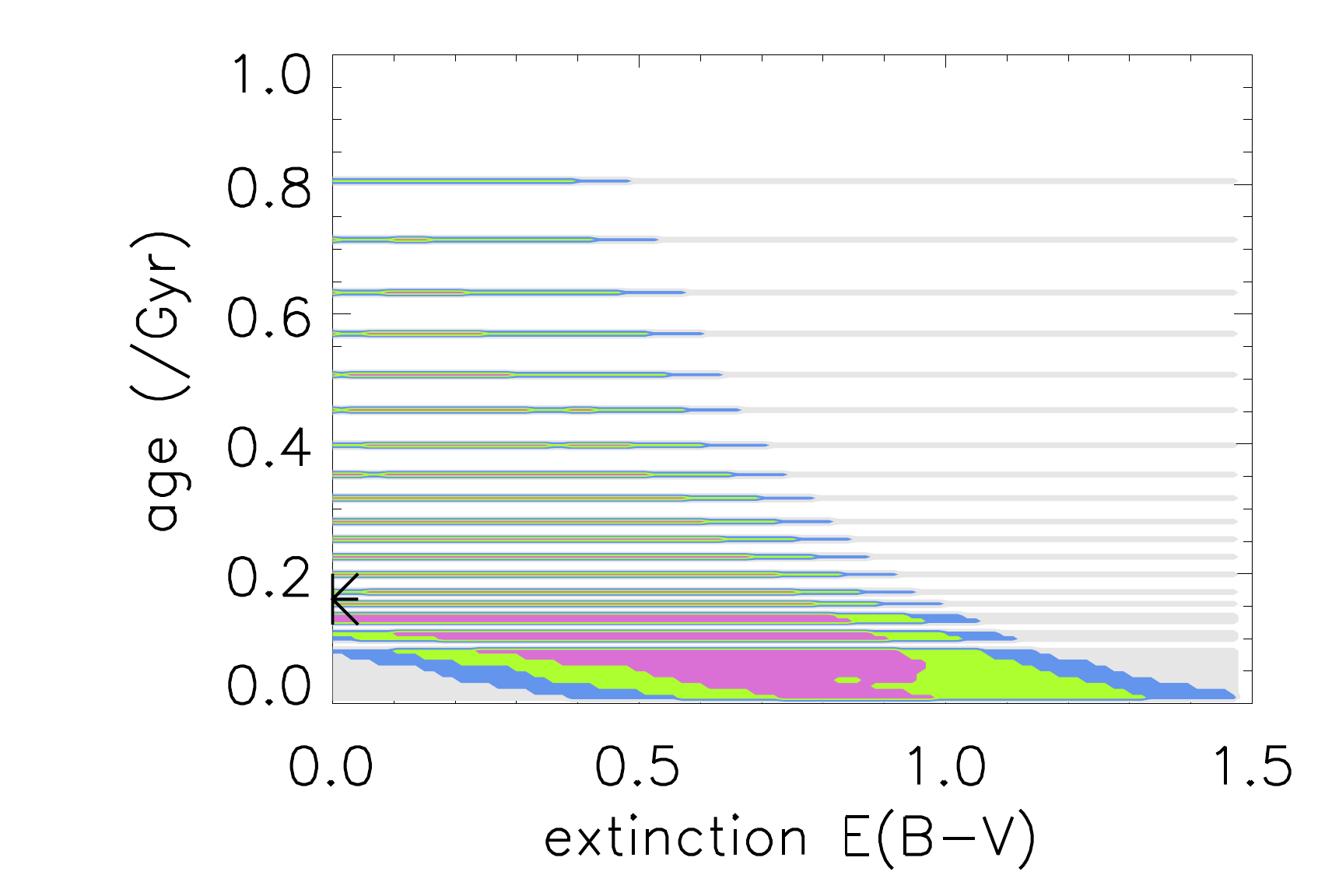}}
   \subfigure[Const SFR models]{\includegraphics[width=2.3in,trim=2cm 13cm 1.5cm 5cm,clip]{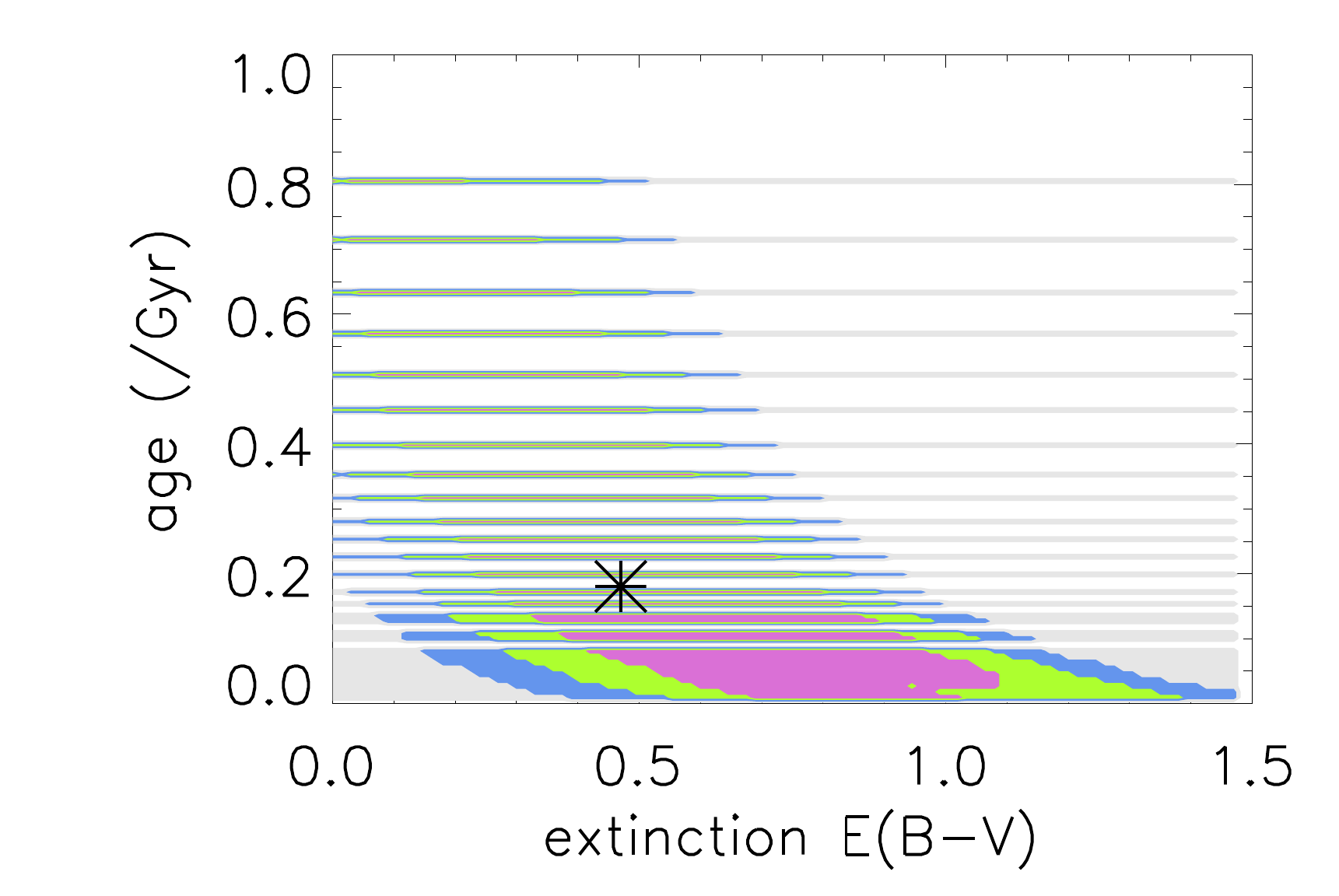}}
   \subfigure[Exponentially increasing SFR models]{\includegraphics[width=2.3in,trim=2cm 13cm 1.5cm 5cm,clip]{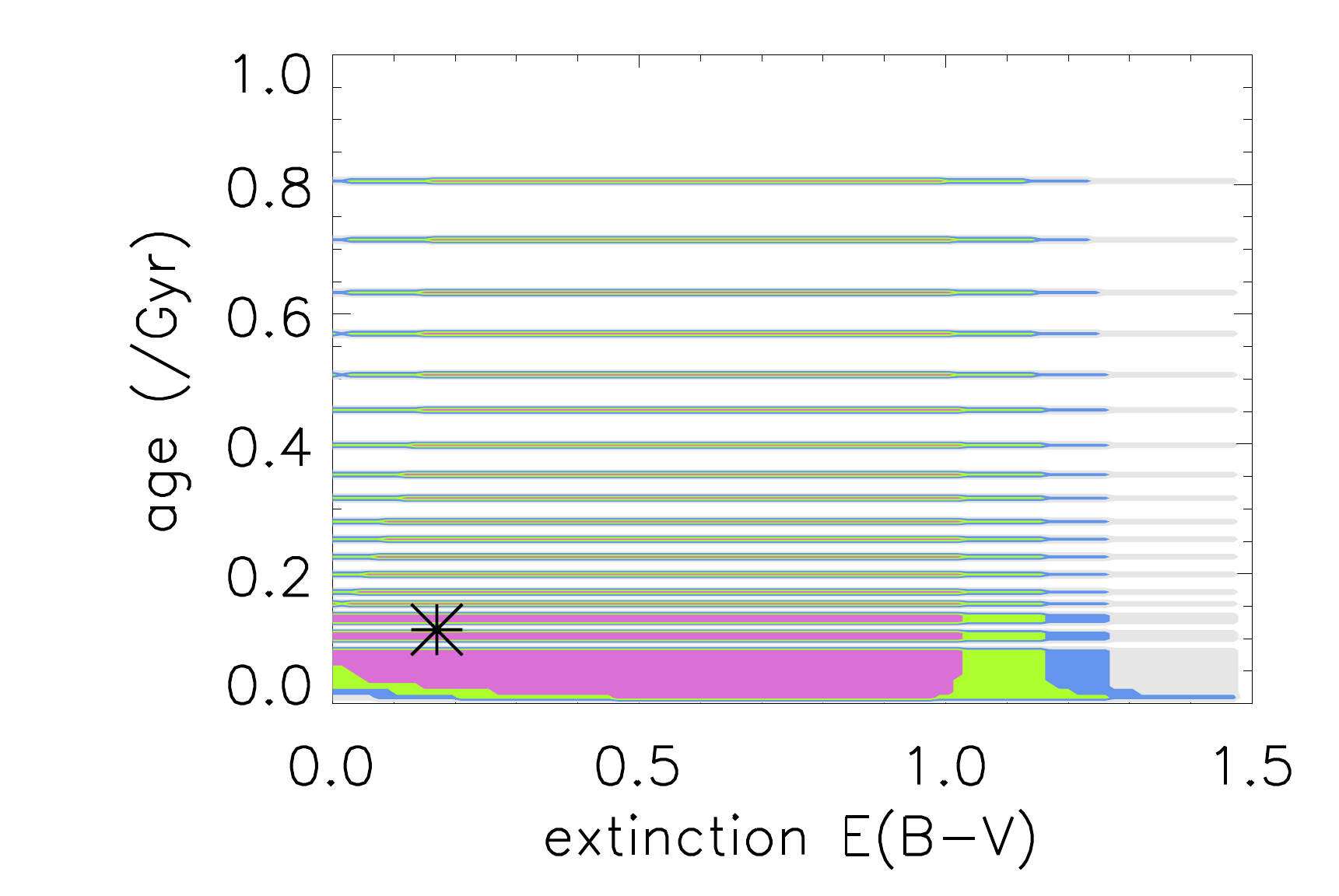}}
   \caption{Fitting to object 2: log(stellar mass) vs log(SFR) of the models (a-c) and model age vs. extinction (d-f) for three different sets of models; $\tau$ models (exponentially decreasing SFR), constant SFR and exponentially increasing SFR.  The greyed areas show the parameter space covered by the best-fitting models.  In the case of age and extinction, these are supplied as a grid of values, while both SFR and mass are the values of the best-fitting models within the grids of parameters.  The pink, green and blue regions show the ranges of parameters with best-fitting model $\chi^2$ values within 
$\Delta \chi^{2}=$ 2.3, 6.2 and 11.8 respectively, of the overall minimum $\chi^2$ value.}
   \label{fig:obj2_1}
\end{figure*}

In general the masses, ages, extinction and metallicity tend to agree between the different models, although burst models tend to give consistently lower best-fitting masses and ages, primarily because younger ages are needed to fit to the rest-frame UV in the absence of on-going star formation in the templates, as described in section 6.1.  We can see this agreement when plotting $\Delta\chi^2$ surfaces for various parameter combinations for a representative object (object 2) in Fig~\ref{fig:obj2_1}.  The 1$\sigma$ limit is plotted in pink and shows the 68\% joint confidence interval ($\Delta\chi^2<2.3$).  These plots show that the parameter ranges overlap very consistently for the different models.  

The differences between the limits can be described mainly by the differences in the allowed ranges of parameter space occupied by different models.  For example, models with a constant star-formation rate must have a minimum normalisation required to match to the luminosity of the sources, thereby the lower SFRs permitted by $\tau$ and EI models are not seen here.  The agreement in overall normalisation of the models is what drives the overall agreement in derived masses.  The resultant SFRs are therefore driven by the range of SFRs permitted within this normalisation which can also reproduce the galaxy's observed rest-frame UV.  

Ages are not well constrained, as expected, because either age or dust reddening can produce the same observed SED, for any SFH that is not increasing with age.  The degeneracies are slightly different for the EI models, due to older galaxies having higher star-formation rates and we see no apparent correlation between age and reddening in the likelihood contours (Fig.~\ref{fig:obj2_1}f).

Most of the galaxies allow for dust within the confidence contours, whether the best-fitting model requires it or not. \cite{Eyles2007} take the absence of dust from the best-fitting models to show little evidence for intrinsic reddening within the galaxies, although their derived ages range from $\sim$ 300 - 700 Myr, requiring older, redder populations to fit to the observed SEDs.  As we have seen from restricting our model set, our results agree.  When looking at all the models, metallicities and reddening values, however, a wide range of ages provide acceptable fits to each object (also remarked on by \cite{Eyles2007} with reference to the minimum and maximum ages derived from burst and constant SFR models), and some objects do show evidence for intrinsic reddening in their SEDs (see Fig.~\ref{fig:new}).  We also note that the older age estimates are at odds with the predictions of various models (e.g. \cite{Finlator2010} predict ages ranging from a few Myr to $\sim$150 Myr at these redshifts).  Consequently it is of interest to investigate whether any of the objects actually require an old stellar population ($\gtrsim300$ Myr) which dominates the stellar mass.  In Section 6.3.2, we consider again the degeneracies inherent to our limited model set and consider the full range of possible models within the marginalised 68\% confidence contours in order to see whether these old ages are truly robust.

\subsubsection{Exponentially increasing SFHs}

As described in Section 4.1, the exponentially increasing SFHs are initiated with a seed mass produced by a burst of initial star formation.  This is just a by-product of the parameterisation, but this seed-mass leaves an imprint on the observed SED by reddening the observed UV-to-optical colours.  

We see that two of the objects are best-fit by EI SFHs, although for a constant sSFR of 2 Gyr$^{-1}$ a galaxy needs to have been growing in mass over the last $\sim890$ Myrs to make the contribution to the SED of the original burst insignificant (i.e. $<10$\% rest-frame optical flux in the absence of nebular emission).  As a consequence, the best-fitting model for one of these objects is essentially an instantaneous burst with some ongoing star-formation.

\subsubsection{Template degeneracies and age determination}

Our results with smoothly-varying SFH models show that three of the sample of 13 galaxies have old best-fitting ages ($>300$ Myr, objects 6, 8 and 248) and all of the objects that are not best-fit by a pure burst template show best-fitting ages of at least 200 Myr.  As discussed previously, the burst templates give very young best-fit ages when fitting to objects that do show star formation, simply because the youngest bursts still have young, hot stars that can fit the rest-frame UV flux.  We therefore look to the lower limits in age (lowest ages allowed within $\Delta\chi^2=1$), when burst models are not included in the template set, which are reported in column 8 of Table~\ref{tab:allModel}.  We see that, when the SEDs are well described by a burst, they also allow for very young ages when the burst models are not present, because very young exponentially increasing SFHs also show a strong burst component, as discussed above.

As for the objects with best fits which indicate an older population, the lower limits in age show that two of the objects, 8, and 248 appear to \textit{require} a population of at least $\sim$300 Myr in their fits.

\begin{figure} 
   \centering
   \subfigure{\includegraphics[width=3in,trim=2cm 13cm 1.9cm 5cm,clip]{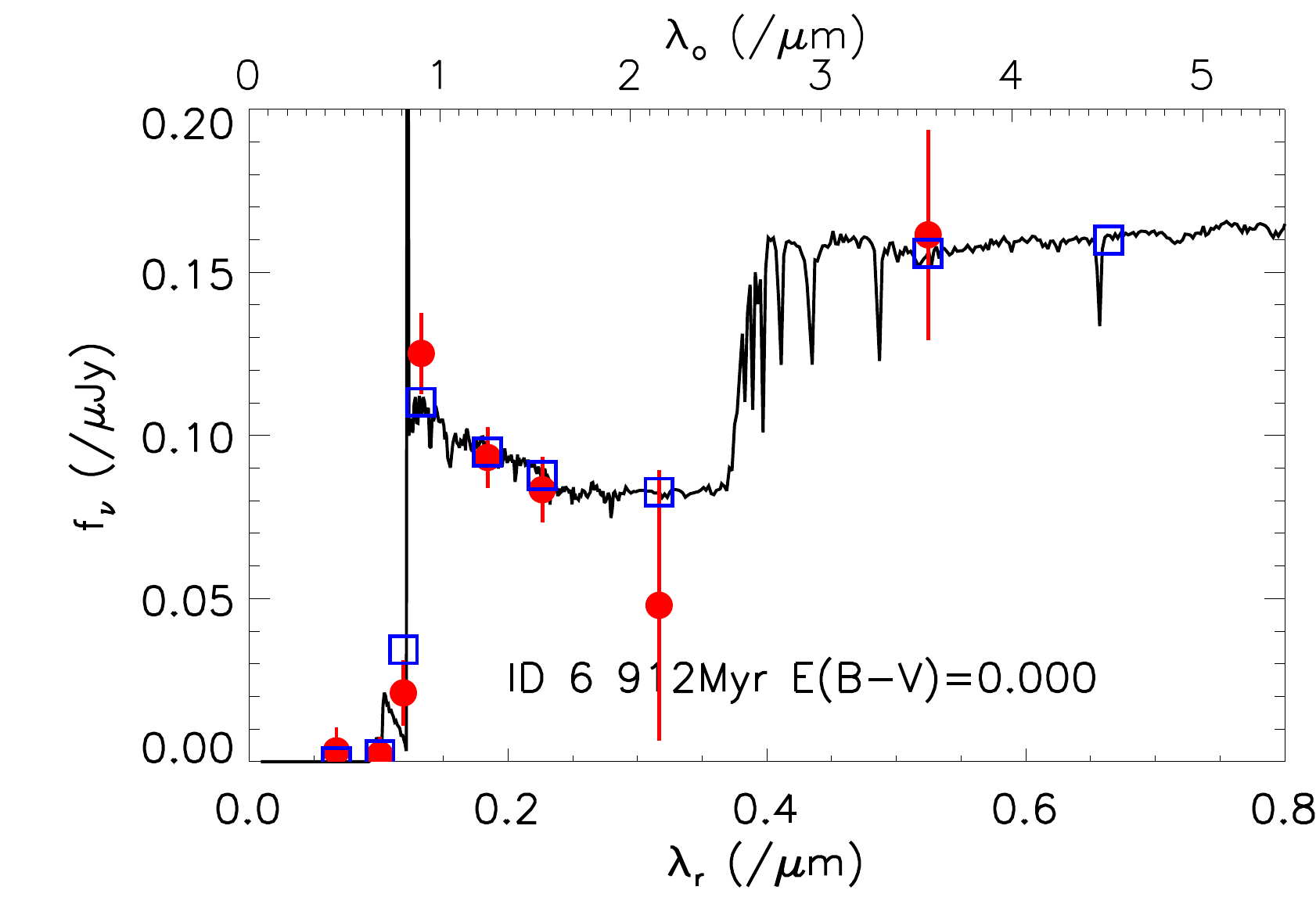}} 
   \subfigure{\includegraphics[width=3in,trim=2cm 13cm 1.9cm 5cm,clip]{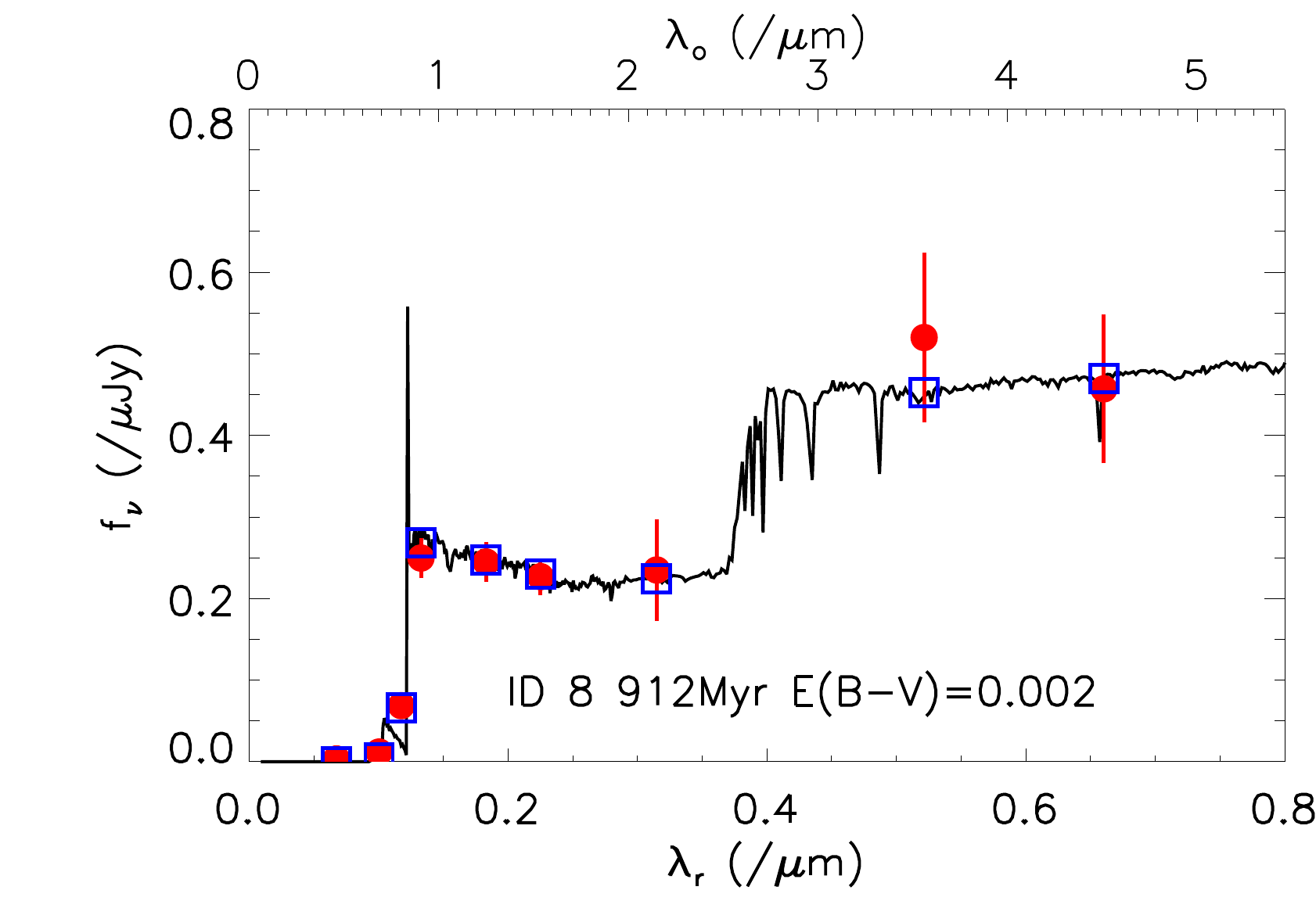}}
   \subfigure{\includegraphics[width=3in,trim=2cm 13cm 1.9cm 5cm,clip]{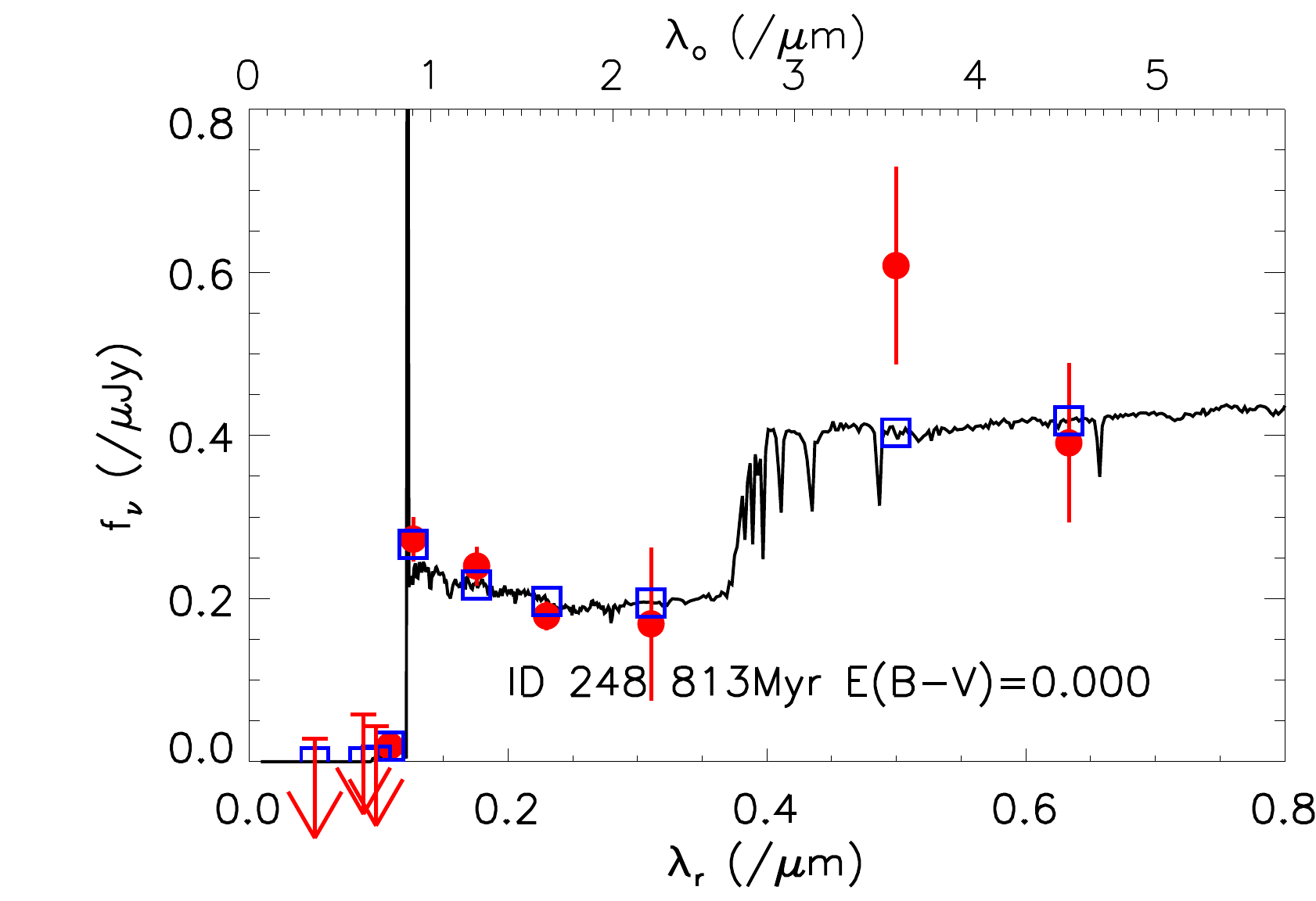}}
   \caption{SEDs and best-fitting models of three galaxies that apparently require old stellar population ages to provide good fits.  The red points plus error bars show the observed SED and the blue points show the synthetic photometry based on the best-fitting model.}
   \label{fig:old}
\end{figure}

\begin{figure} 
   \centering
   \subfigure{\includegraphics[width=3in,trim=2cm 13cm 1.9cm 5cm,clip]{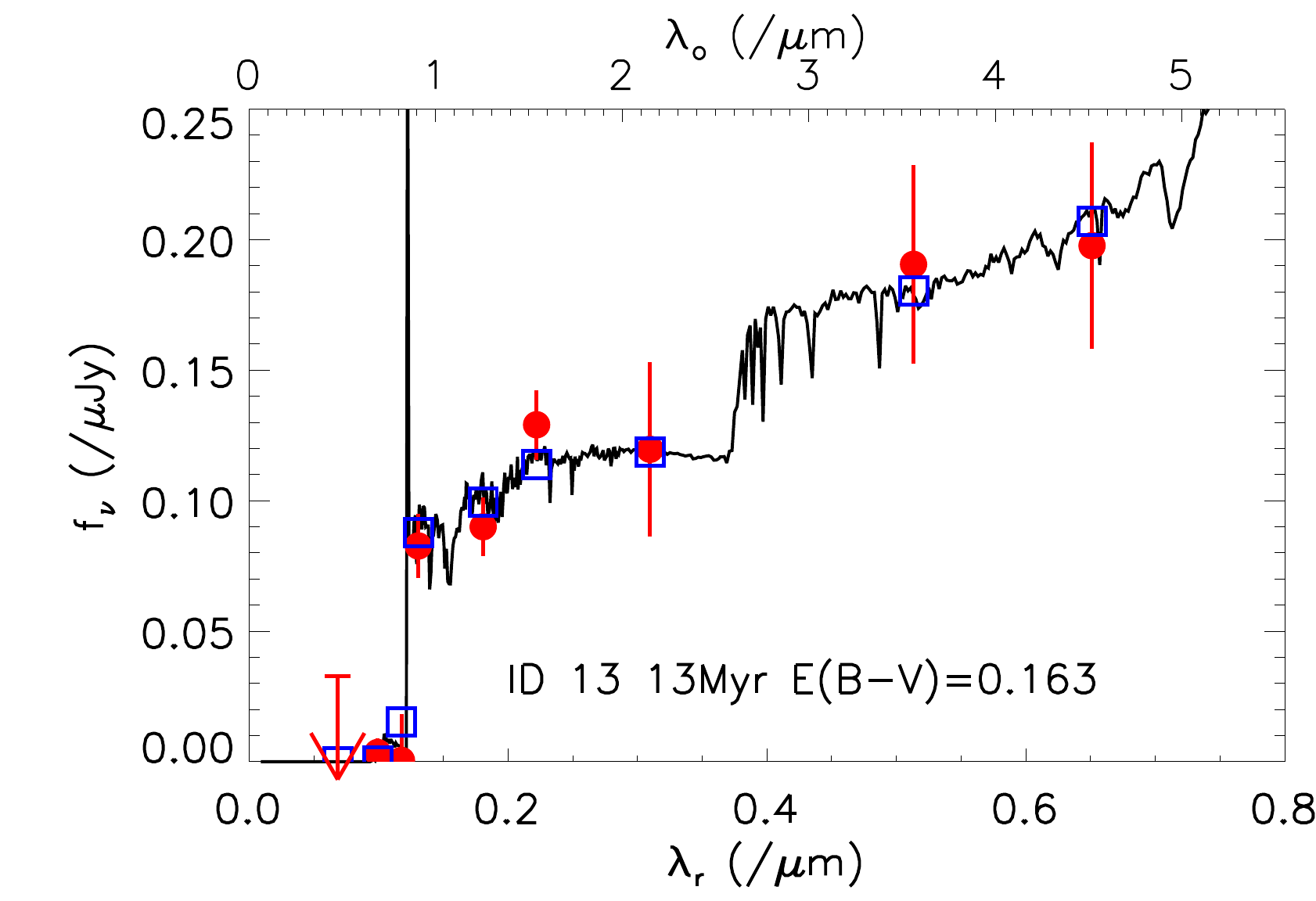}}
   \subfigure{\includegraphics[width=3in,trim=2cm 13cm 1.9cm 5cm,clip]{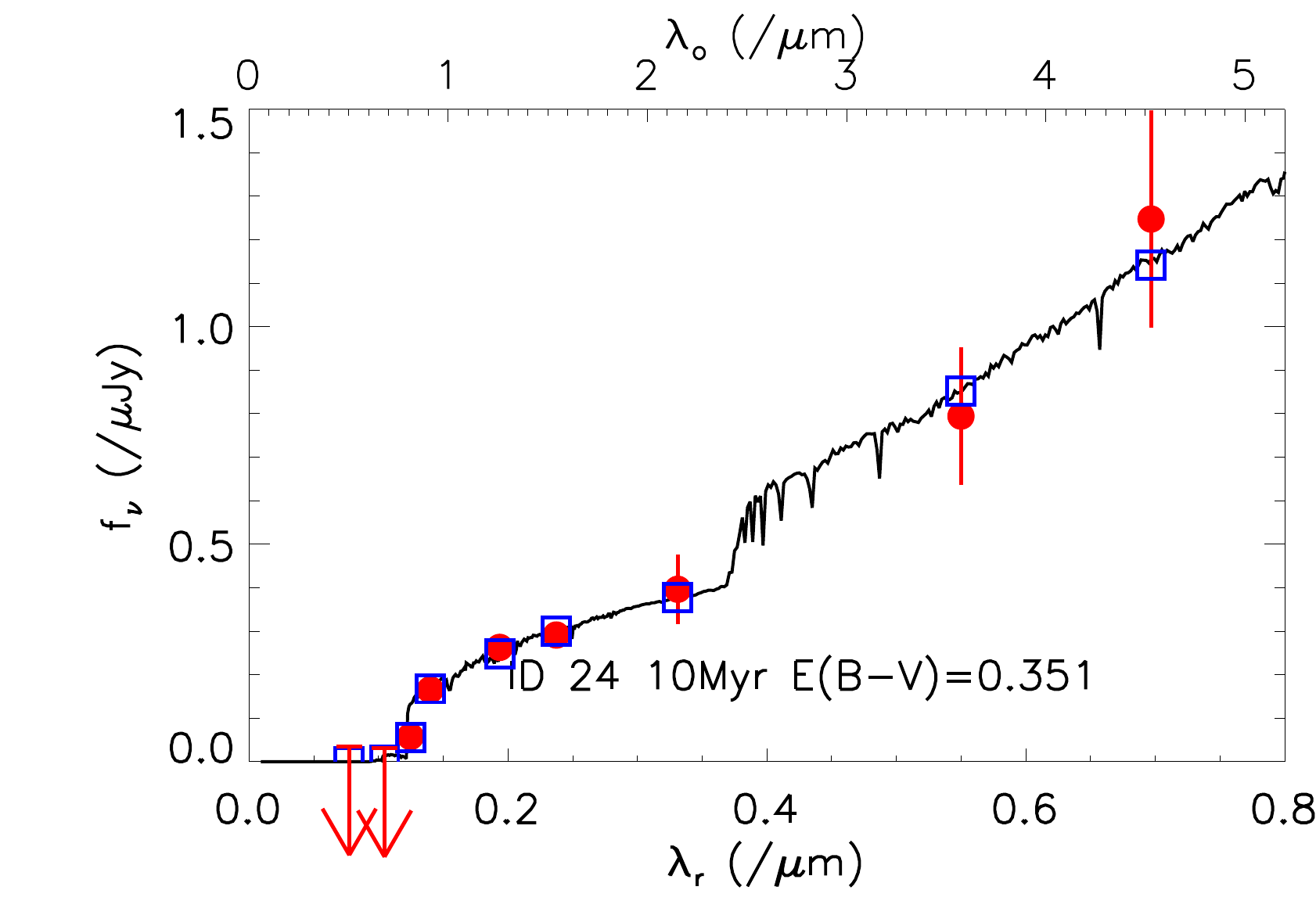}}
   \subfigure{\includegraphics[width=3in,trim=2cm 13cm 1.9cm 5cm,clip]{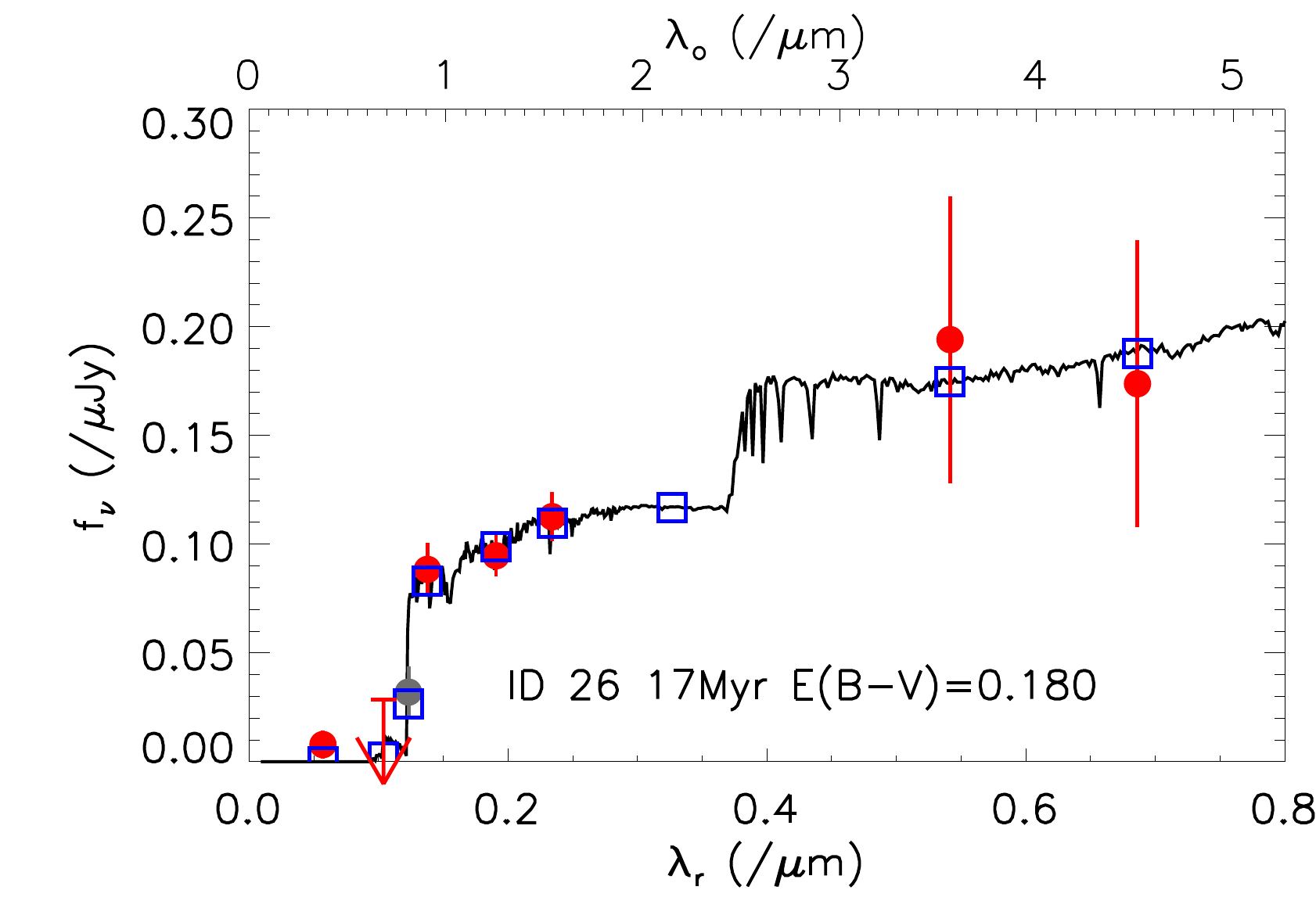}}
   \caption{As for Figure ~\ref{fig:old} but for objects which are consistently best-fit by young ($\lesssim300$ Myr) stellar populations. The grey point in the plot for Object 26 shows the filter excluded from fitting due to the uncertainty in the Ly$\alpha$ contribution.}
   \label{fig:new}
\end{figure}

Within this context it is interesting to examine the fits for typical examples of objects with consistently old and young best-fits (Figs~\ref{fig:old} and \ref{fig:new} respectively).  For most of the SFHs (except EI), the acceptable models with dust have the youngest ages (Fig.~\ref{fig:new}) while the objects being fit with consistently-old stellar populations ($>$ 500 Myr) have bluer UV slopes that allow for less intrinsic reddening (Fig.~\ref{fig:old}). While the objects with the youngest best-fitting ages may still have fairly red UV-to-optical colours, they tend to have redder UV-slopes and evidence of intrinsic reddening from their SEDs.  

\begin{figure} 
   \centering
   \subfigure[]{\includegraphics[width=3in,trim=2.8cm 13cm 1.8cm 5cm,clip]{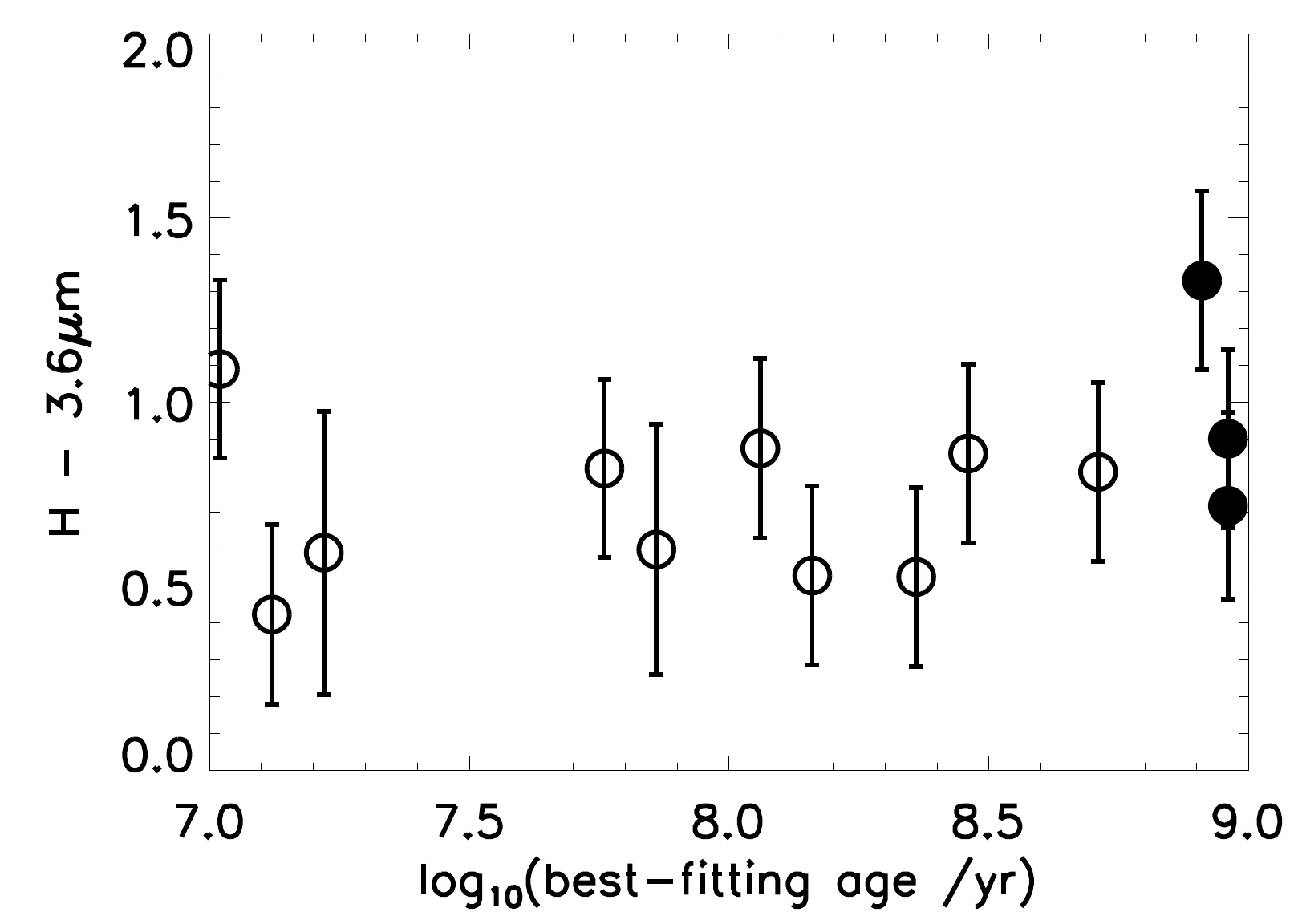}} 
   \subfigure[]{\includegraphics[width=3in,trim=2.8cm 13cm 1.8cm 5cm,clip]{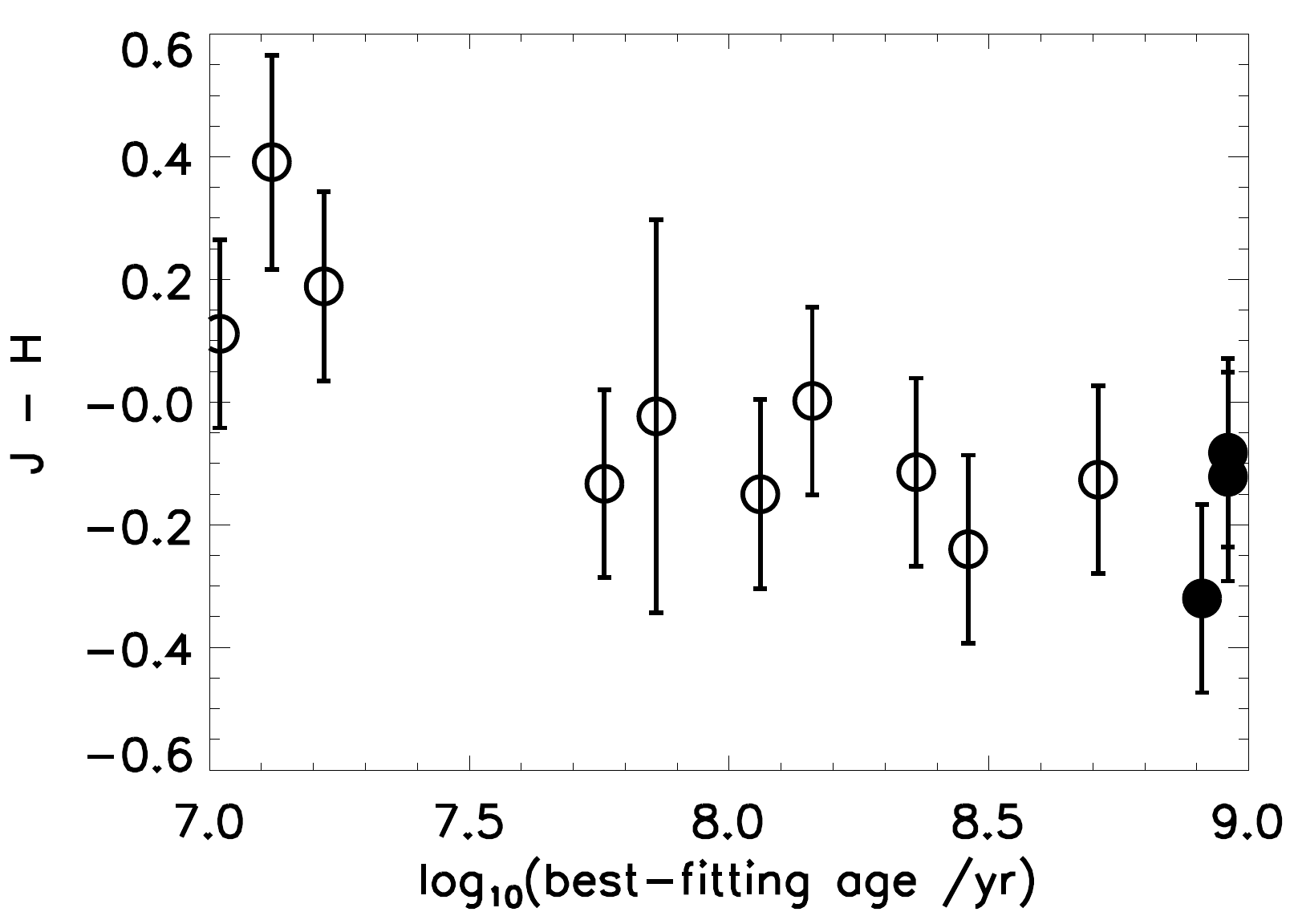}}
   \caption{The best-fitting age vs $H-3.6\mu$m colour, which spans the 4000\AA/Balmer break region (a), and the $J-H$ colour, which samples the near-UV slope (rest-frame) (b).  Filled symbols show best fitting solutions which only allow very small amounts of dust ($E(B-V)<0.05$ within 1$\sigma$ uncertainties).  The three objects with the reddest $J - H$ colours also have the youngest best-fitting ages as they have been fit with burst templates.  The observed correlation between $J-H$ colour (rest-frame UV-slope) and model age may be an indication of inherent model degeneracies producing older age estimates in objects that allow for little reddening by dust.}
   \label{fig:ages}
\end{figure}

Fig.~\ref{fig:ages} (a) and (b) show the UV-to-optical ($H-3.6\mu$m) and UV ($J-H$) colours of the objects plotted against the age of the best-fitting model.  Again, we see that the objects that are assigned the oldest ages from this template set tend to have fairly red UV-to-optical colours, as would be expected if the optical fluxes are dominated by older stellar populations, but they also have very blue UV slopes with very little dust allowed in the fitting (filled symbols).   We note that the object with the youngest best-fit age (object 24) also has a very red $H-3.6\mu$m colour, yet an older stellar population would not have time to build up for such a young galaxy.  This object also has a red UV-slope, however, and the template requires a relatively high level of extinction (E(B$-$V)=0.351, see Table 5) which explains how an object with such a red H $-$ 3.6$\mu$m colour can be fit with such a young model.

Those SEDs that allow for the presence of dust in the shape of their UV-continuum therefore allow for a wider range of ages in the fitting, generally extending the allowed range to younger models, whereas a blue UV-continuum plus a red UV-to-optical colour allows for less freedom when applying dust.  The resultant best fit is then a balance between the age required to fit the UV-optical plus current star formation to fit to the UV.

These slightly counter-intuitive results, with older ages assigned to objects with evidence of current star-formation, are indicative of a conflict between the observed SEDs and the model templates when blue UV slopes are combined with apparently large Balmer breaks.  This is reminiscent of the difficulties reported by \cite{Labbe2010} in simultaneously fitting to a blue UV-slope and red UV-to-optical colours.   In Section 7 we proceed to investigate whether a two component SFH and the addition of nebular emission can resolve this tension.

\section{SED fitting - two component star formation histories (template sets D-E)}

In this section we explore the possibility of explaining the observed SED characteristics with a more general parameterisation of the SFH and, in the process, see whether this solves the conflict between blue UV-slopes and red UV-to-optical colours.  This parameterisation uses a two-component model that decouples the instantaneous SFR from the mass built up over time.  The results both with and without nebular emission are studied (template sets E and D respectively).  These models consist of a constant duration of star formation to describe the current star-forming component and an older burst to fit the rest-frame optical (a full description is provided in Section 4).  

The recent durations of star formation are fairly sparsely sampled with only three different durations (10, 30 and 100 Myr), but by parameterising the star-forming component in this way, rather than by a second burst, we can derive the SFRs more consistently from the models, without using calibrations between UV and SFR based on specific SFHs that do not hold for a very recent burst of star formation (i.e. \citealt{Madau1998}).

Ideally, this parameterisation will give upper and lower limits to the mass \citep{Papovich2001} and age of the dominant stellar component (i.e. the age of the burst with minimum of 85\% of the mass according to this parameterisation).  We have already seen that pure bursts provide a lower limit in age when the UV indicates the presence of young stars.  To be able to place improved constraints on the instantaneous SFR, however, an extra constraint is added to the fits.  The models are also required to produce at least enough ionising photons to explain the Ly$\alpha$ emission for all objects except the two for which we were unable tie the calibration of the spectroscopy to the photometry (objects 18 and 16, see Section 2.3).  This constraint ensures that a very young burst cannot dominate the UV part of the spectrum when the observed Ly$\alpha$ indicates that there should be a contribution from current star formation.  This is important for the SFR and age estimates as it prevents the lower limits being set by models in which the UV flux is dominated by young stars from a young burst, rather than from the star-forming component.

A lower limit on the flux of ionising photons is calculated using the calibrated Ly$\alpha$ flux, which is in any case a lower limit of the intrinsic Ly$\alpha$ due to absorption by the IGM, and assuming Case B recombination \citep{Osterbrock2006} giving:
\begin{equation}
L(Ly\alpha)=24\times4.87\times10^{-13}(1-f_{esc})N_{LyC}
\end{equation}
where $N_{Lyc}$ is the flux of Lyman continuum photons, $L(Ly\alpha)$ is the luminosity of $Ly\alpha$ in erg s$^{-1}$ and $f_{esc}$ is set to $0$ to provide the lower limit.

\subsection{Templates without nebular emission (template set D)}

\subsubsection{Results}

Table~\ref{tab:BCModels} documents the best-fitting parameters with the 1$\sigma$ limits marginalised over all the different models within the set.  In general the derived masses and SFRs agree well with the confidence contours derived from the full set of smoothly-varying SFHs, although when the objects are best described by a burst in template set C, their best-fitting masses are in general lower than when fit with the burst+constant models here.  

Those SEDs with redder UV colours allow for more extinction and so the uncertainties in derived SFR are larger, but this is also true when fitting them with the smoothly varying SFHs.  We find that two of these objects that allow for significant extinction (ID 13, 24) have very high best-fitting SFRs (83 M$_{\odot}$ yr$^{-1}$ and 438 M$_{\odot}$ yr$^{-1}$ respectively).  These high SFRs are due to the rest-frame UV and most of the rest-frame optical flux being dominated by the star-forming component, with the older burst component working to boost the flux in the optical.  However, it is worth noting that younger burst components in combination with lower constant SFR components are also allowable within the uncertainties.

Many of the best-fits indicate old burst components with small star-forming fractions ($<3\%$).  Firstly, taking this at face value, we can say it's incredibly unlikely that these objects formed most of their stars in a large starburst of short duration at very high redshift, given current observational constraints.  For example, a burst just 0.1 Myr old, with 0.02 Z$_{\odot}$ metallicity, that produced 10$^9$ M$_{\odot}$ of stars would be brighter than $H_{AB}=24.4$ when observed at $z\simeq 8$, although extinction with $E(B-V)=0.35$ could bring the brightness below the detection threshold for the CANDELS deep $H-$band imaging (5$\sigma$ point source depth of 27.7, \citealt{GroginNormanA.2011}).  It is possible, however, that these objects could be described by stochastic star formation histories that would leave an imprint on the observed SEDs that would be similar to an old, instantaneous burst.

\addtocounter{footnote}{1}

\begin{table*}
 \centering
 \begin{minipage}{180mm}
  \caption{Results of SED fitting with the set of burst+constant star formation models (template set D).  The first column gives the IDs of the objects and the second indicates the duration of constant star formation of the best-fitting model in Myr.  Columns 3-10 give the best-fitting mass, SFR, sSFR, extinction, age, $1\sigma$ lower limit in the age, fraction of mass in the star-forming population and metallicity (as a fraction of Solar) respectively.  All quoted errors are the 68\% confidence limits, maginalising over all other free parameters.  The final column gives the $\chi^2$ of the best-fitting model.  The typical degrees of freedom for for 9 photometric points is three.}
  \scalebox{0.8}{
  \begin{tabular}{@{}lcccccccccc@{}}
  \hline
  \hline
  ID & Duration & log$_{10}$(Mass /M$_{\odot}$) & SFR & sSFR & E(B-V) & log$_{10}$(Age /yr) & Age lower & Star-forming & Metallicity & $\chi^2$\\
   & /Myr     &                              & /M$_{\odot}$yr$^{-1}$ & (/Gyr$^{-1}$)  & && limit ($1\sigma$ /Myr) & fraction (/\%) & &\\
\hline
2 & 10 & \phantom{0}9.67$^{+0.10}_{-0.26}$ & \phantom{00}8.38$^{+22.55\phantom{0}}_{-7.82\phantom{00}}$ & 1.8 & 0.00$^{+0.08}_{-0.00}$ & 8.06$^{+0.30}_{-0.43}$ & 43 & \phantom{0}1.7$^{+13.3}_{-1.5\phantom{0}}$ & 1.0\phantom{0} & 2.60\\
5 & 10 & \phantom{0}9.96$^{+0.26}_{-0.36}$ & \phantom{0}28.86$^{+96.78\phantom{0}}_{-28.23\phantom{0}}$ & 3.2 & 0.00$^{+0.12}_{-0.00}$ & 8.26$^{+0.70}_{-0.67}$ & 39 & \phantom{0}3.0$^{+12.0}_{-2.9\phantom{0}}$ & 1.0\phantom{0} & 2.27\\
6 & 10 & \phantom{0}9.43$^{+0.50}_{-0.25}$ & \phantom{00}8.46$^{+12.57\phantom{0}}_{-4.44\phantom{00}}$ & 3.2 & 0.00$^{+0.08}_{-0.00}$ & 8.51$^{+0.45}_{-0.40}$ & 128 & \phantom{0}3.0$^{+12.0}_{-2.0\phantom{0}}$ & 0.2\phantom{0} & 3.11\\
8 & 10 & 10.12$^{+0.20}_{-0.21}$ & \phantom{0}19.40$^{+14.36\phantom{0}}_{-9.11\phantom{00}}$ & 1.5 & 0.00$^{+0.04}_{-0.00}$ & 8.71$^{+0.25}_{-0.65}$ & 114 & \phantom{0}1.4$^{+9.6\phantom{0}}_{-0.3\phantom{0}}$ & 0.02 & 5.10\\
13 & 10 & \phantom{0}9.72$^{+0.26}_{-0.69}$ & \phantom{0}83.13$^{+30.70\phantom{0}}_{-83.13\phantom{0}}$ & 15.9 & 0.25$^{+0.12}_{-0.12}$ & 8.81$^{+0.15}_{-1.79}$ & 10 & 15.0$^{+0.0\phantom{0}}_{-15.0}$ & 1.0\phantom{0} & 2.80\\
18 & 30 & \phantom{0}9.02$^{+0.50}_{-0.32}$ & \phantom{00}0.32$^{+29.31\phantom{0}}_{-0.32\phantom{00}}$ & 0.3 & 0.00$^{+0.25}_{-0.00}$ & 7.91$^{+1.05}_{-0.89}$ & 10 & \phantom{0}0.8$^{+14.2}_{-0.8\phantom{0}}$ & 0.02 & 1.23\\
23 & 10 & \phantom{0}9.97$^{+0.54}_{-0.20}$ & \phantom{00}4.89$^{+163.80}_{-4.89\phantom{00}}$ & 0.5 & 0.00$^{+0.21}_{-0.00}$ & 8.21$^{+0.70}_{-0.62}$ & 39 & \phantom{0}0.5$^{+14.5}_{-0.5\phantom{0}}$ & 0.02 & 1.04\\
24 & 10 & 10.92$^{+0.12}_{-1.02}$ & 438.00$^{+598.93}_{-438.00}$ & 5.3 & 0.33$^{+0.08}_{-0.12}$ & 9.01$^{+0.00}_{-1.99}$ & 10 & \phantom{0}5.0$^{+10.0}_{-5.0\phantom{0}}$ & 1.0\phantom{0} & 1.02\\
25 & 10 & 10.04$^{+0.27}_{-0.51}$ & \phantom{0}34.72$^{+88.02\phantom{0}}_{-34.12\phantom{0}}$ & 3.2 & 0.08$^{+0.12}_{-0.08}$ & 8.56$^{+0.45}_{-1.07}$ & 31 & \phantom{0}3.0$^{+12.0}_{-2.9\phantom{0}}$ & 1.0\phantom{0} & 1.97\\
26 & 10 & \phantom{0}9.17$^{+0.83}_{-0.35}$ & \phantom{00}0.00$^{+117.41}_{-0.00\phantom{00}}$ & 0.0 & 0.17$^{+0.12}_{-0.12}$ & 7.26$^{+1.70}_{-0.24}$ & 10 & \phantom{0}0.0$^{+15.0}_{-0.0\phantom{0}}$ & 0.4\phantom{0} & 1.95\\
27 & 30 & \phantom{0}9.37$^{+0.53}_{-0.31}$ & \phantom{00}0.54$^{+58.89\phantom{0}}_{-0.21\phantom{00}}$ & 0.2 & 0.00$^{+0.17}_{-0.00}$ & 7.91$^{+1.05}_{-0.55}$ & 23 & \phantom{0}0.6$^{+14.4}_{-0.4\phantom{0}}$ & 0.02 & 1.03\\
157 & 10 & \phantom{0}9.57$^{+0.27}_{-0.27}$ & \phantom{0}11.62$^{+42.13\phantom{0}}_{-10.87\phantom{0}}$ & 3.1 & 0.00$^{+0.08}_{-0.00}$ & 8.01$^{+0.80}_{-0.38}$ & 43 & \phantom{0}3.0$^{+12.0}_{-2.7\phantom{0}}$ & 0.02 & 5.23\\
248 & 10 & 10.04$^{+0.19}_{-0.18}$ & \phantom{0}18.62$^{+11.34\phantom{0}}_{-7.67\phantom{00}}$ & 1.7 & 0.00$^{+0.04}_{-0.00}$ & 8.71$^{+0.20}_{-0.25}$ & 286 & \phantom{0}1.6$^{+5.4\phantom{0}}_{-0.6\phantom{0}}$ & 0.2\phantom{0} & 4.77\\

  \hline
  \hline
\end{tabular}}
\label{tab:BCModels}
\end{minipage}
\end{table*}

\addtocounter{footnote}{1}

\subsubsection{New age determinations and intrinsic uncertainties}
In this section we investigate the ages derived with the burst+constant models to see whether any of the objects still require an old stellar population once the instantaneous SFR has been decoupled from the accrued mass.  As we have set the constraint that the model must supply enough ionising flux to describe the observed Ly$\alpha$, the lower limits in age (the youngest models supplying fits within $\Delta\chi^2=1$) are not just set by the age of a burst that fits the rest-frame UV, but must also supply the minimum flux of ionising photons required to explain the Ly$\alpha$ emission.  However, caution must be used regarding the results for objects 18 and 26 because we were not able to apply this constraint and we see that the lower age limit is set by the age of a pure burst, with no current star-formation.  We know these two objects have Ly$\alpha$ in emission so these lower limits are likely to be unphysical. 

Many of the objects do not have an upper age constraint, with the upper limits butting up against the age of the Universe.  This is due to the degeneracy between the star-forming fraction and the age of the burst (as well as, to a lesser extent, extinction and metallicity).  Essentially, a model with a fairly young burst ($\gtrsim100$ Myr), which constitutes a majority of the flux in the rest-frame optical, plus a certain star-forming fraction, looks very similar to a model in which the star-forming fraction dominates the rest-frame optical flux but an old, massive burst also makes a significant contribution.  In other words, the older a population is, the easier it is masked by any current star-formation.  We cannot, therefore, rule out old ages for most of the objects although we provide the caveat that the older the burst, the larger the mass of stars required to be built in a shorter amount of time, and hence the less likely it is to represent reality.  Instead we continue to ask the question of whether old ages are \textit{required} to fit to the observed SEDs.

First we see that some of the objects do not have any true age constraints, with lower limits defined by the lowest ages of the models and upper limits approaching the age of the Universe (objects 13, 18, 24 and 26).  These either have no constraint set by Ly$\alpha$, either with poorly calibrated Ly$\alpha$ emission (26) or because the system is an absorption system (24), or the SEDs allow for a lot of intrinsic reddening (13, 24, 26).  We cannot constrain the age when reddening introduces so many degeneracies or if fits are allowed the freedom to fit the majority of the UV with the non-star forming burst component.

The two objects that required an old population from template set C, objects 8 and 248, have constrained lower age-limits of 115 and 290 Myr respectively. Are these ages robust when nebular emission is added to the templates?

\subsection{Templates with nebular emission (template set E)}

The template set with full nebular emission is produced by adding nebular emission (continuum and line emission) to the burst+constant star formation templates, using the method of \cite{Robertson2010a} which is summarised in \cite{McLure2011}.  As we are primarily interested in the effect of nebular emission on the derived ages and masses of the objects, we restrict ourselves to a single value of the escape fraction of ionising photons, $f_{esc}=0.2$, which is motivated by the typical value of $f_{esc}$ which is required for star-forming galaxies to maintain reionisation at $z\sim7$ \citep{Robertson2010a}. 

Nebular continuum and line emission is then added to the burst+constant templates using the number of ionising photons per second, $N_{Lyc}$, provided as an output from the BC03 models.  This number is used to calculate the H$\beta$ luminosity, and other HI line intensities follow from the ratios predicted by standard recombination theory \citep{Osterbrock2006}.  Emission lines of metals are added using the empirically-derived, metallicity-dependent, line intensities (relative to H$\beta$) from \cite{Anders2003}.

\subsubsection{Results}

We find that a subset of the objects (IDs 5, 6, 8, 18, 27, 157, 248) give marginally better fits when nebular emission is added to the models, although for most of these objects the $\chi^2$ improvement is not statistically significant (the minimum $\chi^2$ values are within $\Delta\chi^2=1$).  We cannot say that these objects \textit{require} nebular emission from the templates to provide suitable fits, but we can investigate how the inferred physical properties are affected as well as the properties of the SEDs that are better reproduced when nebular emission is included.  We defer discussion of the second point to Section 8.1 and concentrate here on how the inferred physical properties of the objects can change.

The best-fitting models from this template set give systematically lower best-fitting masses, with a median offset of 0.18 dex, compared to the same template set without nebular emission.  This is comparable to the results of \cite{McLure2011} where, for a subset of their photometrically-selected $6<z_{phot}<8.7$ LBGs detected at 3.6$\mu$m, 16 out of 21 objects show an offset of less than 0.4 dex when comparing results with and without nebular emission (the templates used for fitting to the object SEDs consisted of a single exponentially decaying SFH with two possible metallicities).  The lower mass estimates are caused by additional nebular line contribution to the optical fluxes.  A burst with a deep Balmer break is, in many cases, no longer needed to fit to the observed optical fluxes, as found by other studies (e.g. \citealt{Schaerer2009,Labbe2010}).  

The best-fitting SFRs derived with nebular emission, however, are very similar to those derived from the templates without nebular emission.  This is due to the rest-frame UV constraining the amount of star formation and the inclusion of nebular continuum emission, which acts to redden the rest-frame UV, only minimally affecting the SFR determination.    

The SFRs are actually better constrained for this template set compared to sets D (and C) because the higher mass, higher SFR models no-longer provide adequate fits to the SEDs.  This is because a high SFR produces a large contribution to the rest-frame optical fluxes via nebular emission lines that were not accounted for in the previous template sets.  Nebular emission therefore limits the contribution of the burst component to the IRAC fluxes, therefore limiting the derived masses and SFRs.  For example, the extremely high SFRs from template set D are now not allowed within $\Delta\chi^2=1$ for objects 13 and 24 due to the addition of nebular emission, although the best-fitting SFR of object 24 is still quite high.  It is worth noting here that we assume the old and young stellar populations show the same levels of extinction and that the nebular extinction is the same as the stellar extinction.  Relaxing these constraints would likely show that the SFRs were as uncertain as when estimated from templates without nebular emission, as it would break the direct correspondence between SFR and nebular line contribution to the optical fluxes.

\subsubsection{Age determination from templates with Nebular Emission}

Fig.~\ref{fig:multiSEDs} shows the best-fitting SEDs for each of the objects derived from each of the template sets, C, D and E. In particular, the lower right-hand panel shows the best-fits to object 248 which previously appeared to require an old stellar population.  The lower age limit for 248 is still fairly high at 227 Myr, although it can be seen that its blue 3.6$\mu$m$-$4.5$\mu$m colour is better fit by the models including nebular emission, they have some difficulty in reproducing the very blue colours.  This is discussed in more detail in Section 8.2.

\begin{table*}
 \centering
 \begin{minipage}{180mm}
  \caption{Results of SED fitting using the set of burst+constant star formation models with added nebular emission (template set E).  The first column gives the IDs of the objects and the second indicates the duration of constant star formation of the best-fitting model in Myr.  Columns 3-10 give the best-fitting mass, SFR, sSFR extinction, age, lower age limit, fraction of mass in the star-forming population and metallicity (as a fraction of Solar) respectively.  All quoted errors are the 68\% confidence limits, maginalising over all other free parameters.  The final column give the best-fitting model and $\chi^2$.  The typical degrees of freedom for 9 photometric points is 3.}
  \scalebox{0.8}{
  \begin{tabular}{@{}lcccccccccc@{}}
  \hline
  \hline
  ID & Duration & log$_{10}$(Mass /M$_{\odot}$) & SFR & sSFR & E(B-V) & log$_{10}$(Age /yr) & Age lower & Star-forming & Metallicity & $\chi^2$\\
   & /Myr     &                              & /M$_{\odot}$yr$^{-1}$ & (/Gyr$^{-1}$)  & && limit ($1\sigma$ /Myr) & fraction (/\%) & &\\
\hline
2 & 10 & \phantom{0}9.52$^{+0.16}_{-0.30}$ & \phantom{0}6.96$^{+42.29\phantom{0}}_{-6.21\phantom{00}}$ & 2.1 & 0.00$^{+0.12}_{-0.00}$ & 7.91$^{+1.05}_{-0.34}$ & 37 & \phantom{0}2.0$^{+13.0}_{-1.8\phantom{0}}$ & 1.0\phantom{0} & 2.61\\
5 & 10 & \phantom{0}9.73$^{+0.36}_{-0.42}$ & 28.63$^{+50.07\phantom{0}}_{-24.58\phantom{0}}$ & 5.3 & 0.00$^{+0.08}_{-0.00}$ & 8.16$^{+0.80}_{-0.90}$ & 18 & \phantom{0}5.0$^{+10.0}_{-4.0\phantom{0}}$ & 1.0\phantom{0} & 2.16\\
6 & 10 & \phantom{0}9.07$^{+0.72}_{-0.36}$ & \phantom{0}8.61$^{+4.96\phantom{00}}_{-4.69\phantom{00}}$ & 7.4 & 0.00$^{+0.04}_{-0.00}$ & 8.46$^{+0.50}_{-0.45}$ & 102 & \phantom{0}7.0$^{+8.0\phantom{0}}_{-5.7\phantom{0}}$ & 0.2\phantom{0} & 3.01\\
8 & 10 & 10.02$^{+0.25}_{-0.45}$ & 19.78$^{+14.61\phantom{0}}_{-9.56\phantom{00}}$ & 1.9 & 0.00$^{+0.04}_{-0.00}$ & 8.76$^{+0.20}_{-0.50}$ & 181 & \phantom{0}1.8$^{+12.2}_{-0.7\phantom{0}}$ & 0.02 & 5.00\\
13 & 10 & \phantom{0}9.25$^{+0.17}_{-0.33}$ & \phantom{0}1.89$^{+23.13\phantom{0}}_{-1.08\phantom{00}}$ & 1.1 & 0.21$^{+0.04}_{-0.12}$ & 7.16$^{+0.41}_{-0.14}$ & 10 & \phantom{0}1.0$^{+14.0}_{-0.0\phantom{0}}$ & 0.4\phantom{0} & 2.89\\
18 & 10 & \phantom{0}8.73$^{+0.67}_{-0.21}$ & \phantom{0}8.40$^{+6.52\phantom{00}}_{-7.55\phantom{00}}$ & 15.8 & 0.04$^{+0.08}_{-0.04}$ & 8.96$^{+0.00}_{-0.95}$ & 102 & 15.0$^{+0.0\phantom{0}}_{-14.0}$ & 0.2\phantom{0} & 0.53\\
23 & 30 & \phantom{0}9.93$^{+0.38}_{-0.17}$ & \phantom{0}2.56$^{+106.93}_{-2.56\phantom{00}}$ & 0.3 & 0.00$^{+0.17}_{-0.00}$ & 8.16$^{+0.75}_{-0.15}$ & 102 & \phantom{0}0.8$^{+14.2}_{-0.8\phantom{0}}$ & 0.02 & 1.06\\
24 & 10 & \phantom{0}9.96$^{+0.78}_{-0.16}$ & \phantom{0}9.78$^{+393.42}_{-3.42\phantom{00}}$ & 1.1 & 0.29$^{+0.04}_{-0.08}$ & 7.08$^{+1.93}_{-0.06}$ & 10 & \phantom{0}1.0$^{+14.0}_{-0.0\phantom{0}}$ & 1.0\phantom{0} & 1.29\\
25 & 10 & \phantom{0}9.90$^{+0.24}_{-0.45}$ & 33.60$^{+44.24\phantom{0}}_{-33.22\phantom{0}}$ & 4.2 & 0.08$^{+0.08}_{-0.08}$ & 8.61$^{+0.40}_{-0.70}$ & 81 & \phantom{0}4.0$^{+11.0}_{-3.9\phantom{0}}$ & 1.0\phantom{0} & 2.13\\
26 & 10 & \phantom{0}9.32$^{+0.39}_{-0.20}$ & 33.28$^{+0.00\phantom{00}}_{-29.23\phantom{0}}$ & 15.9 & 0.17$^{+0.00}_{-0.12}$ & 8.96$^{+0.00}_{-0.95}$ & 102 & 15.0$^{+0.0\phantom{0}}_{-11.0}$ & 1.0\phantom{0} & 2.24\\
27 & 10 & \phantom{0}9.19$^{+0.52}_{-0.32}$ & 24.54$^{+1.99\phantom{00}}_{-23.57\phantom{0}}$ & 15.8 & 0.08$^{+0.00}_{-0.08}$ & 8.96$^{+0.00}_{-1.70}$ & 18 & 15.0$^{+0.0\phantom{0}}_{-14.5}$ & 0.4\phantom{0} & 0.92\\
157 & 10 & \phantom{0}9.17$^{+0.41}_{-0.15}$ & 20.01$^{+15.10\phantom{0}}_{-7.46\phantom{00}}$ & 13.7 & 0.00$^{+0.04}_{-0.00}$ & 8.01$^{+0.95}_{-0.38}$ & 43 & 13.0$^{+2.0\phantom{0}}_{-8.0\phantom{0}}$ & 0.2\phantom{0} & 3.39\\
248 & 10 & \phantom{0}9.97$^{+0.14}_{-0.36}$ & 18.62$^{+10.38\phantom{0}}_{-6.24\phantom{00}}$ & 2.0 & 0.00$^{+0.04}_{-0.00}$ & 8.81$^{+0.10}_{-0.45}$ & 227 & \phantom{0}1.9$^{+3.1\phantom{0}}_{-0.6\phantom{0}}$ & 0.2\phantom{0} & 3.57\\

  \hline
  \hline
\end{tabular}}
\label{tab:nebular}
\end{minipage}
\end{table*}

\section{Properties of UV-bright LBGs}

\subsection{Observed SED characteristics driving inferred physical parameters}

\begin{figure*} 
   \centering
   \subfigure{\includegraphics[width=2in,trim=2cm 13cm 1.9cm 5cm,clip]{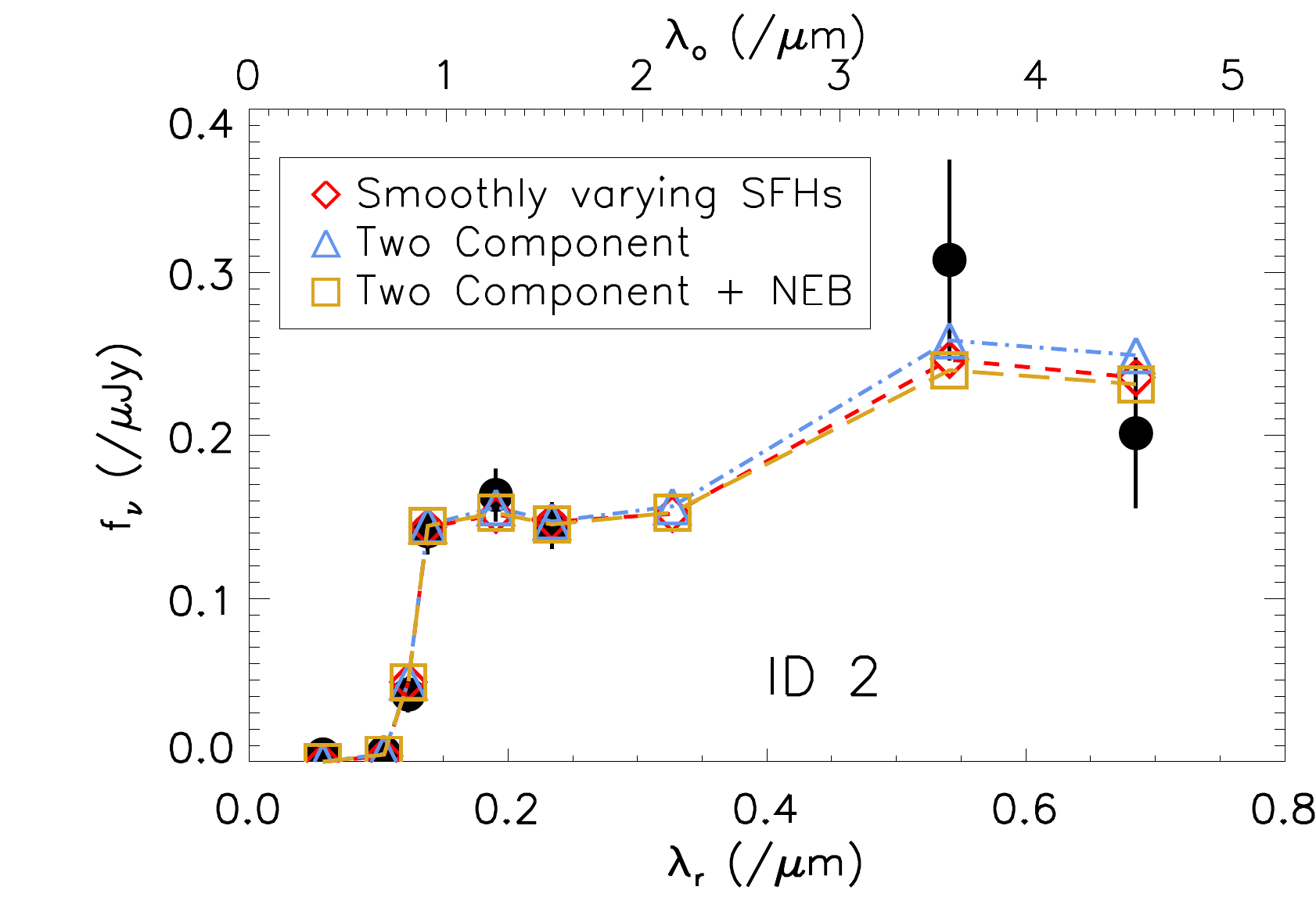}} 
   \subfigure{\includegraphics[width=2in,trim=2cm 13cm 1.9cm 5cm,clip]{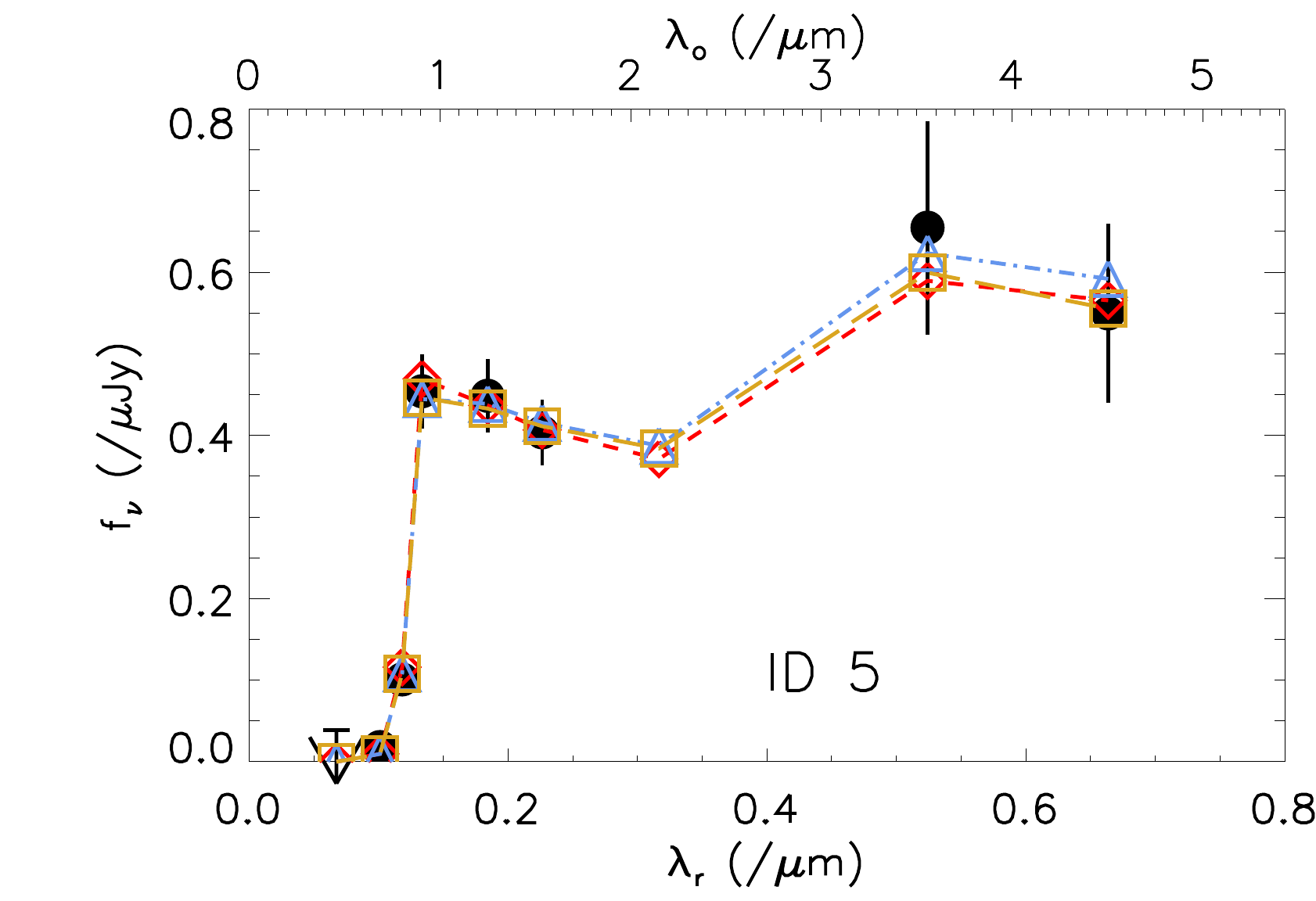}}
   \subfigure{\includegraphics[width=2in,trim=2cm 13cm 1.9cm 5cm,clip]{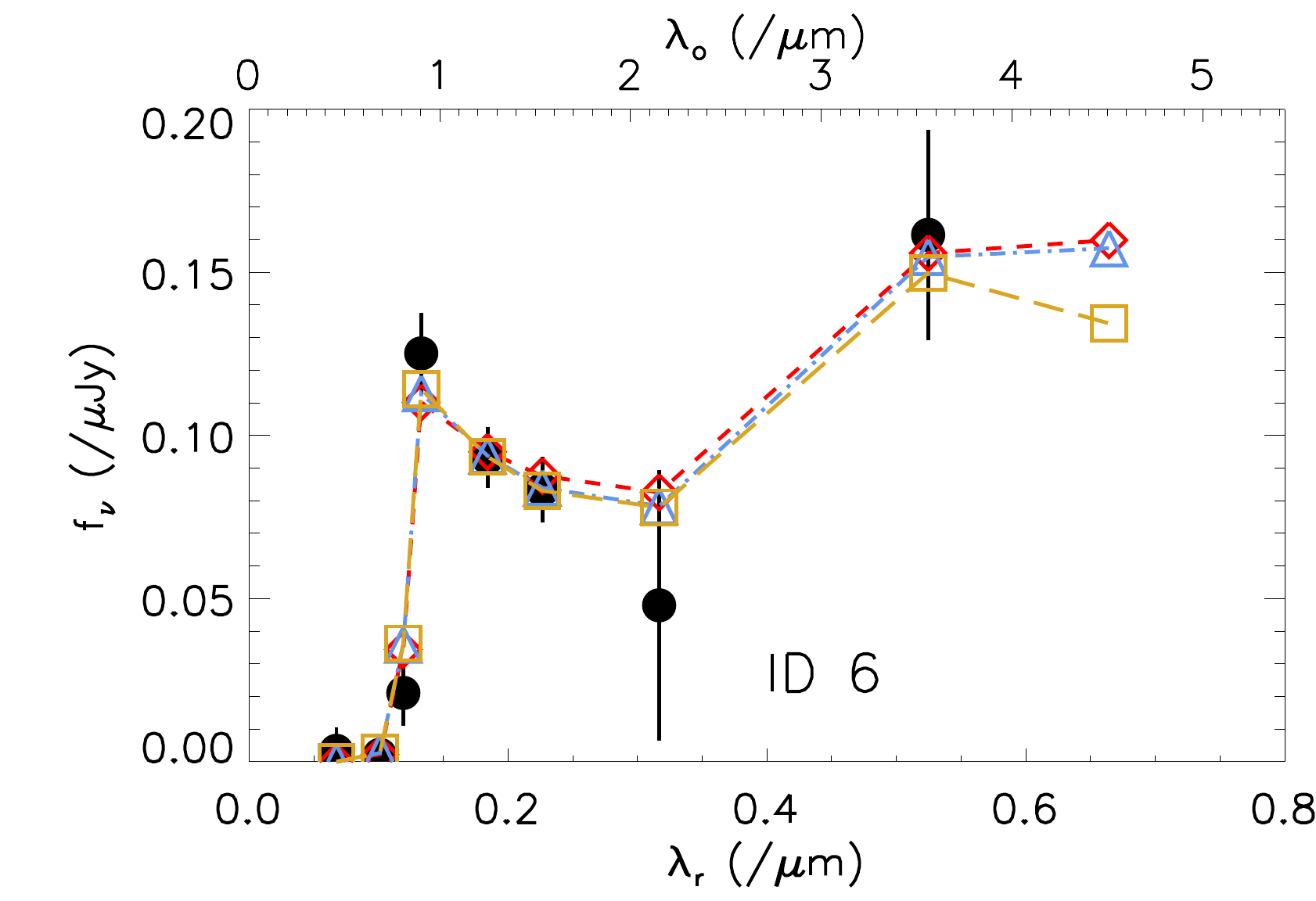}}
   \subfigure{\includegraphics[width=2in,trim=2cm 13cm 1.9cm 5cm,clip]{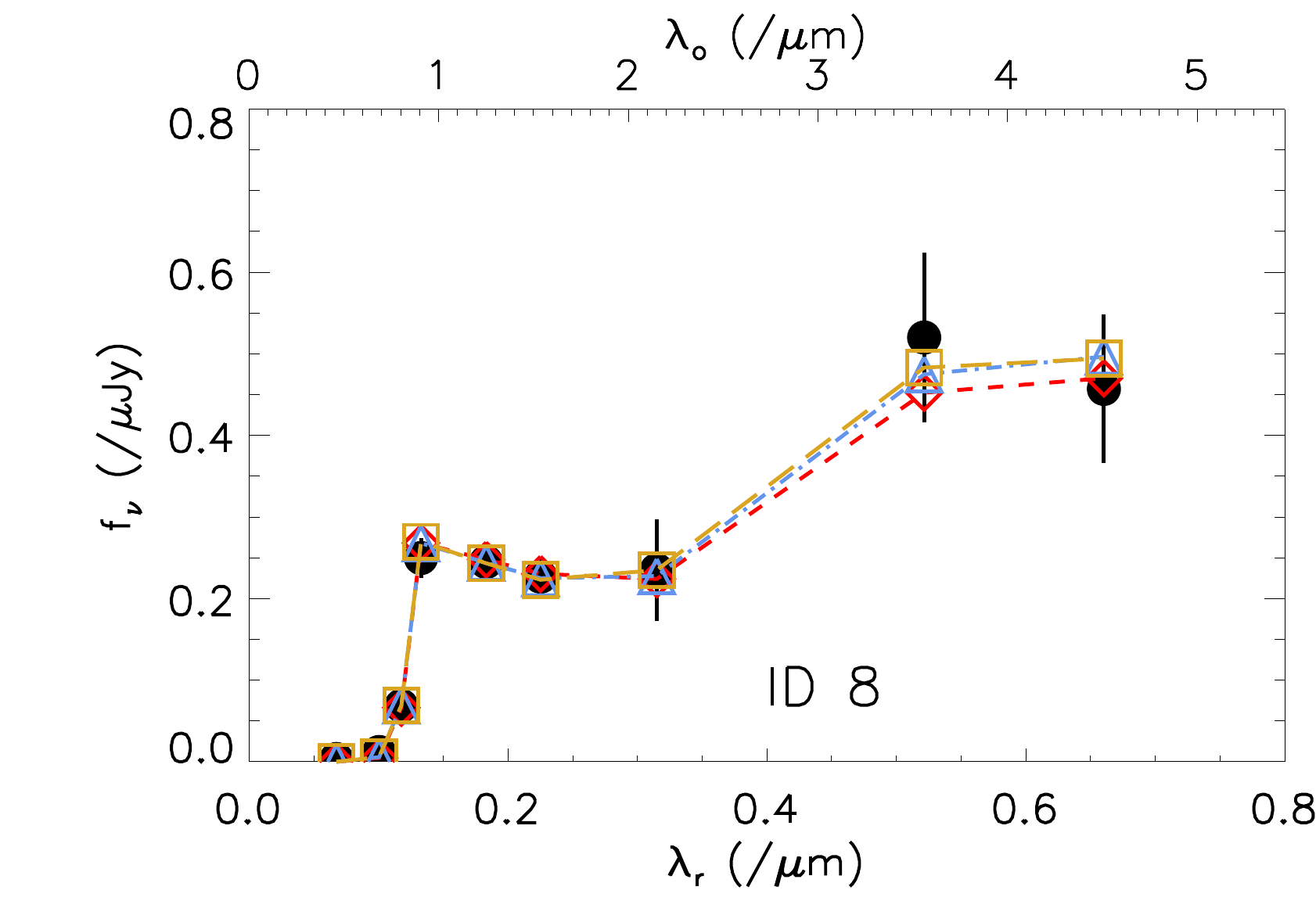}}
   \subfigure{\includegraphics[width=2in,trim=2cm 13cm 1.9cm 5cm,clip]{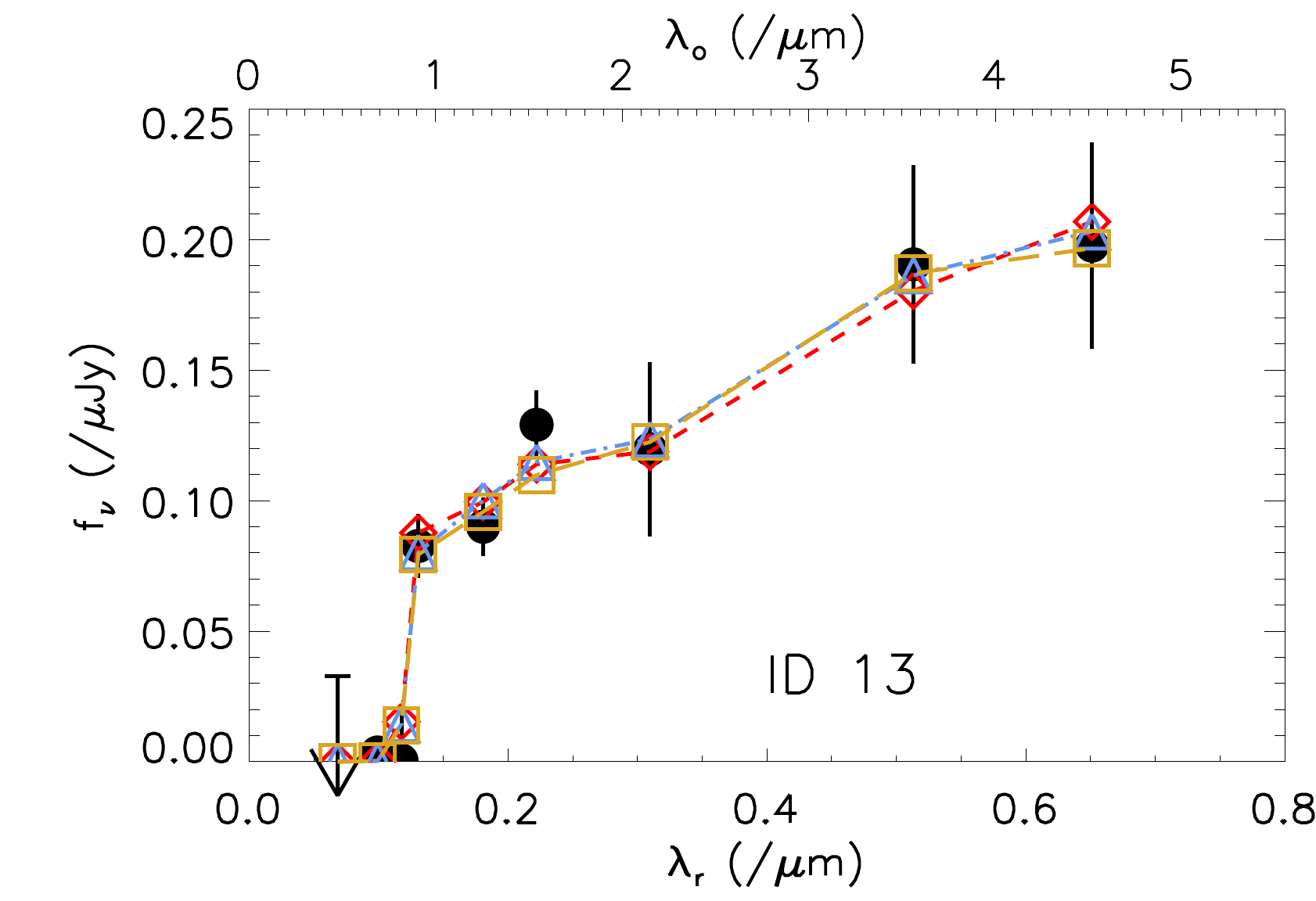}}
   \subfigure{\includegraphics[width=2in,trim=2cm 13cm 1.9cm 5cm,clip]{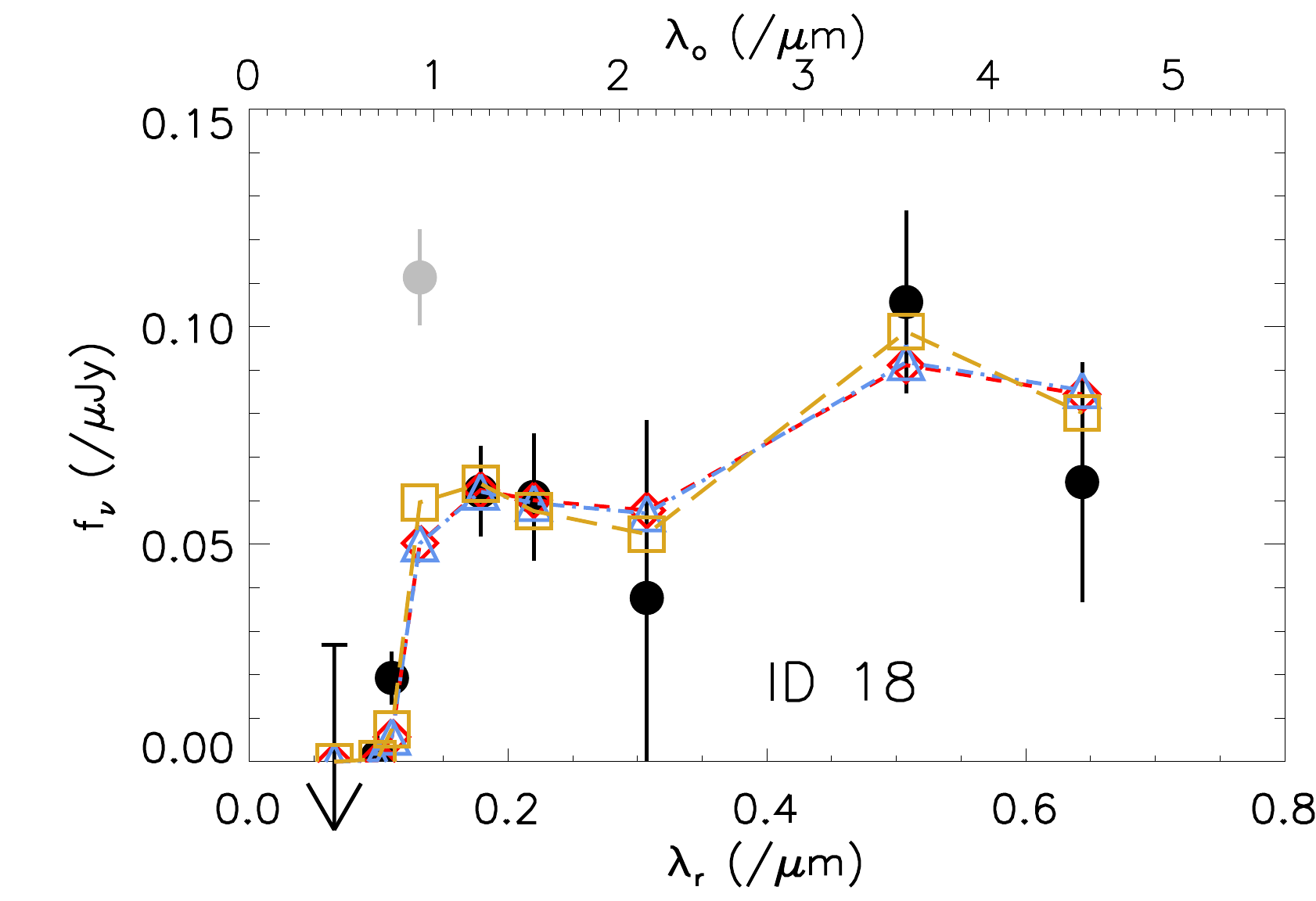}}
   \subfigure{\includegraphics[width=2in,trim=2cm 13cm 1.9cm 5cm,clip]{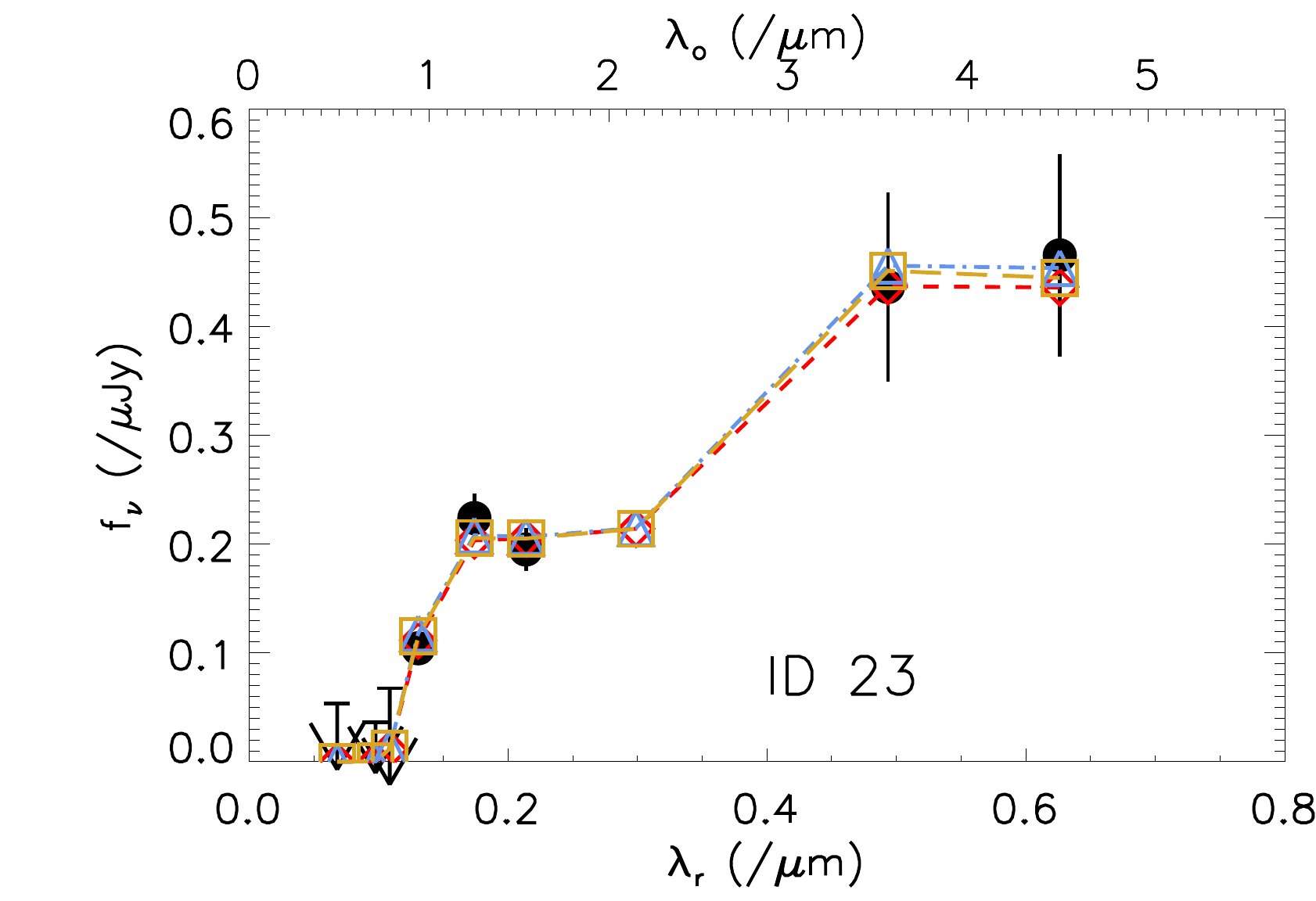}}
   \subfigure{\includegraphics[width=2in,trim=2cm 13cm 1.9cm 5cm,clip]{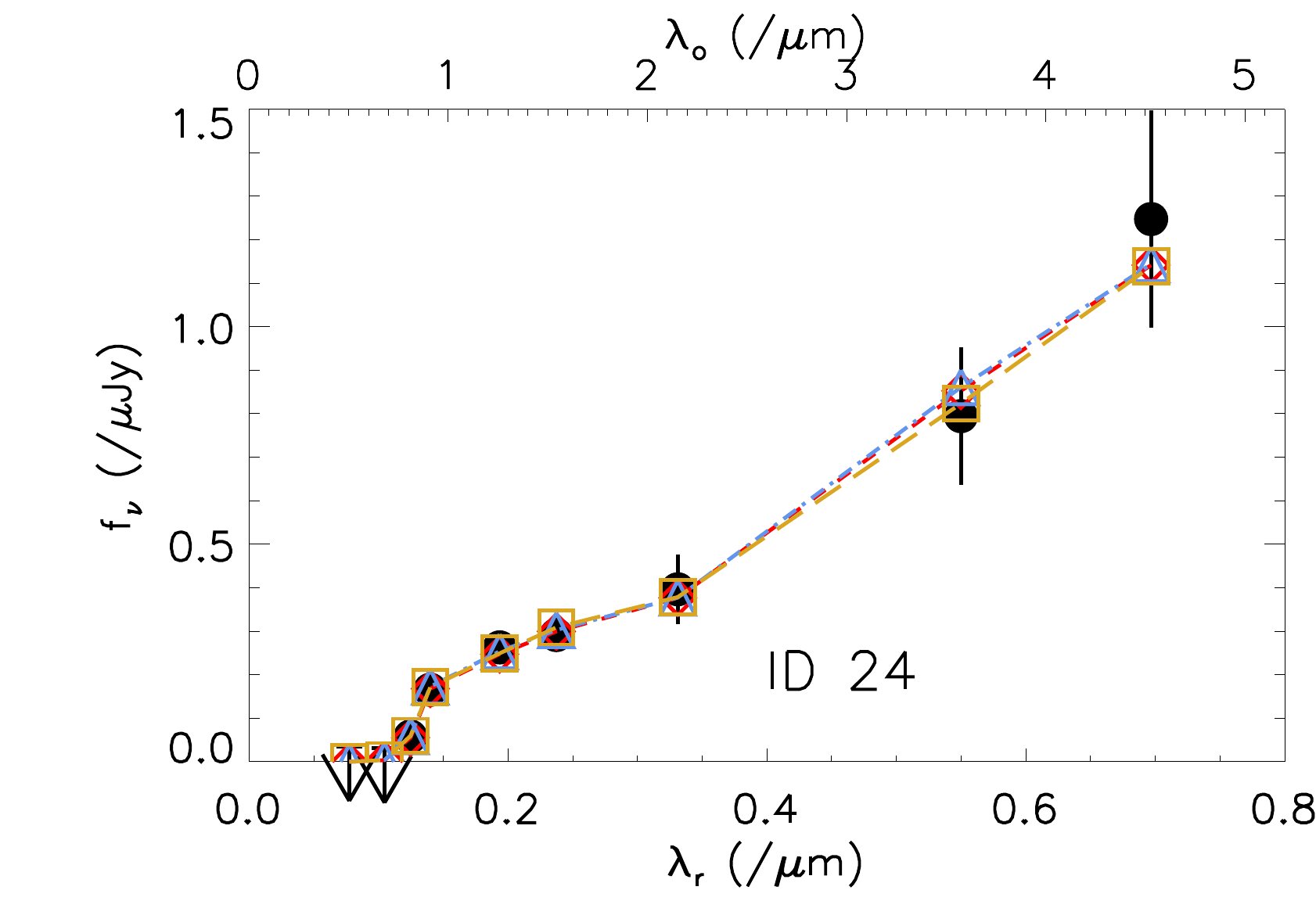}}
   \subfigure{\includegraphics[width=2in,trim=2cm 13cm 1.9cm 5cm,clip]{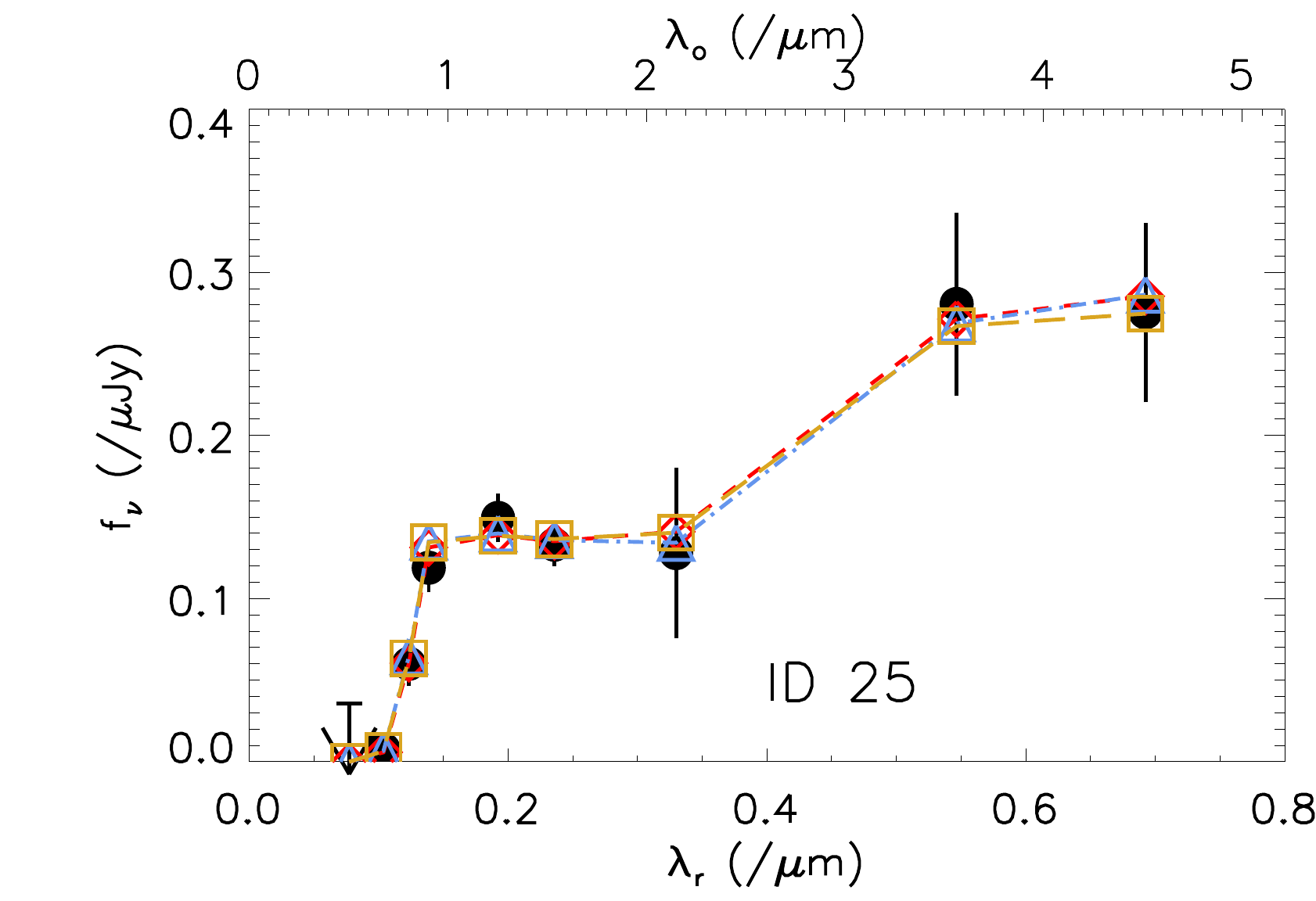}}
   \subfigure{\includegraphics[width=2in,trim=2cm 13cm 1.9cm 5cm,clip]{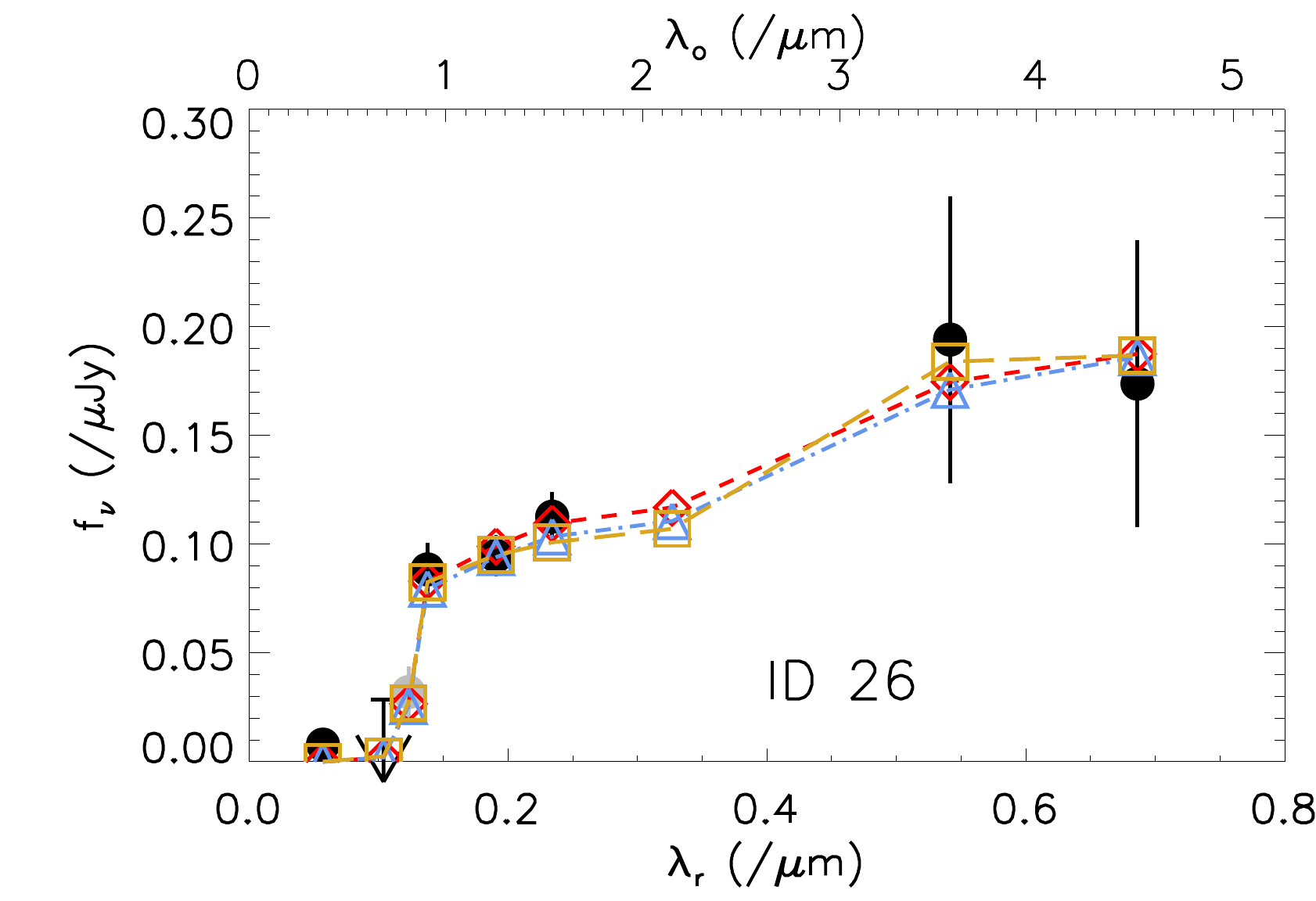}}
   \subfigure{\includegraphics[width=2in,trim=2cm 13cm 1.9cm 5cm,clip]{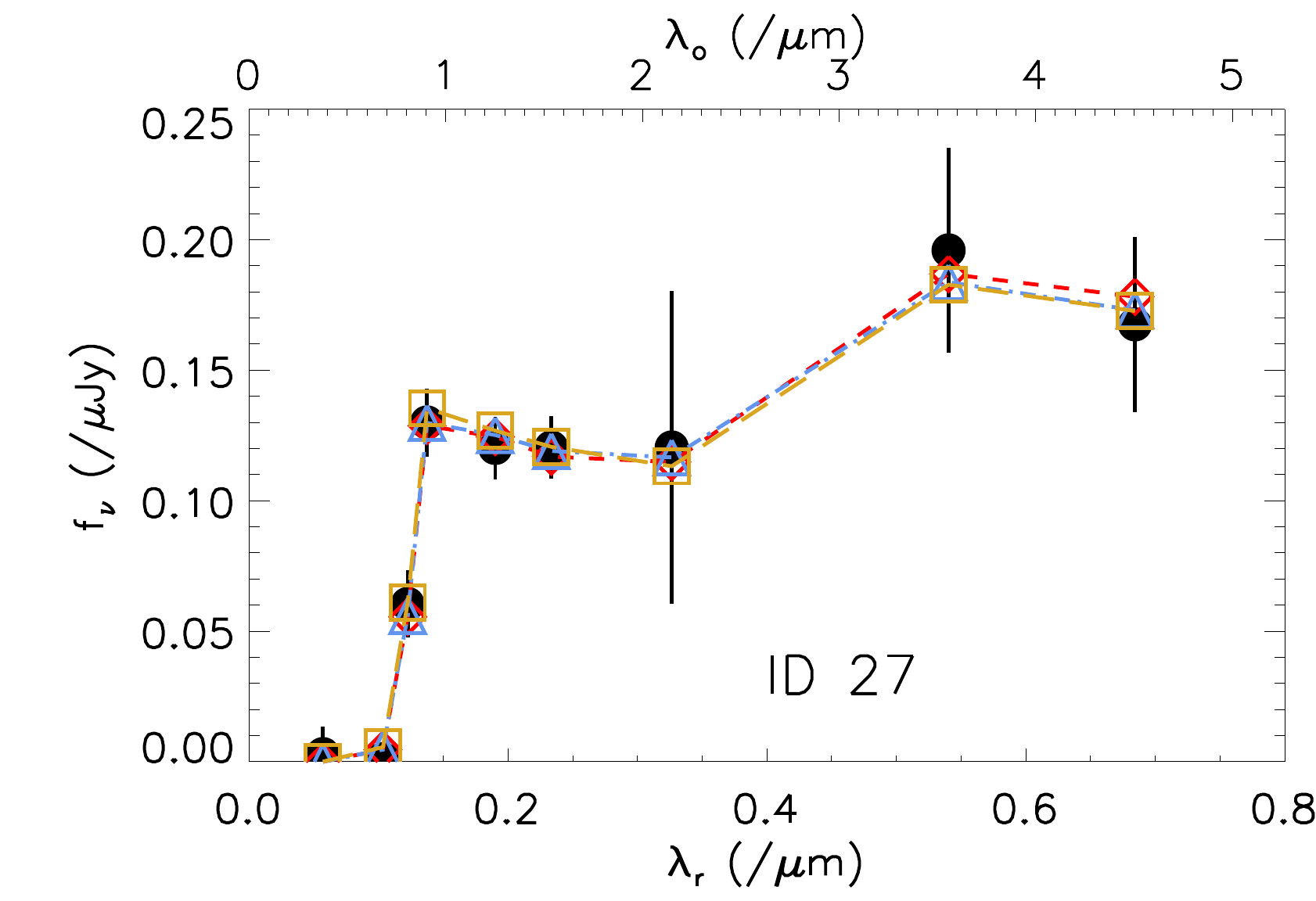}}
   \subfigure{\includegraphics[width=2in,trim=2cm 13cm 1.9cm 5cm,clip]{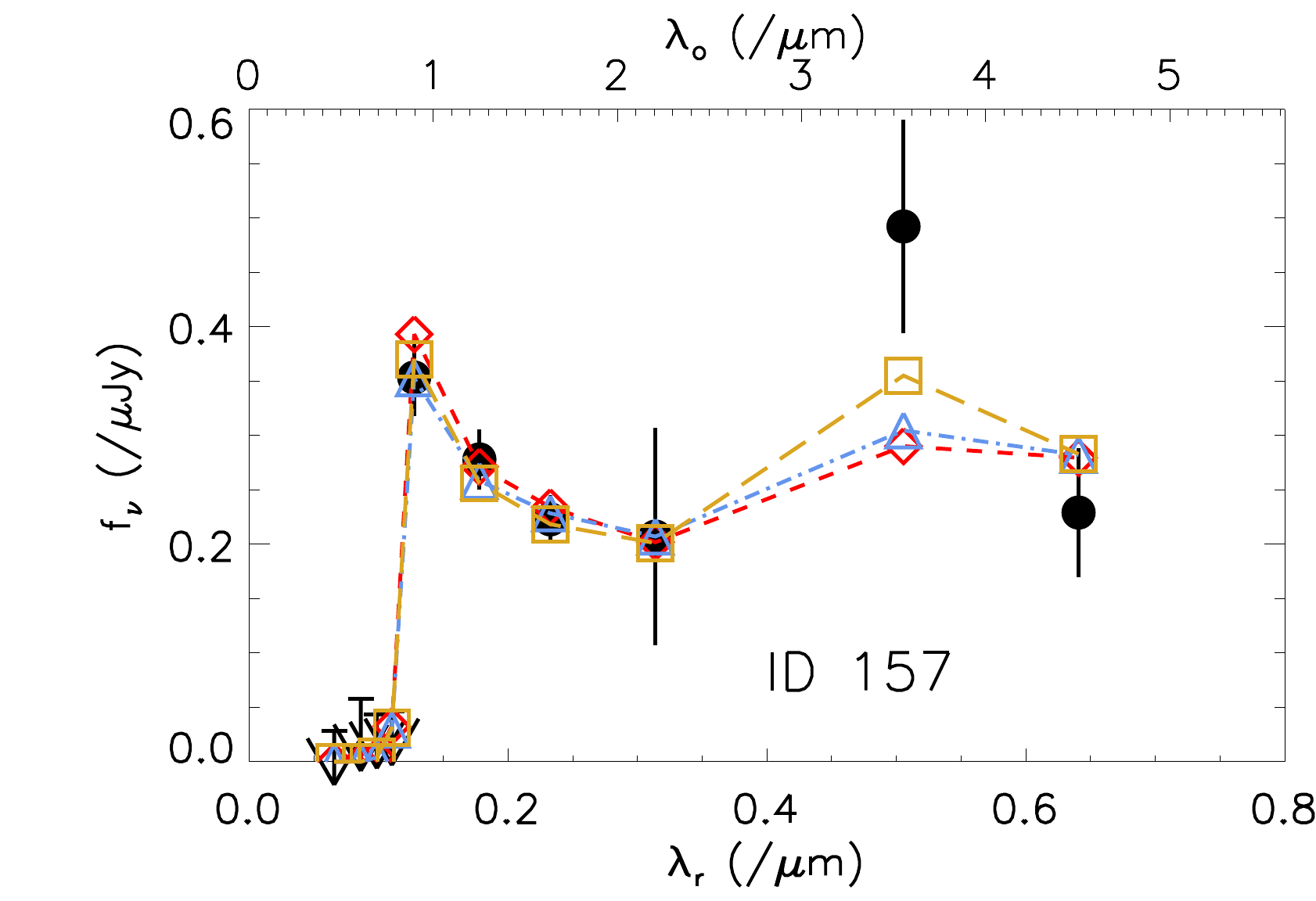}}
   \subfigure{\includegraphics[width=2in,trim=2cm 13cm 1.9cm 5cm,clip]{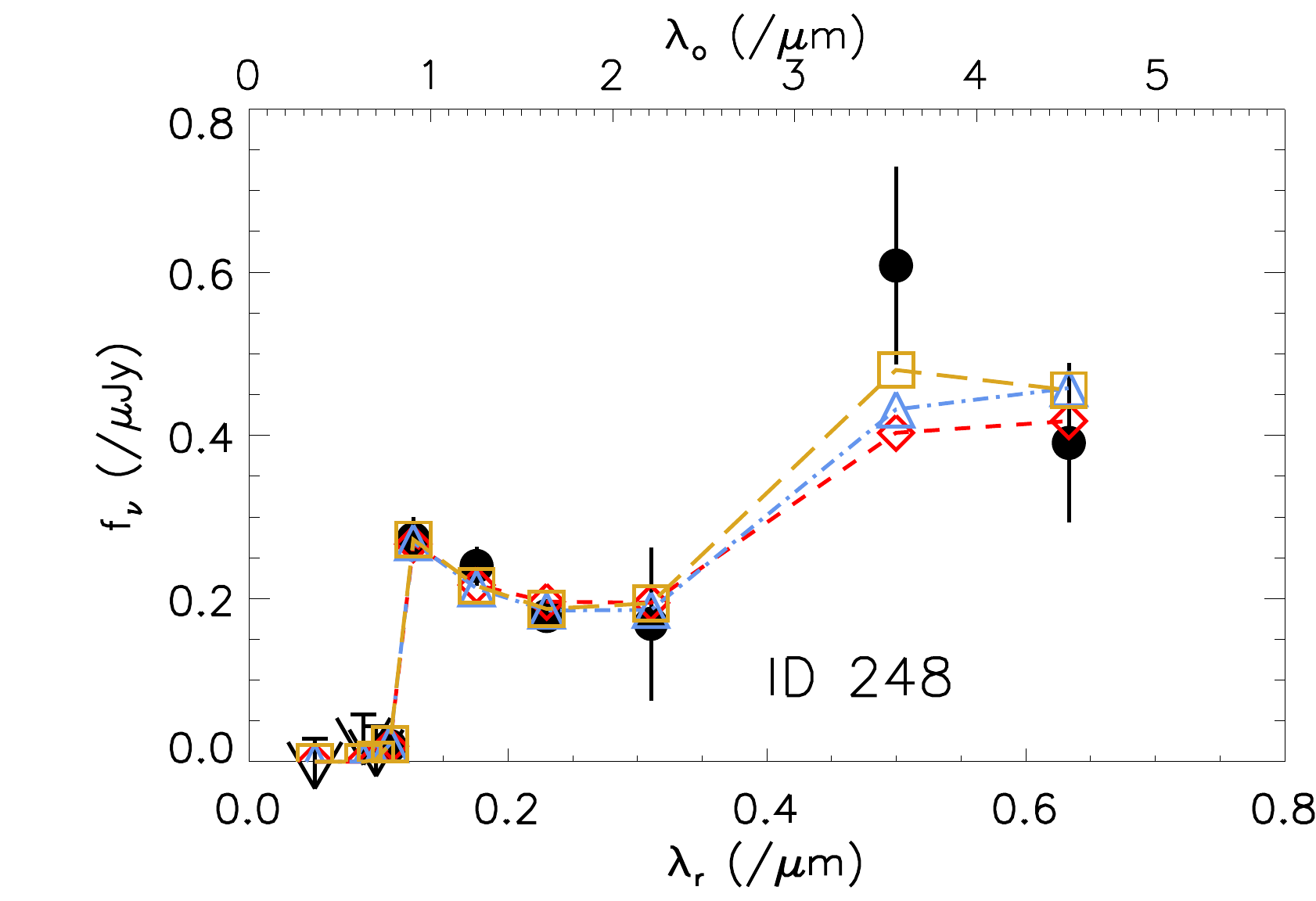}}
   \caption{The best-fitting SEDs for each object from the following template sets: Smoothly varying SFHs (template set C, plotted as red diamonds), two component SFHs without nebular emission (template set D, plotted as blue triangles) and with nebular emission (template set E, plotted as yellow squares).  The observed SEDs and corresponding errors are plotted as black circles, with any filters that have been excluded from the fitting due to an uncertain contribution from Ly$\alpha$ being plotted in grey.}
   \label{fig:multiSEDs}
\end{figure*}

We have used a number of different template sets for SED-fitting to our sample of spectroscopically confirmed LBGs.  The differences between the inferred physical properties from each template set illustrates how restraining our priors in SFHs, extinction, metallicity and nebular emission can influence our conclusions.  As has been well established, the most uncertain parameters are the ages and metallicities of the objects.  When nebular line emission, in particular, is omitted from templates then the inability to pin down the age and metallicity of the object does not impact significantly on the derived masses and allowing for a full range of each in the fitting allows for a realistic determination of the uncertainties.  

When nebular line emission is added, however, we see that it is possible for a recent starburst with strong nebular emission lines to mimic the signal of an old stellar population.  For certain redshift and filter combinations (in particular, for our sample at $z\sim6$ with the only two filters deep enough to sample the rest-frame optical region of the SED being the IRAC 3.6 and 4.5$\mu$m channels) this can lead to a case where the nebular emission contribution cannot be constrained, or equivalently the contribution of any underlying old population is also not well constrained, leading to marginally larger uncertainties on the derived stellar masses.

The SFRs derived from different template sets show good agreement, in general, because the deep near-IR imaging provides strong constraints in the rest-frame UV.  For objects that show evidence of intrinsic reddening in their SEDs, the SFRs are unconstrained because the level of extinction is very uncertain.  Adding nebular emission to the templates limits the maximum estimates of SFR as they would be required to contribute to the rest-frame optical fluxes via nebular line emission, but this relies on our assumption that the nebular and stellar components suffer the same extinction.

The observed SEDs and best-fitting model SEDs for the template sets C (all smoothly varying SFHs), D and E (two component SFHs without and with nebular emission, respectively) are plotted in Fig.~\ref{fig:multiSEDs}.  We can see that there is very little difference between the fits derived from any of the template sets, with the small errors from the deep HST imaging providing very good constraints in the UV.  There are differences in the contributions to the optical fluxes, although for the most part they are also marginal and in general can be attributed to the poorer constraints.  For the objects showing blue 3.6$\mu$m $-$ 4.6$\mu$m fluxes, however, this is not the case and they are slightly better reproduced by the templates including nebular emission.  This may indicate that there is indeed nebular emission contributing to the broad-band fluxes of these objects but the improvement in the fits are not statistically significant.

\begin{figure*}
  \centering
  \subfigure[]{\includegraphics[width=2.1in,trim=2cm 13cm 2cm 5cm,clip]{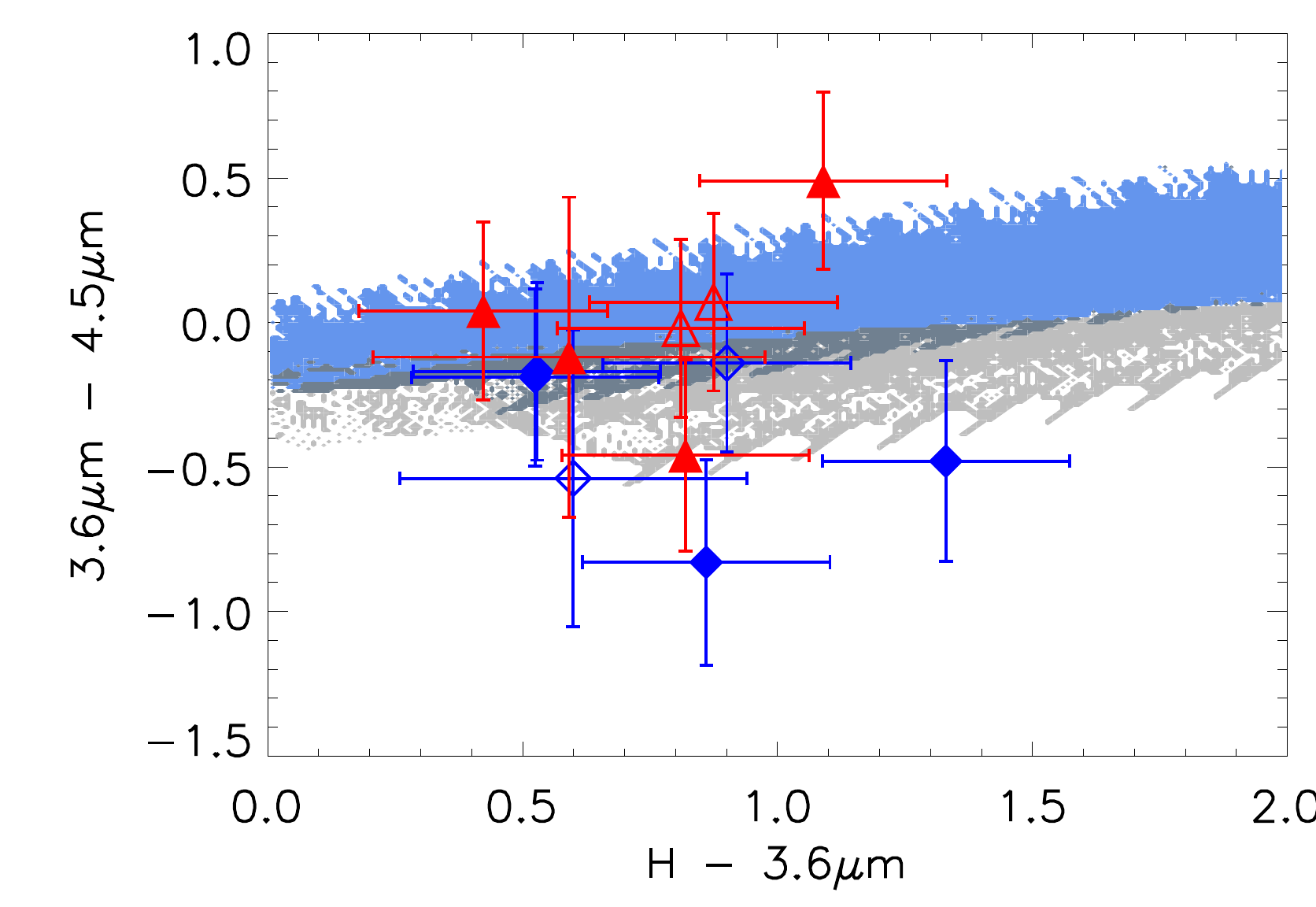}}
  \subfigure[]{\includegraphics[width=2.1in,trim=2cm 13cm 2cm 5cm,clip]{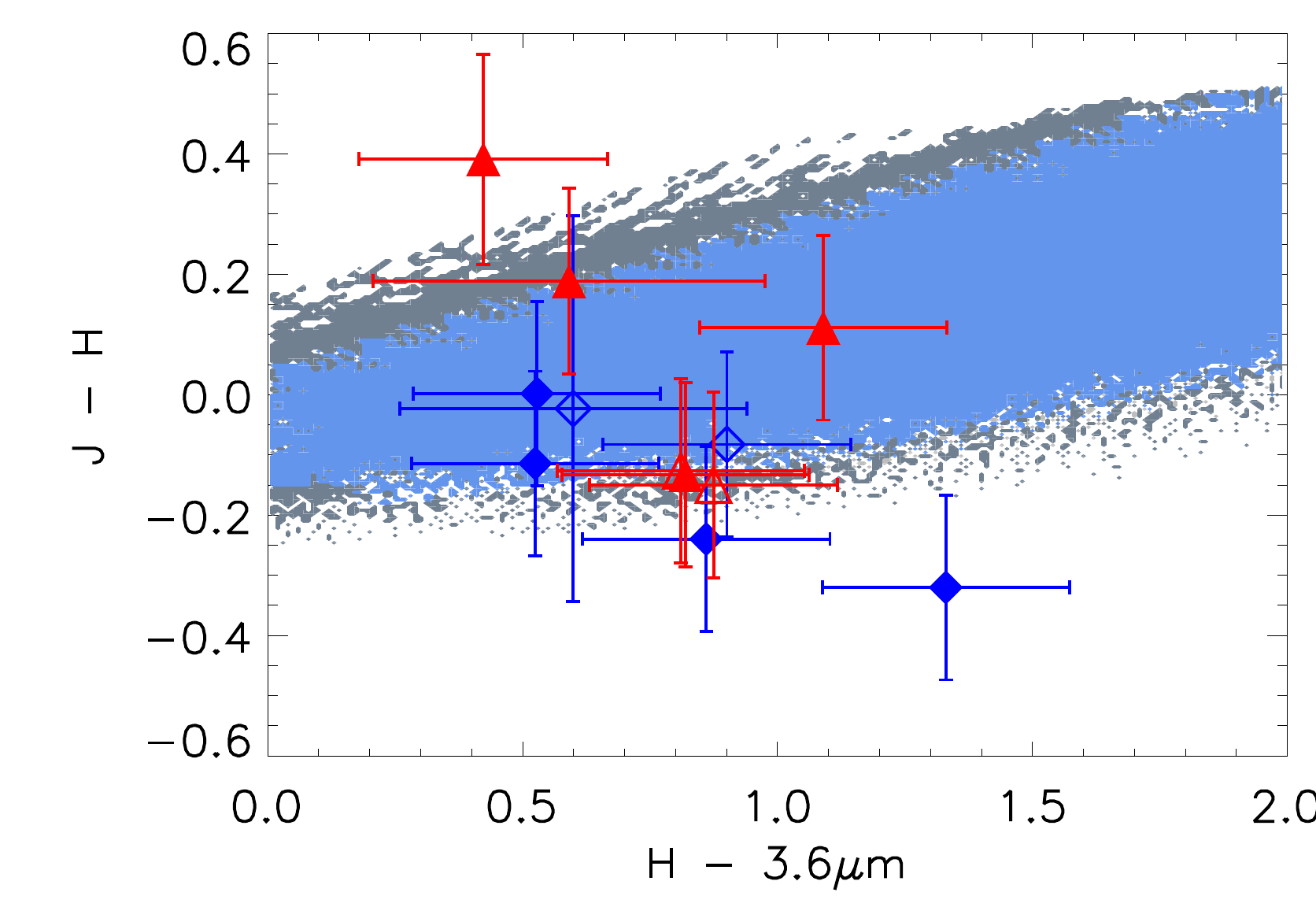}}
  \subfigure[]{\includegraphics[width=2.1in,trim=2cm 13cm 2cm 5cm,clip]{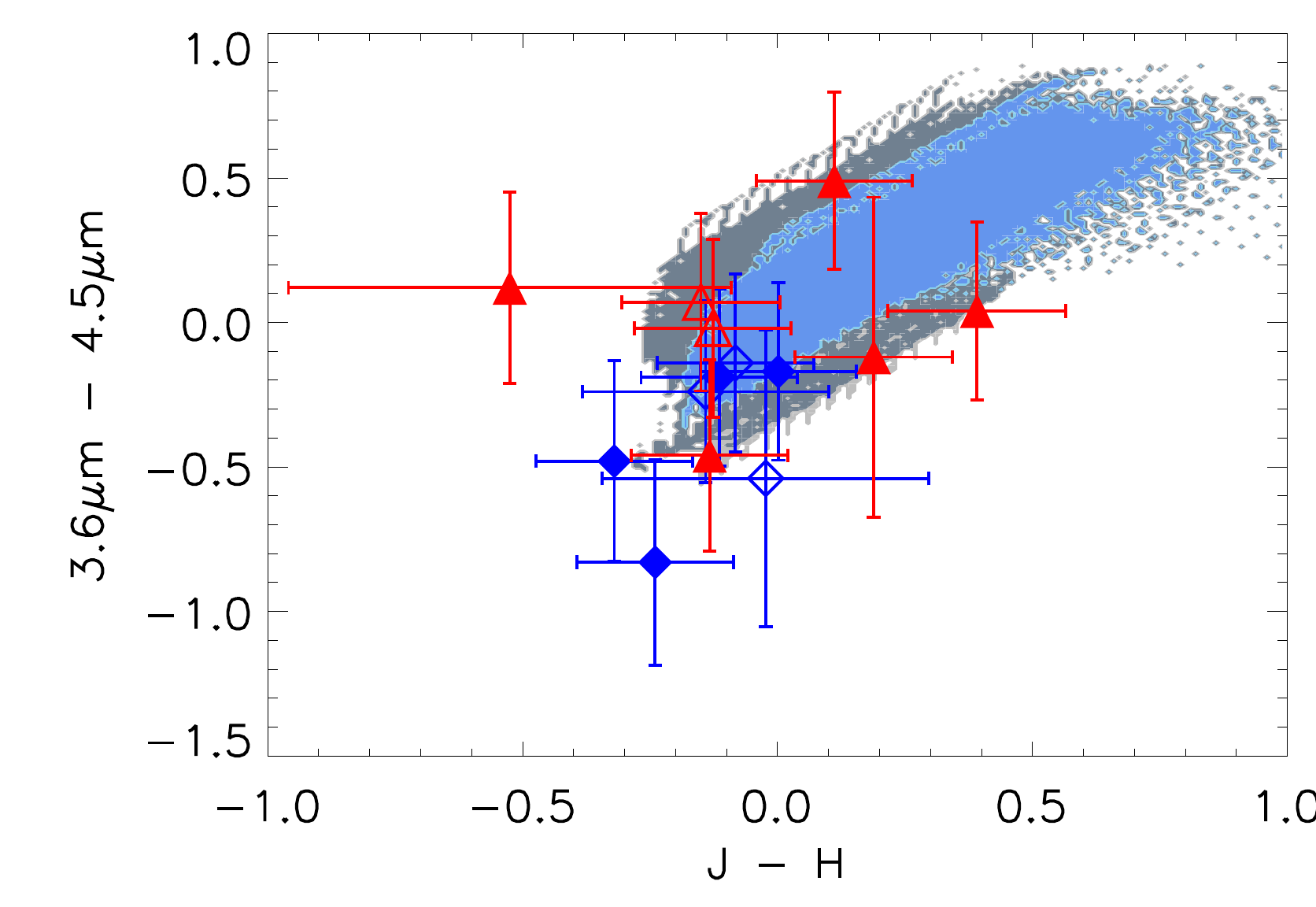}}
  \caption{Observed colours of our $z\sim6$ LBG sample plotted against the two-component model colours.  From left to right the plots show: The IRAC 3.6$\mu$m - 4.5$\mu$m vs. $H- 3.6\mu$m colours, the $J-H$ vs. $H-3.6\mu$m colours and the 3.6$\mu$m $-$ 4.5$\mu$m vs. $J-H$ colours.  Objects that are better fit by models without nebular emission are plotted as red triangles and the objects that are better fit by models with nebular emission are plotted as blue diamonds. The filled symbols represent objects which were originally confused in the IRAC images, whereas the open symbols show objects that are relatively isolated. The dark grey regions show the colours covered by the models with nebular emission and the light blue regions show the colours covered by the models without nebular emission.  The model colours are all plotted at $z=6$, although they are minimally affected by allowing the full range of redshifts within our sample ($5.5<z<6.2$).  The lighter grey regions show how the colours are affected by boosting the [OIII]/H$\beta$ ratio (see text for details).}
  \label{fig:modelsVsColours}
\end{figure*}

\begin{figure*}
  \centering
  \subfigure[Constant SFR models]{\includegraphics[width=2.1in,trim=2cm 13cm 2cm 5cm,clip]{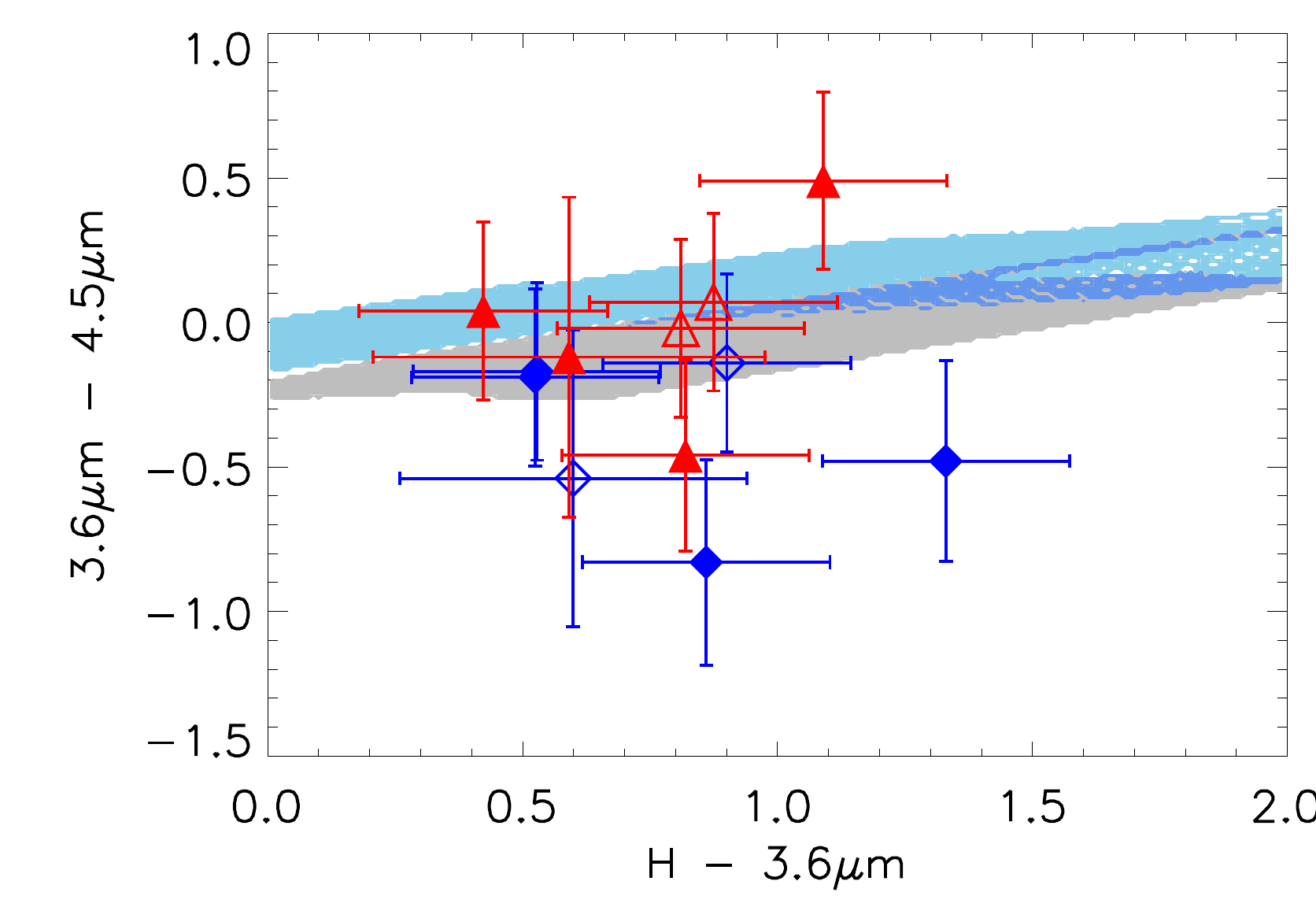}}
  \subfigure[EI models]{\includegraphics[width=2.1in,trim=2cm 13cm 2cm 5cm,clip]{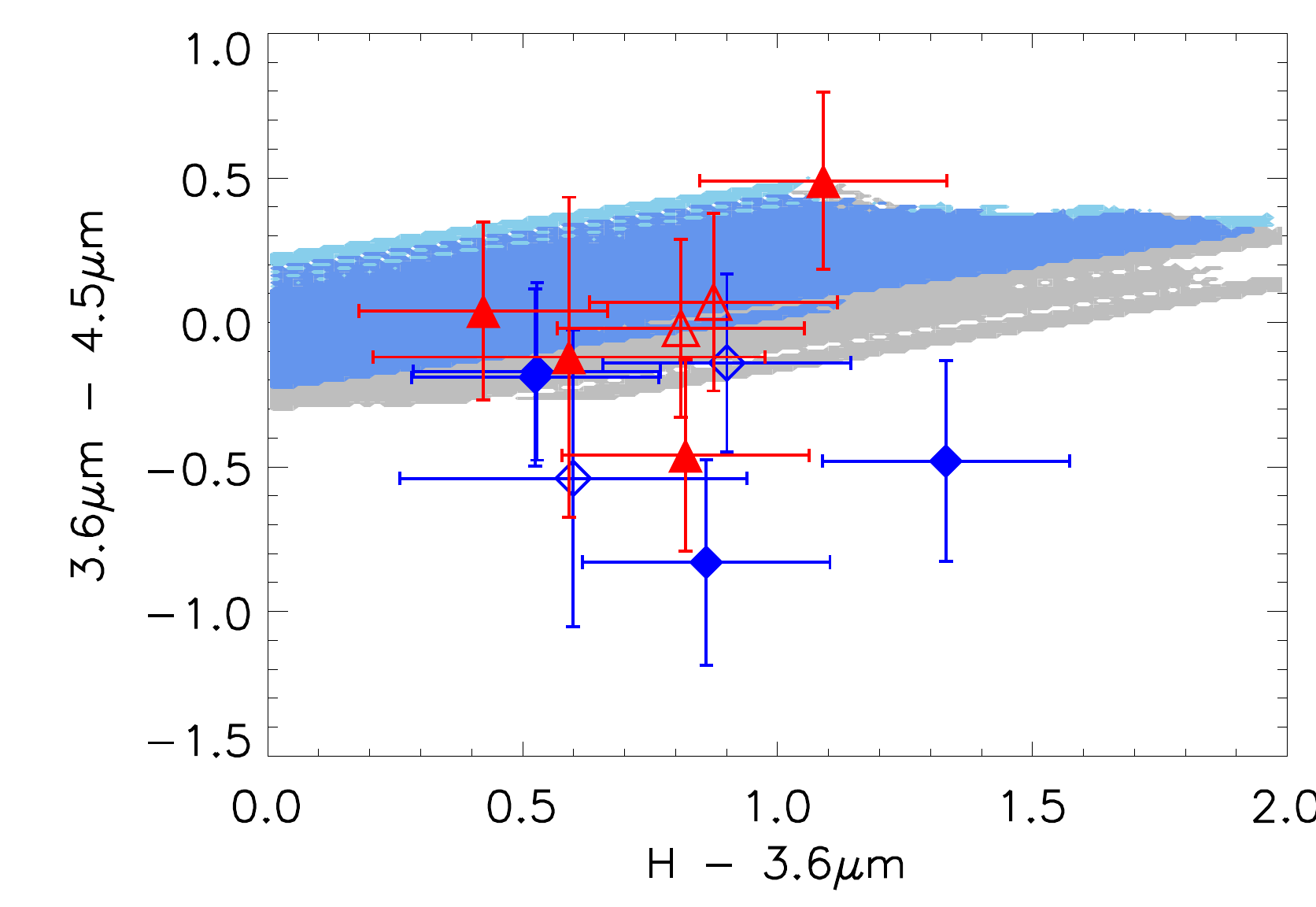}}
  \subfigure[$\tau$ models]{\includegraphics[width=2.1in,trim=2cm 13cm 2cm 5cm,clip]{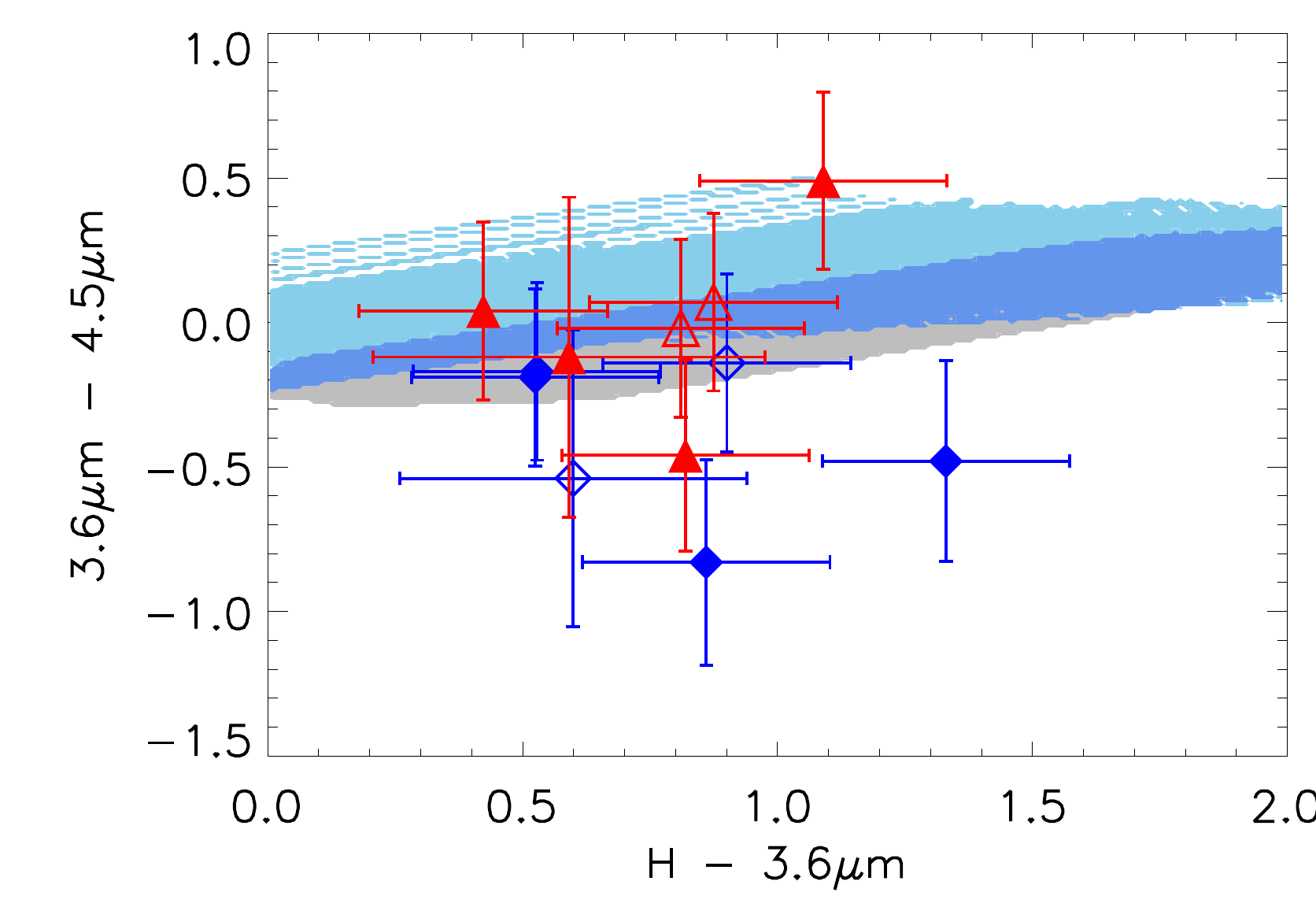}}
  \subfigure[Constant SFR models]{\includegraphics[width=2.1in,trim=2cm 13cm 2cm 5cm,clip]{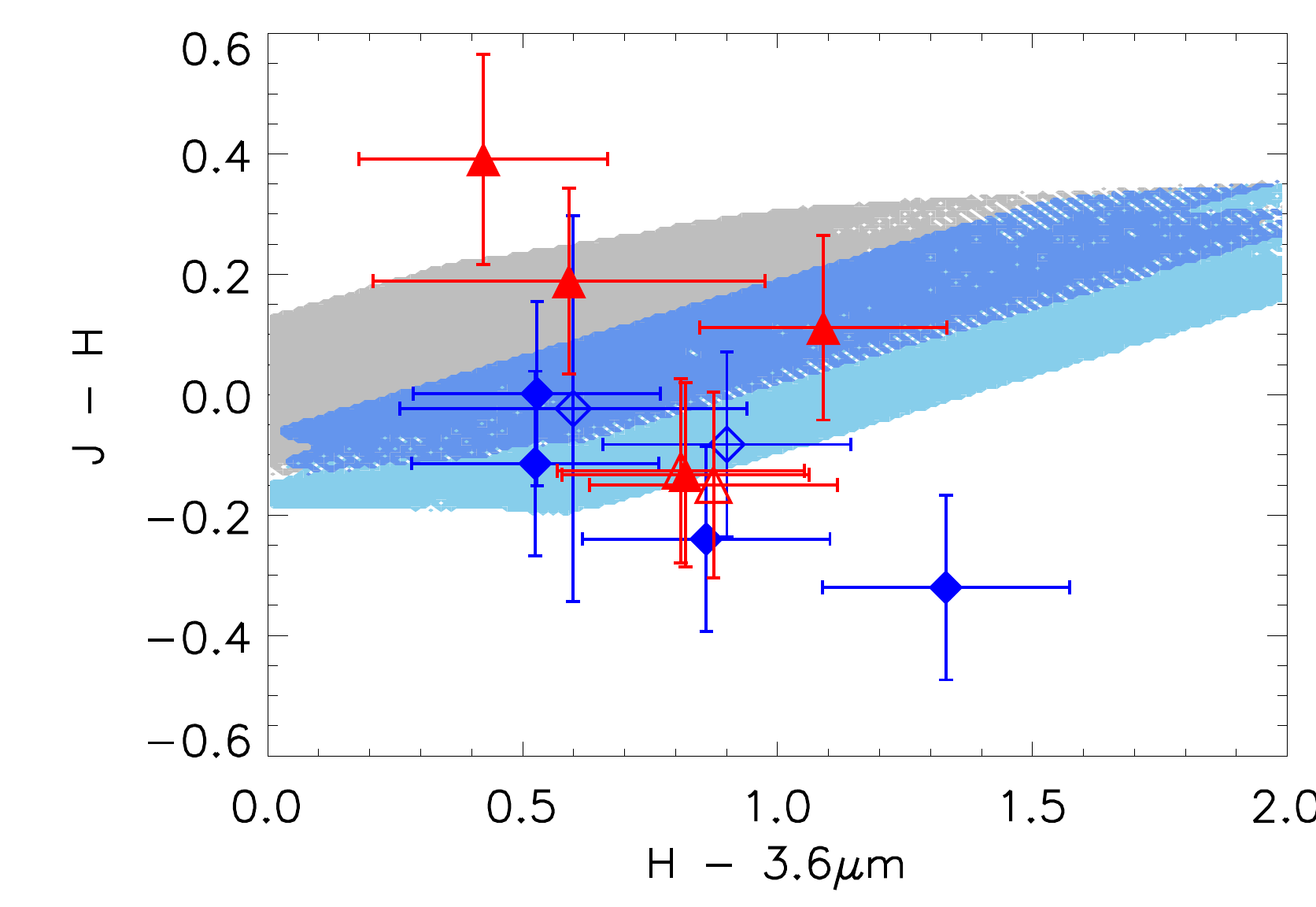}}
  \subfigure[EI models]{\includegraphics[width=2.1in,trim=2cm 13cm 2cm 5cm,clip]{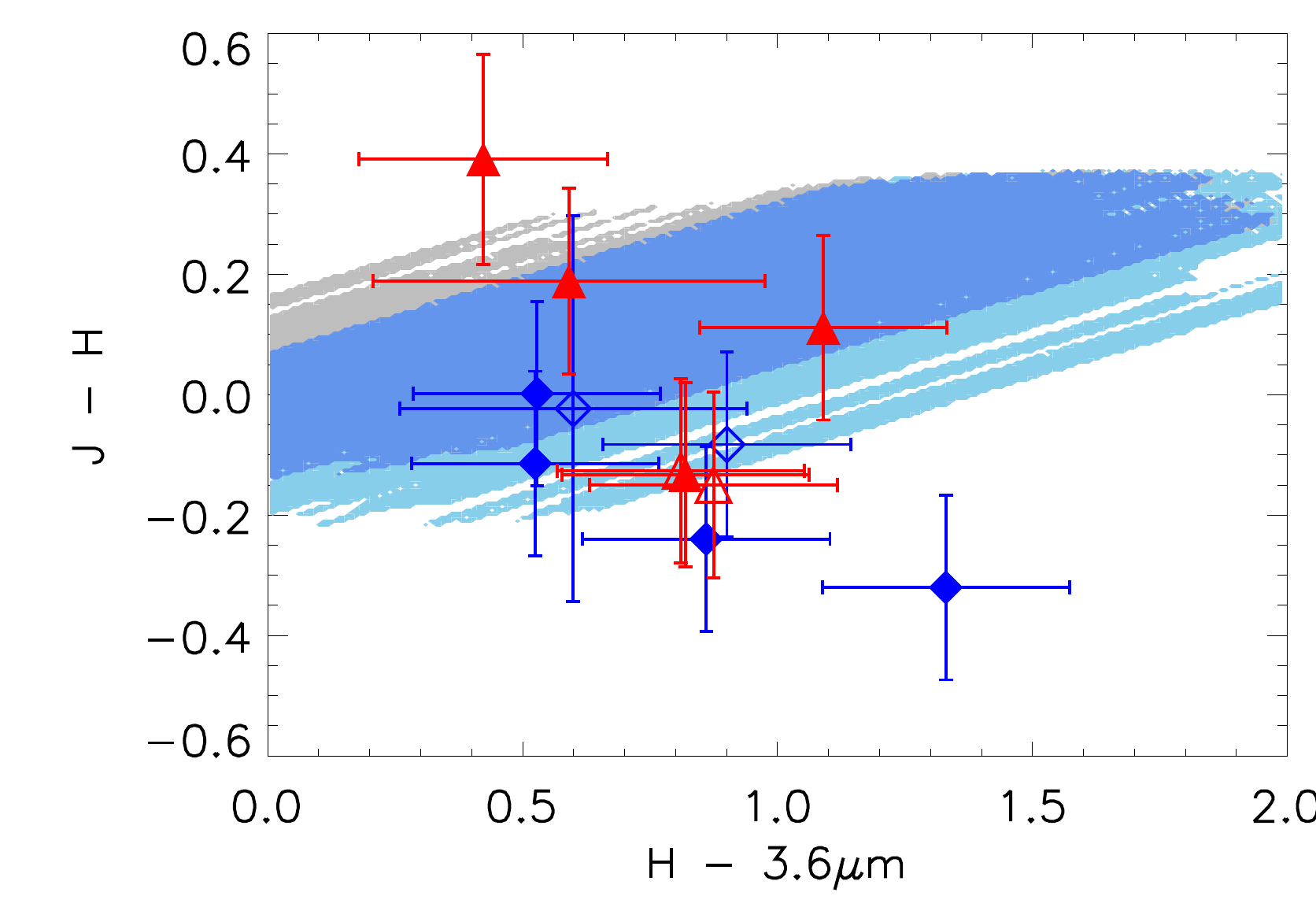}}
  \subfigure[$\tau$ models]{\includegraphics[width=2.1in,trim=2cm 13cm 2cm 5cm,clip]{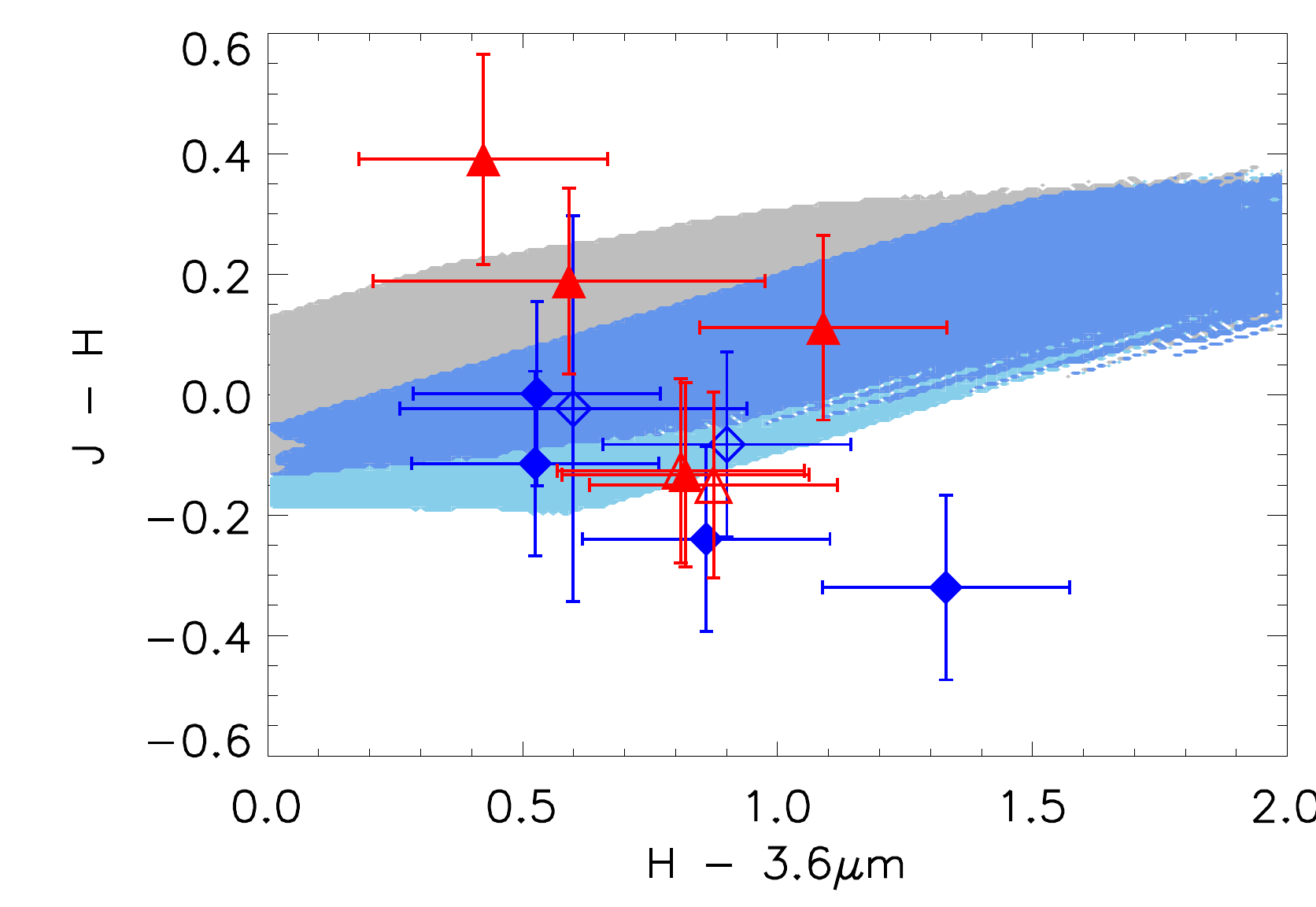}}
  \caption{The 3.6$\mu$m $-$ 4.5$\mu$m vs. $H-3.6\mu$m and $J-H$ vs. $H-3.6\mu$m colours of the models with smoothly-varying SFHs at $z=6$, both with (grey regions) and without (blue regions) nebular emission.  The darker blue regions show where these colours overlap.  The observed colours are plotted as described in Fig.~\ref{fig:modelsVsColours}.  The plotted colours include very young ($<10$ Myr) populations.}
  \label{fig:smoothSFHcolours}
\end{figure*}

Plotting the colour-colour space covered by the models and observations for a range of different SFHs in Figs~\ref{fig:modelsVsColours} and \ref{fig:smoothSFHcolours}, we see that templates including nebular emission (dark grey regions) extend to bluer colours in 3.6$\mu$m $-$ 4.5$\mu$m because, at this redshift, [OIII] and H$\beta$ are contributing to the 3.6$\mu$m flux and H$\alpha$ is contributing to the 4.5$\mu$m flux.  None of the models are able to reproduce blue IRAC colours with the appropriate range in $H-3.6\mu$m colour even with a contribution from nebular emission lines.  Even allowing for ages $<$10 Myrs in the smoothly varying SFHs, the bluest observed 3.6 $-$ 4.5$\mu$m colours are not reproduced by any of the SFHs plotted in the figure.  A potential reason for this is discussed further in the next section.

\subsubsection{IRAC colours}

Fig.~\ref{fig:modelsVsColours} (c) shows that the two-component models follow the observed correlation between the 3.6$\mu$m $-$ 4.5$\mu$m and $J-H$ colours.  This correlation is expected when considering red 3.6$\mu$m $-$ 4.5$\mu$m and $J-H$ colours, as extinction would act to redden both the rest-frame UV and the rest-frame optical.  In this projection, we can see that the models both with and without nebular emission do quite well at reproducing the blue 3.6$\mu$m $-$ 4.5$\mu$m colours. 

Fig.~\ref{fig:modelsVsColours}(a) shows the 3.6$\mu$m$-4.5\mu$m vs. $H-3.6\mu$m colours for the two component models both with and without nebular emission (dark grey and blue regions respectively) with the observed colours over-plotted.  The observed 3.6$\mu$m $-$ 4.5$\mu$m colours extend to much bluer values than covered by the models.  Although it is possible that there may be systematic errors introduced by the IRAC deconfusion that are not included in the errors plotted, it is important to question whether something is missing from the models as the high 3.6$\mu$m flux could be compensated for by allowing a larger Balmer-break from an old underlying stellar population, as we found to be the case when nebular emission is not included in the templates.  

One plausible limitation of the nebular models is that the average line ratios used to assign line fluxes to collisionally excited lines in the nebular templates may not be representative of the ratios expected for these objects.  In particular, a higher $f([OIII])/f(H\beta)$ ratio would boost the [OIII] flux relative to H$\alpha$ and hence produce bluer IRAC colours.  Rest-frame optical spectroscopy of LBGs at lower redshift ($z\sim3$) give a spread in measured $f([OIII])/f(H\beta)$ ratio, although the derived metallicities are all close to solar \citep{Mannucci2009}.  \cite{Finkelstein2011} also obtained spectroscopy of two LAEs at $z\sim2.3$ that show strong [OIII] emission, with no indication of any AGN contribution to the ionising flux.  Given the strong temperature dependence of [OIII] emission compared to that of H$\beta$, this ratio would be sensitive to HII region geometry within the galaxies and so could present a wide range of values within the population.  

In Fig.~\ref{fig:modelsVsColours}(a) we illustrate how the IRAC colours are affected by boosting the [OIII] flux of the 0.2Z$_{\odot}$ metallicity models (the metallicity that gives the highest intrinsic $f([OIII])/f(H\beta)$ ratio) by a factor of 1.5 giving a flux ratio comparable to the highest flux ratio measured for the objects at $z\sim3$ \citep{Mannucci2009}.  This is purely illustrative, as without rest-frame optical spectroscopy we cannot constrain the $f([OIII])/f(H\beta)$ ratio at these redshifts, although it can be seen that boosting the [OIII] flux in this way does improve the coverage of colour space of the models.

\subsubsection{X-ray emission}

Given the possibility that the blue observed 3.6$\mu$m $-$ 4.5$\mu$m fluxes can potentially be explained by boosted [OIII] EWs, it is important to consider the likelihood that any of these objects harbour an AGN that could be driving the strong [OIII] emission. First we note that there has, as yet, been no direct evidence of an AGN contribution from spectroscopic observations of low-redshift LBGs that also show high $f([OIII])/f(H\beta)$ ratios, as discussed in the previous section.

We further note that none of these objects have matches in the 4Ms Chandra point source catalogues.  The new Chandra 2$-$8 keV X-ray data has on-axis, S/N$>$3 sensitivity of $5.5\times10^{-17}$ erg s$^{-1}$ cm$^{-2}$ \citep{Xue2011} and, following the argument laid out in \cite{Vanzella2010b}, this can be used as an upper limit to the AGN-driven $H\alpha$ flux using the observed correlation between X-ray luminosity and $H\alpha$ luminosity \citep{Panessa2006}.  

The limiting $H\alpha$ luminosity can be related to the observed 3.6$\mu$m $-$ 4.5$\mu$m colours by finding the $f([OIII])/f(H\beta)$ ratio that would be required to boost the flux at 3.6$\mu$m, assuming the continuum is flat and the lines are not affected by extinction (extinction would only act to redden the derived 3.6$\mu$m $-$ 4.5$\mu$m colours).  We assume standard line ratios of $f(H\alpha)/f(H\beta)=2.85$ (case B recombination, \citealt{Osterbrock1989}) and $f([OIII]\lambda4959\AA)/f([OIII]\lambda5007\AA)=2.9$ (\citealt{Anders2003}, close to the measured value of 2.99 for a sample of 62 AGN by \citealt{Dimitrijevic2007}).

The correlation between $H\alpha$ flux and X-ray flux indicates that if the objects all had X-ray fluxes at the sensitivity limit, their AGN-driven $H\alpha$ fluxes\footnote{The derived correlation between X-ray luminosity and $H\alpha$ luminosity used here is taken from the total Panessa et al. (2006) sample plus low-redshift bright Seyfert 1 galaxies (Tot+QSO sample, see their Table 3).} would be of order $2.3\times10^{18}$ erg s$^{-1}$ cm$^{-2}$.  The $f([OIII])/f(H\beta)$ ratio that would be required to reproduce the blue 3.6$\mu$m $-$ 4.5$\mu$m colours for four of the objects (IDs 2, 18, 157 and 248) are shown in Table \ref{tab:XrayRatio}, where the errors in $f([OIII])/f(H\beta)$ ratio are derived from the measured photometric errors.    

The required $f([OIII])/f(H\beta)$ for objects 2, 157 and 248 are, in particular, high compared with measurements of the order $\sim10-15$ for local AGN \citep{Kauffmann2003b}.  Although it is possible that a central AGN may be obscured in the X-ray, allowing for higher AGN-driven line fluxes for the Chandra 4Ms flux limit and hence a lower $f([OIII])/f(H\beta)$ ratio to produce the observed 3.6$\mu$m $-$ 4.5$\mu$m colour, this would still require the objects to have X-ray fluxes sitting at the flux limit of the Chandra survey.  It is therefore unlikely, if the blue 3.6$\mu$m $-$ 4.5$\mu$m colours are driven by emission lines, that the dominant ionising flux is supplied by a central AGN.

\begin{table}
 \centering
  \caption{A table showing the $f([OIII])/f(H\beta)$ ratio that would be required to produce the observed blue 3.6$\mu$m $-$ 4.5$\mu$m colours for four objects in our sample if the line emission were fueled by an AGN with X-ray luminosity at the sensitivity limit of the Chandra 4Ms data.  The errors are determined from the photometric uncertainties in 3.6 $\mu$m and 4.5 $\mu$m fluxes.}
  \begin{tabular}{@{}lc@{}}
  \hline
  \hline
  ID  & $f([OIII])/f(H\beta)$\\
  \hline
  2   & 28$\pm$13\\ 
  18  & 11$\pm$6\phantom{0}\\
  157 & 55$\pm$21\\
  248 & 57$\pm$28\\
  \hline
  \hline
\end{tabular}
\label{tab:XrayRatio}
\end{table}

\begin{figure*}
   \centering
   \includegraphics[width=7in,trim=0.7cm 11cm 1cm 4cm,clip]{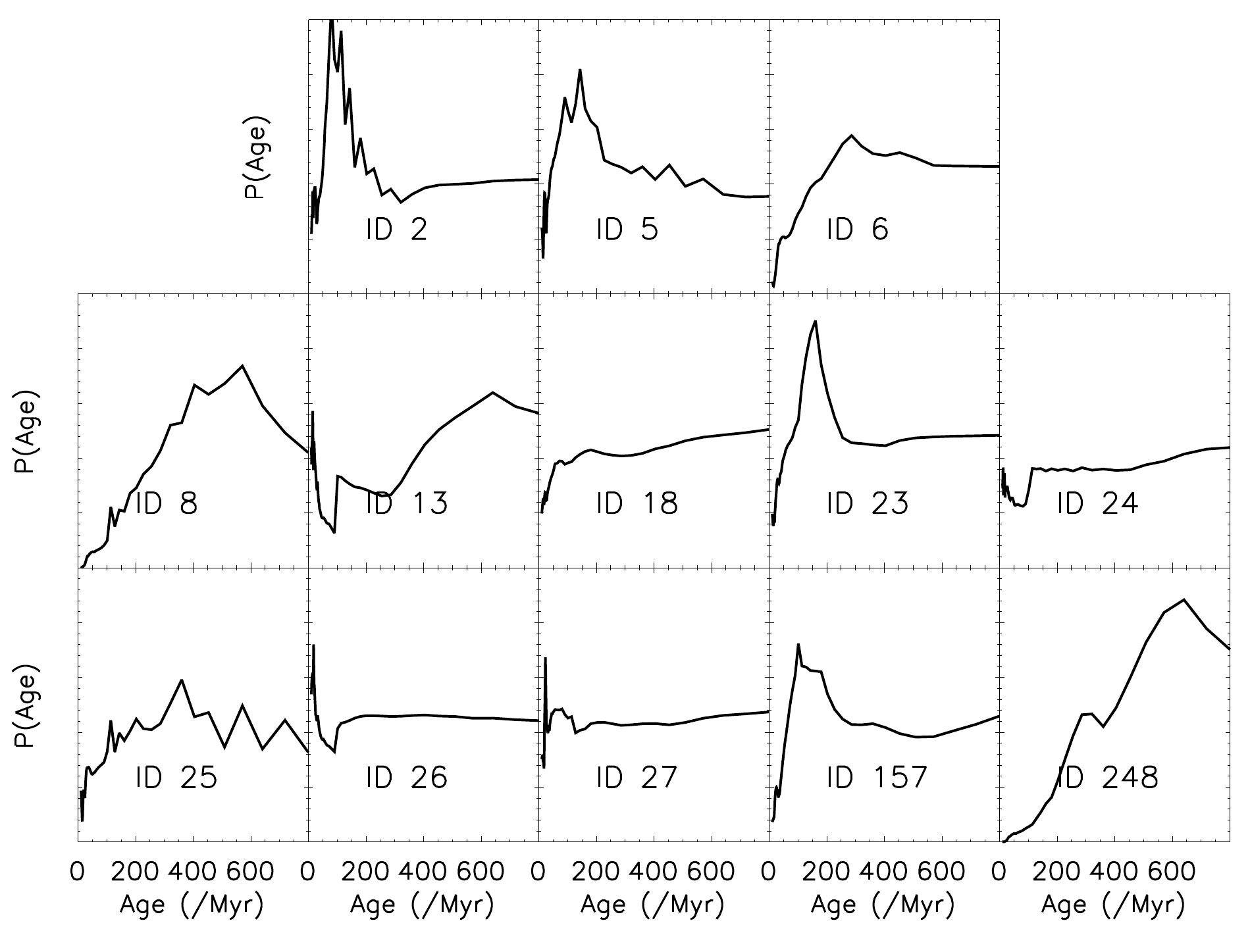}
   \caption{The probability distribution functions (PDF) for the age of each object combining the results for the two-component SFHs with and without nebular emission (template sets D+E).}
   \label{fig:agePdfs}
\end{figure*}

\subsection{Ages}
The median best-fitting ages derived for each of the template sets are 145 Myr (template set C), 182 Myr (D) and 288 Myr (E) respectively.  The median age derived using template set C is lower than for the other two template sets because just under half of the sample are best-fit by young bursts.  Fig.~\ref{fig:agePdfs} shows the constraints placed on the ages for each object from the two-component templates combining the results obtained with and without nebular emission.  

The three objects with the best age constraints are objects 2, 5, and 23 which show peaks covering the age range $\sim50-200$ Myr. The age constraints are, however, quite poor for many of the objects in the sample, in particular those objects without a full complement of filters (6, 18, 26), which have insufficient data points once these filters are removed from the fitting to provide meaningful constraints (number of filters $\leq7$).  Without the 4.5$\mu$m filter, the nebular emission for object 6 is completely unconstrained, and without the filters that have contributions from Ly$\alpha$, as well as a minimum SFR constraint, the other three objects have very flat probability density functions (PDFs).

The old median ages ($\sim$200 Myr) derived from the different template sets are at odds with the extremely young ages often found from fitting with smoothly varying star formation histories with added nebular emission.  To consider more carefully what this means, we need to understand whether these median age estimates describe the typical age of the galaxies, or whether they are driven by any other uncertainties.  Also, we need to understand what the young ages from smoothly-varying SFHs reveal.

\subsubsection{Do any of the object SEDs \textit{require} old stellar populations?}

To address the first consideration, we note that there is an inherent degeneracy in the two-component SFHs which prevent one from being able to distinguish between a fairly young (of order 100 Myr) burst that dominates the rest-frame optical flux, and an older burst that is mostly masked by the young star-forming component.  The old median ages derived from the two component templates are therefore likely biased by this degeneracy and it is more meaningful the ask the question: \textit{Do we require old stellar populations to adequately describe the SEDs?}

It is clear from the object PDFs that most of the objects do not actually require stellar populations with particularly large ages.  A few of the objects do show quite low probabilities of harbouring stellar populations much younger than 100 Myr (e.g. 6, 8, 248), however.  Objects 8 and 248, in particular, consistently give fairly high minimum ages (lower age limit within 1$\sigma$ uncertainties) from fitting with the different template sets.  Plotting the object colours against the model colours (Figs~\ref{fig:modelsVsColours} and \ref{fig:smoothSFHcolours}) shows that the 3.6$\mu$m $-$ 4.5$\mu$m colour of object 248 is further than 1$\sigma$ from the template colours (the object with the bluest 3.6$\mu$m $-$ 4.5$\mu$m colour).  Given that the templates can try to reproduce red UV-to-optical colours by either boosting the nebular contribution or allowing for an old underlying population, it is possible that, if these objects have stronger [OIII] emission than accounted for in the models (as discussed in Section 8.1.1), the models would strive to reproduce this flux with an older underlying population.

\subsubsection{Do any of the objects show evidence of being dominated by very young stellar populations?}

It is also interesting to note that three of the objects show distinctive peaks at extremely young ages (13, 26, 27) where the SEDs are well described by a burst with some recent star formation.  With the current parameterisation, the peak for object 26 is likely to be unphysical given that the minimum $N_{LyC}$ limit could not be applied (see Section 6).  Object 13 has a best-fitting model that requires reddening by dust and so the minimum $N_{LyC}$ limit is likely to be underestimated.  Each of these three PDFs also allow the full range of ages within 1$\sigma$ confidence.  Raising the fraction of mass allowed within the star-forming component, however, would likely produce a high enough flux of Lyman-continuum photons to describe the Ly$\alpha$ emission, and so it is important to check whether other objects might show such peaks at young ages using a constant SFR model with added nebular emission.  We find slightly increased probability of very young ages ($<50$ Myrs) for many of the objects, although only objects 18, 24 and 26 show a more significant peak at these young ages.  However, these three objects have very uncertain derived parameters, either due to added degeneracies due to large amounts of reddening (24) or because the filters containing any Ly$\alpha$ emission had to be excluded from the fit (18, 26).  

Using a template set consisting of only the smoothly-varying SFHs with added nebular emission, we find that two of the objects with very blue 3.6$\mu$m - 4.5$\mu$m colours prefer very young ages ($\sim10$ Myr).  This is reminiscent of the possibly un-physically young ages ($\sim3$ Myr) fitted by \cite{Ono2010} to stacked observations of LAEs at $z\sim5.7$.  In fact, four of the objects with the reddest J - H colours would also show young best-fitting ages but with broader uncertainties that extend to $\sim100$ Myr, attributable to uncertainties in extinction.  

It is worth noting, however, that by insisting on a prior of smoothly varying SFHs, the only way to fit to the blue 3.6$\mu$m $-$ 4.5$\mu$m colours observed in a subset of these objects would be with an extremely young population, as the nebular emission line equivalent widths fall quickly as the underlying population builds up.  This would suggest an age dichotomy in the LBG population, with some of the objects having just been formed with very low masses and very high sSFRs, and some at higher masses (those without blue 3.6$\mu$m $-$ 4.5$\mu$m colours), still building up their mass.  It is more likely, however, that the smoothly varying SFHs are primarily fitting to the dominant component by luminosity, the recent starburst.  The young ages might be consistent with the expected duration of such a starburst, but the two component SFHs show that the SEDs also allow for an underlying population that can dominate the mass of the object.

Although we cannot wholly discount the possibility that some of these objects have very young stellar populations with extreme [OIII] EWs ($>1000$\AA), as observed in lower redshift galaxies such as Green Peas \citep{Cardamone2009} at $z\sim0.3$ or the extreme emission line galaxies observed at $z\sim1.7$ by \cite{VanderWel2011}, we find that only two of the four objects with the bluest 3.6$\mu$m - 4.5$\mu$m colours show high probability of young ages when fit with smoothly-varying SFHs with added nebular emission.  With a boosted $f([OIII])/f(H\beta)$ ratio in the nebular models, such young ages may no-longer be required.  New evidence suggests that even Green Peas (low redshift systems showing evidence of a strong starburst and young ages) show older, underlying stellar components with low surface brightness that cannot be accounted for in SED fitting with a smooth SFH \citep{AmorinRicardo2012}.

\subsubsection{Summary}

In summary, our estimates of the median best-fitting ages for our sample of $z\simeq6$ LBGs are of order 200-300 Myr, although the objects displaying the best age constraints give ages of $\sim50-200$ Myr.  We find no firm evidence for extremely young stellar populations $\lesssim50$ Myr and any objects showing old best-fitting ages ($>300$ Myr), that also have colours that are well described by the template set, can also be described by stellar populations as young as $\sim100$ Myr within the 1$\sigma$ confidence contours.

We note that \cite{Schaerer2009} investigated the importance of nebular emission on the derived ages of the \cite{Eyles2007} sample, of which four had spectroscopic redshifts at the time, and found that the best-fitting ages from templates with nebular emission gave a mean age of 120 Myr (with some ages of order $\sim20$ Myr) compared to 500 Myr without nebular emission.  Here, for a larger sample of spectroscopically confirmed objects with improved near-Infrared photometry from CANDELS, we do not see such a dramatic decrease in average ages between models without and models with nebular emission.  This we attribute to the two component parameterisation of the SFH that allows for an underlying older stellar population that may dominate the mass, as discussed in the previous section.

\subsection{The Stellar mass-SFR relation}

\begin{figure*} 
   \centering
   \subfigure[All smoothly varying SFHs]{\includegraphics[width=2.2in,trim=3cm 13cm 1.5cm 5cm,clip]{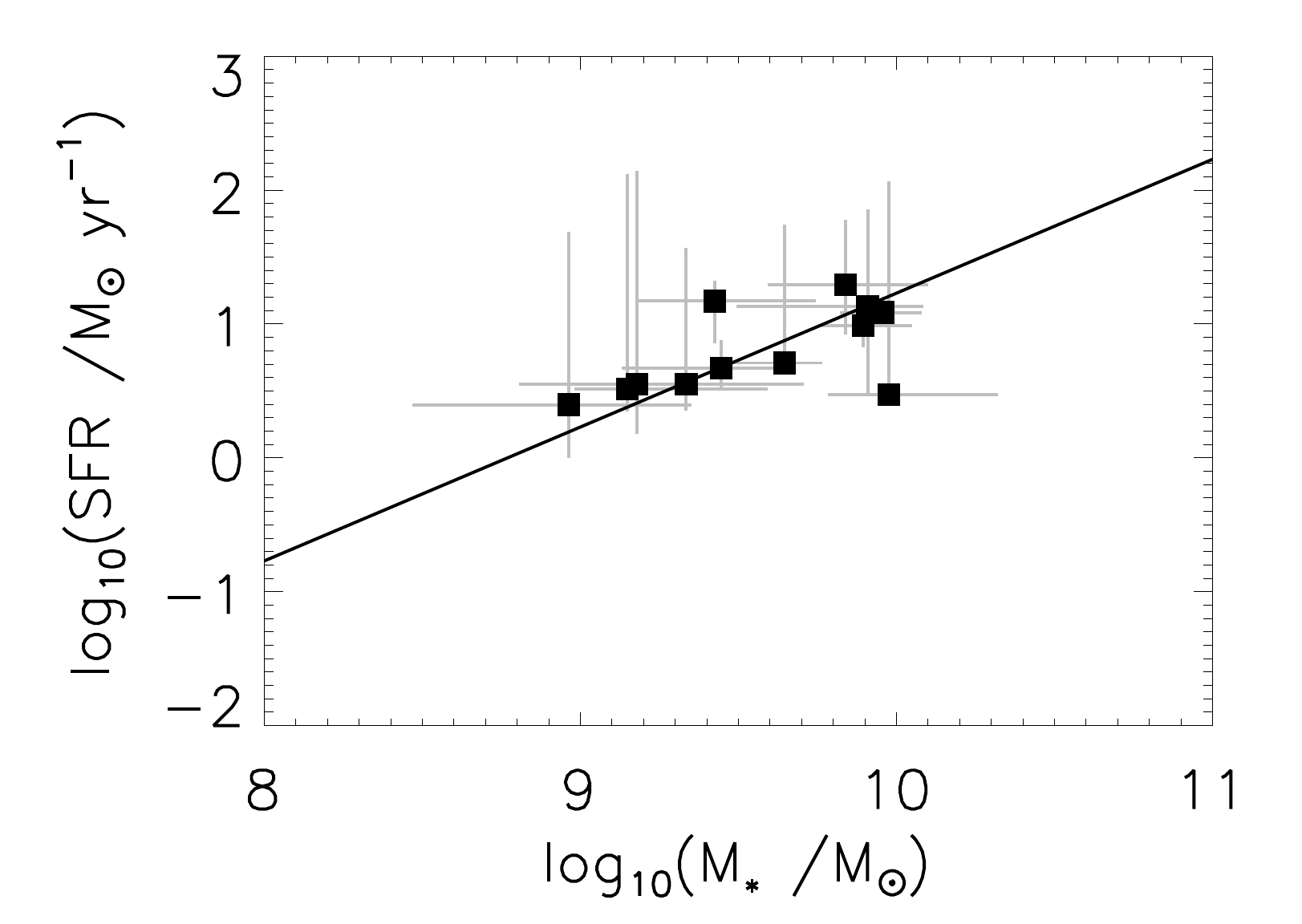}}
   \subfigure[Constant star formation plus burst models]{\includegraphics[width=2.2in,trim=3cm 13cm 1.5cm 5cm,clip]{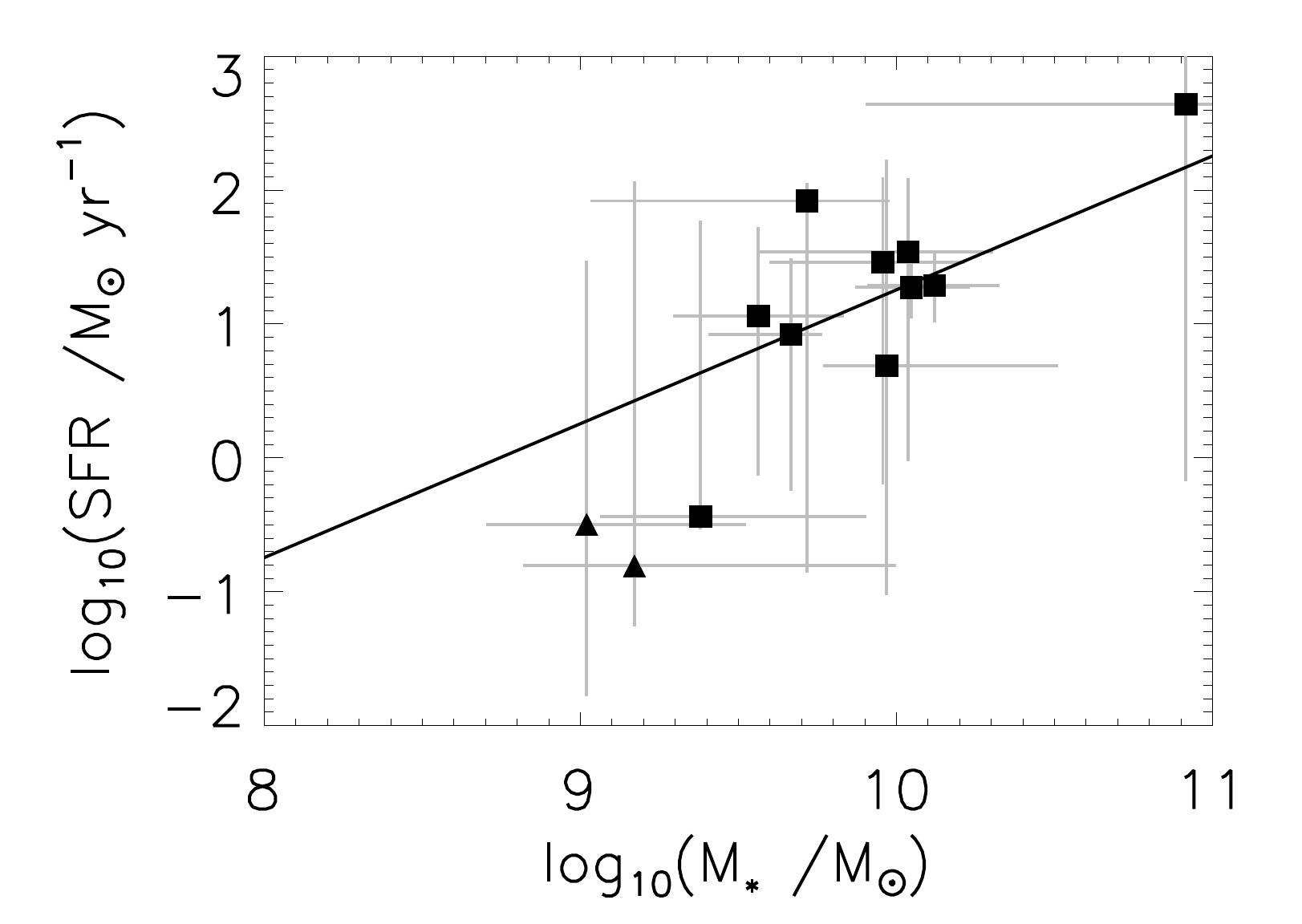}}
   \subfigure[Constant star formation plus burst models with added nebular emission]{\includegraphics[width=2.2in,trim=3cm 13cm 1.5cm 5cm,clip]{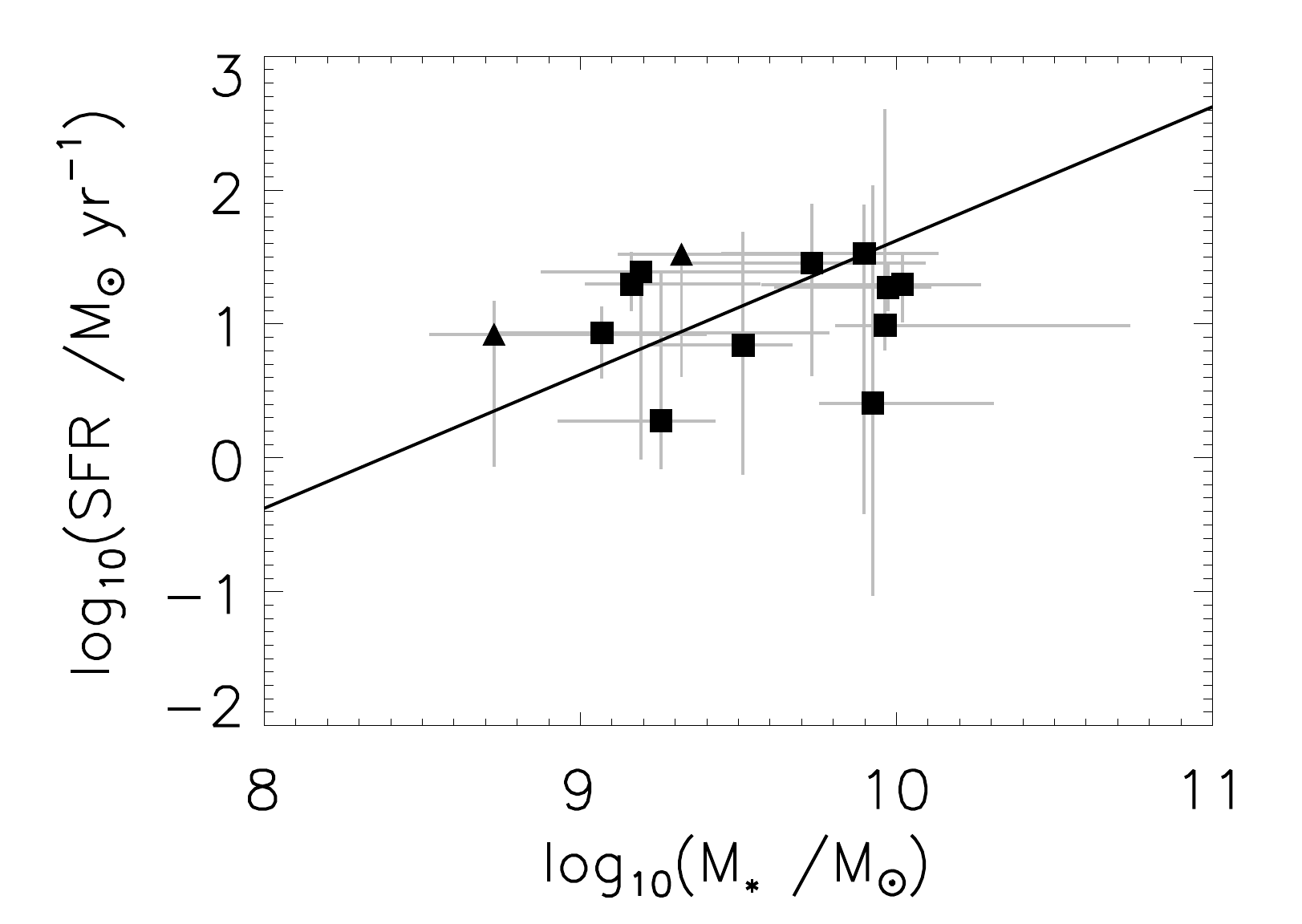}}
   \caption{SFR vs. Mass derived using the different template sets as described in the captions.  Errors shown are 68\% confidence limits derived from $\Delta\chi^2=1$ from all models with SFR$>0$ M$_{\odot}$yr$^{-1}$.  The three objects for which the SFR constraint from Ly$\alpha$ was not applied (see Section 6) are plotted as triangles in panels (b) and (c).  The line in each panel shows the median sSFR derived from that template set; 1.7 Gyr$^{-1}$, 1.8 Gyr$^{-1}$ and 4.2 Gyr$^{-1}$ in (a), (b) and (c) respectively.  The median sSFR for the two component models are derived excluding the three objects without Ly$\alpha$ constraints (triangle symbols).}
   \label{fig:massSFR}
\end{figure*}

Fig.~\ref{fig:massSFR} shows stellar mass vs. SFR for the full sample, derived using the full set of smoothly varying SFHs (a), as well as the constant+burst templates without and with nebular emission ((b) and (c) respectively).   In producing these plots, any models without current star formation (bursts) are discounted when determining the new best-fitting parameters and errors, which are the full range of masses and SFR of models with fits within $\Delta\chi^2=1$. 

The median sSFRs are plotted as the solid lines from the plot and are 1.7 Gyr$^{-1}$, 1.8 Gyr$^{-1}$ and 4.2 Gyr$^{-1}$ for template sets C, D and E respectively.  For template sets D and E this median excludes the objects without any Ly$\alpha$ constraints, because the minimum SFRs are not restricted.

The derived parameters from using template set D (Fig.~\ref{fig:massSFR} (b)) have much larger uncertainties in both mass and SFR, mainly introduced by the degeneracies between fitting the rest-frame UV light primarily by the burst component or the star-forming component.  They do, however, illustrate that if the optical flux is dominated by low-mass, longer-lived, stars that also dominate the stellar mass, then the set of smoothly varying SFHs (i.e. Fig.~\ref{fig:massSFR}a) limit the lower possible values of SFR.  This can contribute to imposing a fairly well defined correlation when plotting sSFR against $M_{*}$.  If episodes of star formation occur over very short timescales, however, then the instantaneous SFR may not be closely coupled to the stellar mass on an object-by-object basis, even if the average properties of a population suggest otherwise.

There is still the possibility, however, that red UV to optical colours are driven by emission-line contributions to the IRAC bands and not by the older stellar population.  We see from Fig.~\ref{fig:massSFR} (c) that, compared to the results from the smoothly varying SFHs, there is greater scatter in the best-fitting parameters when nebular emission is included, as recently observed by \cite{Schaerer2011}.  This is because there is a wider range of allowed masses, with the range of possible masses extending to lower values due to the uncertainty in the contribution of the nebular emission to the optical fluxes.  

We find that the estimated typical sSFR of the sample does increase  using templates with added nebular emission, although by less than a factor of two.  It is important to note, however, that our observations suggest that the ingredients of the nebular emission models may not be wholly representative of the sample, given that the colour space of the observations is not fully covered by the templates.  A higher $f([OIII])/f(H\beta)$ ratio would better represent the bluer IRAC colours in the sample, potentially leading to a higher mass and lower SFR estimate (because the colour relies on the EW of [OIII]).

\subsection{The stellar mass-UV luminosity relation}

\begin{figure*} 
   \centering
   \subfigure[All smoothly varying SFHs]{\includegraphics[width=2.2in,trim=3cm 13cm 1.5cm 5cm,clip]{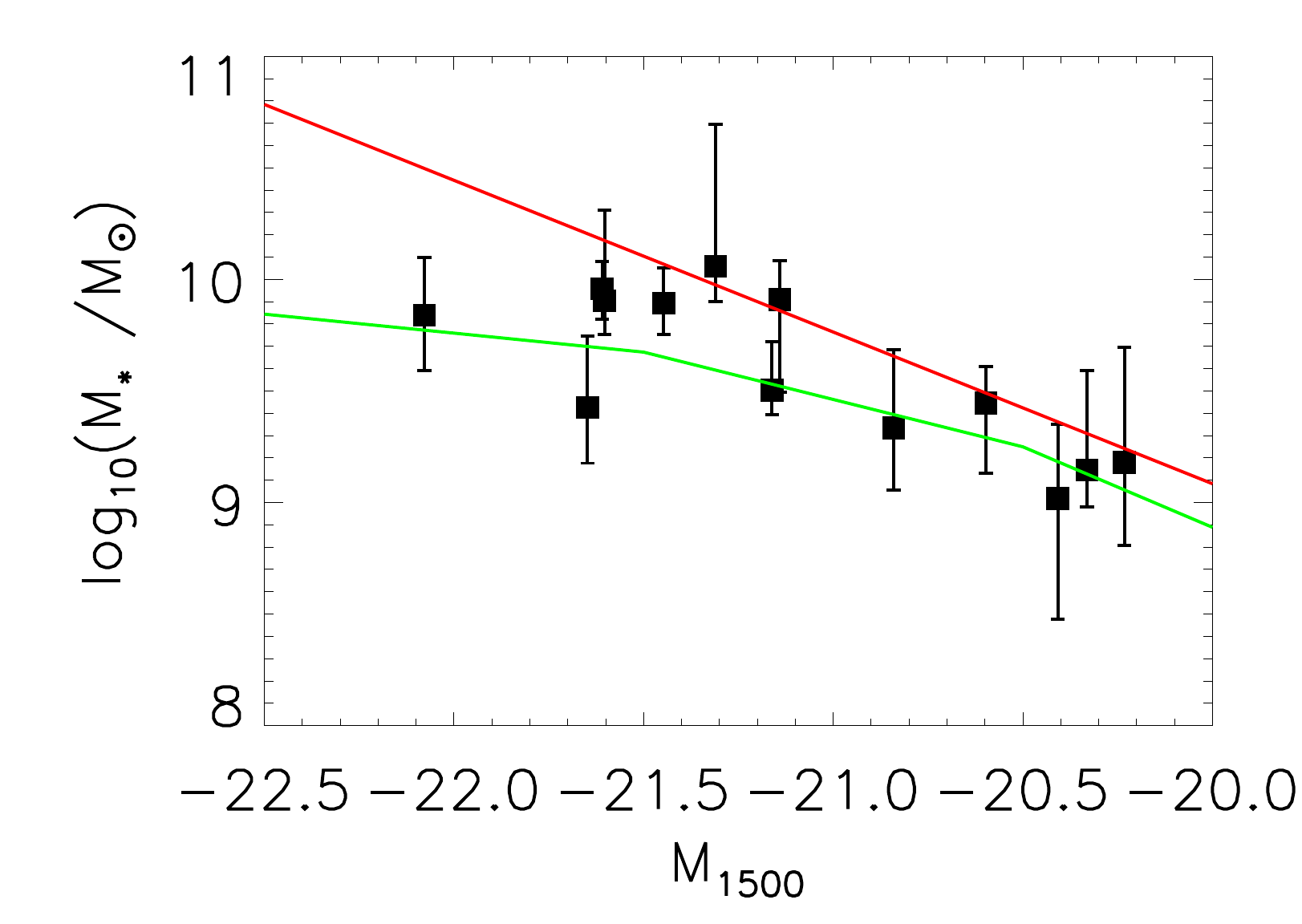}}
   \subfigure[Constant star formation plus burst models]{\includegraphics[width=2.2in,trim=3cm 13cm 1.5cm 5cm,clip]{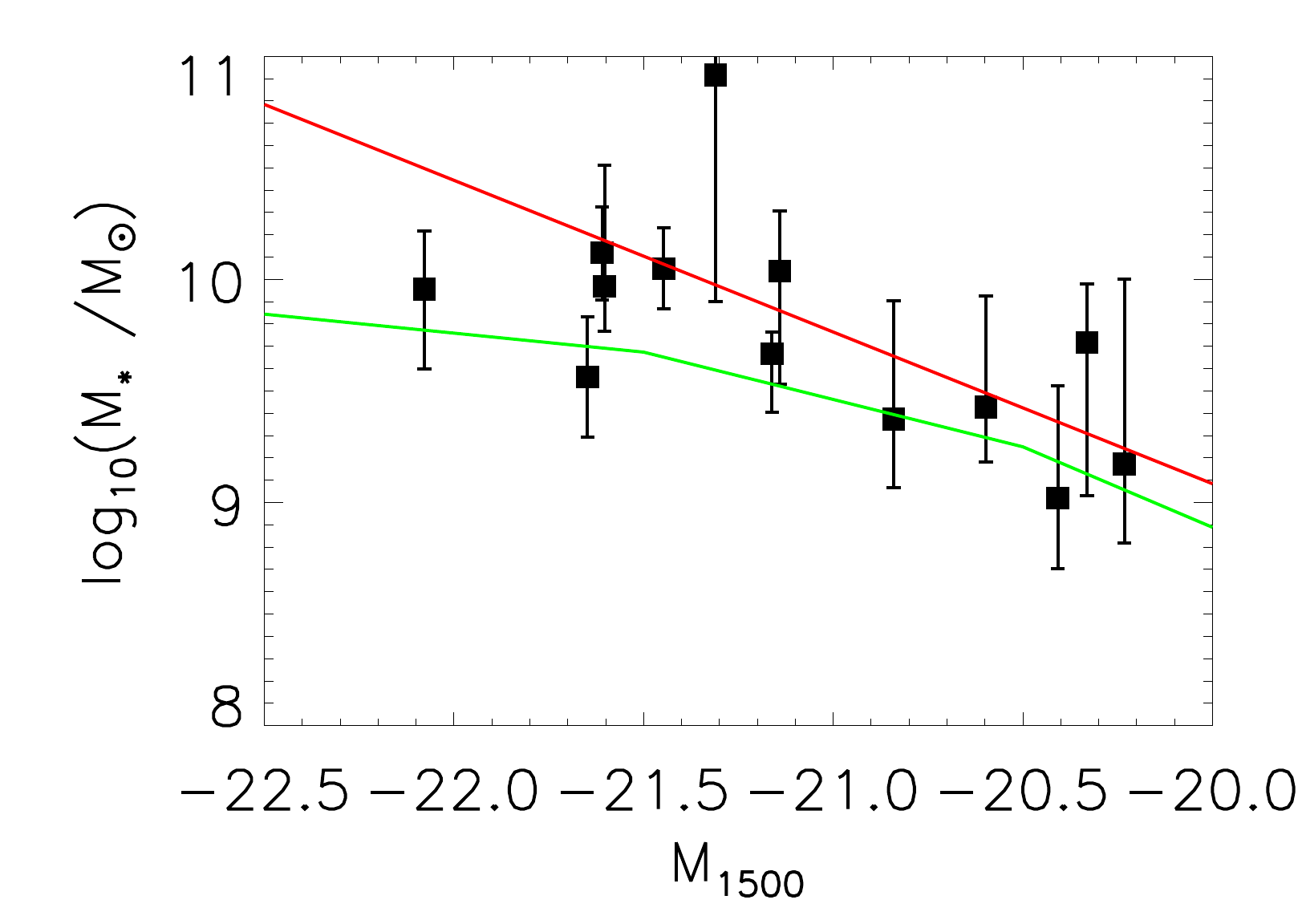}}
   \subfigure[Constant star formation plus burst models with added nebular emission]{\includegraphics[width=2.2in,trim=3cm 13cm 1.5cm 5cm,clip]{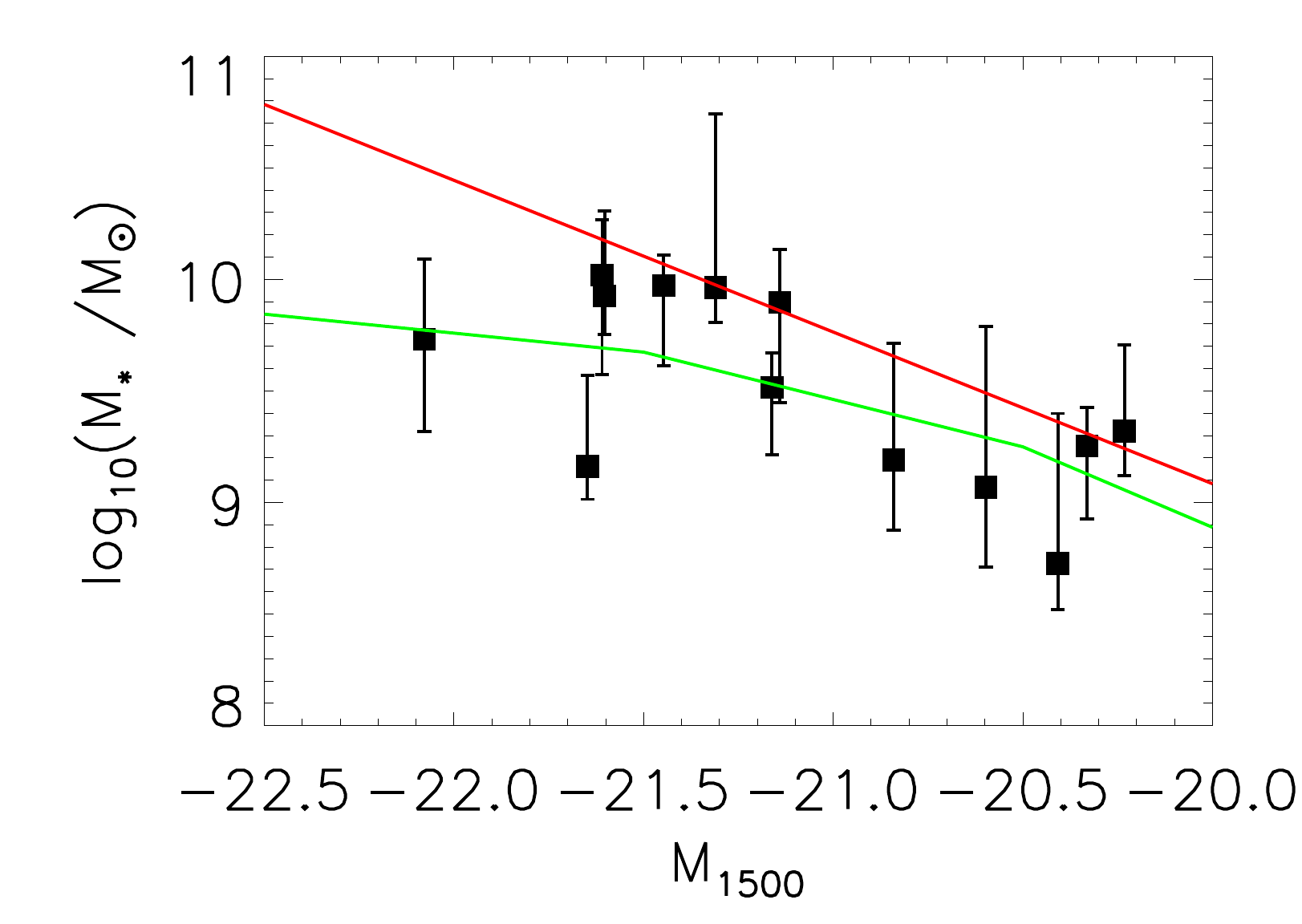}}
   \caption{Mass vs. UV absolute magnitude derived using different template sets (as indicated by the panel captions).  Errors shown are 68\% confidence limits derived from $\Delta\chi^2=1$.  Over-plotted in green is the relation derived from the $z\sim4$ sample of Stark et al. (2009) and the red line shows the relation derived at $z\sim4$ in Gonz\'{a}lez et al. (2011)}
   \label{fig:massUV}
\end{figure*}

We have seen that the errors involved in determining the SFR for these galaxies are large given the degeneracies between SFR and extinction.  \cite{Stark2009} chose to investigate any evidence of mass growth for successive populations of LBGs between $4<z<6$ via the $M_*-M_{1500}$ relation in order to distinguish between various scenarios of LBG evolution. This can also reveal whether the typical sSFR and stellar masses change significantly over this time. 

Fig.~\ref{fig:massUV} shows the best-fitting masses derived from template sets C, D and E (panels (a), (b) and (c) respectively) vs. UV absolute magnitudes for this sample.  Over-plotted in green is the median stellar mass-UV relation derived from the \cite{Stark2009} $z\sim4$ sample (from masses derived using an exponentially decreasing SFH with $\tau=100$ Myr and corrected to the Chabrier IMF; see their Fig. 9) and the red line shows the relation derived in \cite{Gonzalez2011}, also from an LBG sample at $z\sim4$.  

The errors in mass are taken from the range of acceptable mass within $\Delta\chi^2=1$.  
Our results agree qualitatively with the flattening of the slope evident in the \cite{Stark2009} relation, while at magnitudes fainter than $M_{1500}=-21.5$, the relationships from both papers agree.  Any flattening to high UV luminosities could not be observed in the \cite{Gonzalez2011} results, however, because their template set was restricted to a constant SFR model. 

The masses derived from the templates that include nebular emission extend to lower values than estimates from the other two template sets, although they still quantitatively agree with the $z\sim4$ relationship.  We do, however, find a marked increase in derived sSFRs from fitting with template set E compared to the results derived using template set C (Section 7.2) that is not so readily observed in the $M_*-M_{1500}$ relation, especially when full uncertainties in mass are considered.

From the apparent lack of evolutionin the $M_*-M_{1500}$ relation, \cite{Stark2009} argue that it rules out a steady growth scenario in which the LBGs from the higher redshift populations continue to grow at a constant SFR, as the derived ages imply that they would populate successive generations, shifting the normalisation of the lower redshift populations to higher masses at constant UV luminosity.  They suggest instead that short duty cycles implied by exponentially decreasing SFHs fit the observations better, although this would mean that each of the objects observed would have had a higher SFR in the past.  Our results also display little evolution between our $z\sim6$ sample and the \cite{Stark2009} $z\sim4$ sample, consistent with the theory that LBGs selected at each redshift are an independent population.  We also see, however, that simple two component SFHs describe the SEDs well and note that more stochastic SFHs would also fulfill the requirement for short duty-cycles.

\section{Conclusions}

By performing SED fitting to a sample of spectroscopically confirmed LBGs, we are able to remove the uncertainty in the redshift from the derived parameters, as well as the contribution of Ly$\alpha$ to the observed photometry in all but two cases.  Moreover, deep near-IR imaging supplied by the CANDELS survey and deep IRAC imaging from \textit{Spitzer} SEDS has allowed us to accurately constrain the rest-frame UV to optical photometry.  Fitting with templates covering a wide range of SFHs, metallicity, extinction, as well as whether or not they include nebular emission, we find that:

\begin{enumerate}
  \item{When fitting with a constricted template set with a single metallicity and no extinction, the derived best-fitting ages show 3/13 of the objects with ages $>500$ Myr.  Expanding the template set to contain a wide range of SFHs, metallicities and extinction, however, shows large uncertainties in the derived ages, although the derived masses are shown to be robust to within a factor of two, in accordance with other studies (e.g. \citealt{Yabe2009, Pacifici2012}).}
  \item{Motivated by an observed tension between the observed photometry and models that caused objects with strong apparent Balmer-breaks but blue UV slopes to be fit with extremely old templates, simple two-component models were used to decouple the assembled stellar mass from the instantaneous SFR.  Fitting these templates, both with and without nebular continuum+line emission, shows that 6 of the 13 objects are marginally better fit without nebular emission, while the remaining objects show better fits to their blue 3.6$\mu$m - 4.5 $\mu$m colours when nebular emission is included.  The bluest 3.6$\mu$m - 4.5 $\mu$m colours are not well described by any of the models, although a moderate increase in the flux of [OIII] compared to H$\beta$, consistent with measured flux ratios in some $z\sim3$ LBGs, would better describe the observed colours.  Masses derived using templates with nebular emission are systematically lower than those from templates without nebular emission, with a median offset of 0.18 dex.}
  \item{Full analysis of the individual probability distribution functions demonstrates the poor constraints that can be placed on stellar population age on an object by object basis.  Without spectroscopic redshifts and the extra information provided by the Ly$\alpha$ flux these constraints would be poorer still.  The combined probability distributions for the two-component models, both with and without nebular emission, show no firm evidence for extremely young stellar populations ($<50$ Myr) and no objects that \textit{require} populations with ages $>300$ Myr to describe their SEDs.  The three objects with the strongest constraints give ages in the range $\sim50-200$ Myr.}
  \item{Results from all of the different template sets show that specific SFRs derived using templates with nebular emission are moderately higher than those derived without nebular emission (median sSFRs of 4.2 Gyr$^{-1}$ compared to 1.8 Gyr$^{-1}$ respectively), although the uncertainties are large and the scatter is greater for the models including nebular emission, in agreement with recent results by \cite{Schaerer2011} and \cite{DeBarrosStephane2011}.}  
  \item{The derived $M_{*}-L_{\rm UV}$ relation displays little evidence of evolution in the LBG population between $4<z<6$ which can naturally be described either by a more stochastic parameterisation of the SFH, with short durations of star-formation, or an exponentially increasing SFH.}
\end{enumerate} 

\section*{Acknowledgments}
ECL would like to acknowledge financial support from the UK Science and Technology Facilities Council (STFC) and the Leverhulme Trust.  RJM would like to acknowledge the funding of the Royal Society via the
award of a University Research Fellowship and the Leverhulme Trust via the award of a Philip Leverhulme research prize. JSD acknowledges the support of the Royal Society via a Wolfson Research Merit award, and also the support of the European Research Council via the award of an Advanced Grant. MC acknowledges the award of an STFC Advanced Fellowship. HJP, ABR, RC, and EB acknowledge the award of STFC PhD studentships.  WGH acknowledges the award of an STFC PD
RA.  This work is based on observations taken by the CANDELS Multi-Cycle Treasury Program with the NASA/ESA HST, which is operated by the Association of Universities for Research in Astronomy, Inc., under NASA contract NAS5-26555. This work is based [in part] on observations made with the Spitzer Space Telescope, which is operated by the Jet Propulsion Laboratory, California Institute of Technology under a contract with NASA.

\bibliographystyle{mn2e}
\bibliography{/home/efcl/Documents/library}

\appendix

\section[]{Photometry}

In this appendix we provide the multi-wavelength photometry for each of the objects in our sample.  In Table A1 we list the photometry for the eleven objects within the GOODS-S field and the photometry for the two objects within the UDS is listed in Table A2.

\begin{table*}
 \centering
  \caption{Photometry of the objects in GOODS-S field.  Column 1 gives the object ID while columns 2-10 give the AB magnitudes and corresponding errors in the following filters: the HST ACS $B_{435}$, $V_{606}$, $i_{775}$ and $z_{850}$-band filters,  HST WFC3 $J_{125}$ and  $H_{160}$ filters, the HAWK-I $K_{S}-$band and IRAC 3.6$\mu$m and 4.5$\mu$m band filters.  For the purposes of the SED fitting the photometric errors for the optical and near-infrared bands have been allocated a  minimum error of 10\%, and the IRAC fluxes have been allocated a  minimum error of 20\%.  For undetected objects in any of the bands, the 2$\sigma$ limiting magnitudes are reported. All magnitudes have been aperture corrected to the same enclosed flux provided by a 0.6\asec-diameter circular aperture at the spatial resolution of the HST WFC3 H$_{160}$ imaging. The aperture corrections needed to convert the photometry into total magnitudes (based on the $H_{160}$ measurements) are listed in column 11.}
  \scalebox{0.9}{
  \begin{tabular}{@{}ccccccccccc@{}}
  \hline
  \hline
  ID & $B_{435}$ & $V_{606}$ & $i_{775}$ & $z_{850}$ & $J_{125}$ & $H_{160}$ & $K_{s}$ & IRAC1 & IRAC2 & aperture\\
   &           &           &           &           &           &           &         &       &       & correction\\
\hline
2  & $>$28.44 & $>$28.68 & 27.38$\pm$0.30 & 26.04$\pm$0.10 & 25.89$\pm$0.10 & 26.02$\pm$0.10 & n/a            & 25.20$\pm$0.20 & 25.66$\pm$0.25 & 0.60\\
5  & $>$28.43 & $>$28.37 & 26.41$\pm$0.12 & 24.78$\pm$0.10 & 24.79$\pm$0.10 & 24.90$\pm$0.10 & n/a            & 24.38$\pm$0.20 & 24.57$\pm$0.20 & 0.20\\
6  & $>$28.53 & $>$28.78 & $>$28.15       & 26.18$\pm$0.10 & 26.50$\pm$0.10 & 26.62$\pm$0.13 & 27.22$\pm$0.94 & 25.90$\pm$0.20 & 25.37$\pm$0.20 & 0.32\\
8  & $>$28.65 & $>$29.13 & 26.83$\pm$0.15 & 25.43$\pm$0.10 & 25.45$\pm$0.10 & 25.53$\pm$0.10 & 25.49$\pm$0.29 & 24.63$\pm$0.20 & 24.77$\pm$0.20 & 0.31\\
13 & $>$28.62 & $>$28.80 & $>$27.53       & 26.63$\pm$0.16 & 26.53$\pm$0.13 & 26.14$\pm$0.10 & 26.22$\pm$0.30 & 25.72$\pm$0.20 & 25.68$\pm$0.20 & 0.25\\
18 & $>$28.84 & $>$28.65 & $>$28.29       & 26.30$\pm$0.10 & 26.94$\pm$0.18 & 26.96$\pm$0.26 & 27.48$\pm$1.18 & 26.36$\pm$0.20 & 26.90$\pm$0.47 & 0.20\\
23 & $>$28.08 & $>$28.51 & $>$27.83       & 26.37$\pm$0.14 & 25.54$\pm$0.10 & 25.69$\pm$0.10 & n/a            & 24.82$\pm$0.20 & 24.75$\pm$0.20 & 0.48\\
24 & $>$28.54 & $>$28.68 & 27.01$\pm$0.19 & 25.87$\pm$0.10 & 25.37$\pm$0.10 & 25.26$\pm$0.10 & 24.92$\pm$0.22 & 24.17$\pm$0.20 & 23.68$\pm$0.20 & 0.49\\
25 & $>$28.52 & $>$28.61 & $>$27.82       & 26.23$\pm$0.14 & 25.98$\pm$0.10 & 26.11$\pm$0.10 & 26.15$\pm$0.44 & 25.30$\pm$0.20 & 25.32$\pm$0.20 & 0.73\\
26 & $>$28.63 & $>$28.76 & $>$27.96       & 26.55$\pm$0.15 & 26.48$\pm$0.10 & 26.29$\pm$0.10 & n/a            & 25.70$\pm$0.37 & 25.82$\pm$0.41 & 0.24\\
27 & $>$28.10 & $>$29.16 & $>$27.89       & 26.14$\pm$0.10 & 26.22$\pm$0.10 & 26.22$\pm$0.10 & 26.22$\pm$0.54 & 25.69$\pm$0.20 & 25.86$\pm$0.20 & 0.41\\

  \hline
  \hline
\end{tabular}}
\end{table*}

\begin{table*}
 \centering
  \caption{Photometry of the objects in UDS field.  Column 1 gives the object ID while columns 2-11 give the AB magnitudes and corresponding errors in the following filters: the Subaru $B, V, R, i$ and $z$-band filters,  HST WFC3 $J_{125}$ and $H_{160}$ filters, the UDS WFCAM $K-$band and IRAC 3.6$\mu$m and 4.5$\mu$m band filters. 
For the purposes of the SED fitting the photometric errors for the optical and near-infrared bands have been allocated a  minimum error of 10\%, and the IRAC fluxes have been allocated a  minimum error of 20\%. For undetected objects in any of the bands, the 2$\sigma$ limiting magnitudes are reported. All magnitudes have been aperture corrected to the same enclosed flux provided by a 0.6\asec-diameter circular aperture at the spatial resolution of the HST WFC3 H$_{160}$ imaging. The aperture corrections needed to convert the photometry into total magnitudes (based on the $H_{160}$ measurements) are listed in column 12.}

  \scalebox{0.9}{
  \begin{tabular}{@{}cccccccccccc@{}}
  \hline
  \hline
  ID & B & V & R & $i$ & $z$ & $J_{125}$ & $H_{160}$ & $K$ & IRAC1 & IRAC2 & aperture\\
   &   &   &   &     &     &           &           &         &       &       & correction\\
\hline
157 & $>$28.78 & $>$28.00 & $>$28.31 & $>$28.24 & 25.05$\pm$0.10 & 25.31$\pm$0.10 & 25.55$\pm$0.10 & 25.63$\pm$0.53 & 24.69$\pm$0.20 & 25.52$\pm$0.28 & 0.21\\
248 & $>$28.78 & $>$28.00 & $>$28.31 & $>$28.04 & 25.33$\pm$0.10 & 25.47$\pm$0.10 & 25.79$\pm$0.10 & 25.85$\pm$0.60 & 24.46$\pm$0.20 & 24.94$\pm$0.27 & 0.20\\

  \hline
  \hline
\end{tabular}}
\end{table*}

\label{lastpage}

\end{document}